\documentclass[structabstract]{aa}  
\usepackage{graphicx}
\usepackage{natbib}
\usepackage{float}
\usepackage{amsmath}
\usepackage{rotating}
\usepackage{lscape}
\usepackage{color}
\usepackage{latexsym}
\usepackage{amsfonts}
\usepackage{multirow}
\usepackage{rotating}
\usepackage{amssymb,amsbsy}
\bibpunct{(}{)}{;}{a}{}{,}
\newcommand{\ie}{$i.e.,\;$}
\newcommand{\eg}{$e.g.,\;$}

\newcommand{\cf}{$cf.,\;$}

\usepackage{txfonts}
\begin{document}
    \title{Low-frequency radio observations of Seyfert galaxies: A test to the unification scheme}
%   \subtitle{A test to the unification scheme}

   \author{Veeresh Singh\inst{1,2,3}\fnmsep\thanks{veeresh.singh@ias.u-psud.fr}, Prajval Shastri\inst{2}, 
Ishwara-Chandra C. H.\inst{3} \and Ramana Athreya\inst{3,4}}
   \institute{Institut d'Astrophysique Spatiale, Bât. 121, Université Paris-Sud, 91405 Orsay Cedex, France   
         \and
            Indian Institute of Astrophysics, Bangalore - 560034, India
           \and
             National Center for Radio Astrophysics - TIFR, Pune, India
            \and
            Indian Institute of Science Education and Research, Pune, India}

   \date{Received xxxx xx, xxxx; accepted xxxx xx, xxxx}

% \abstract{}{}{}{}{} 
% 5 {} token are mandatory
 
  \abstract
{}
%    {To investigate the physical }
{We present low-frequency radio imaging and spectral properties of a well defined sample of Seyfert galaxies using GMRT 240/610 MHz dual 
frequency observations. Radio spectra of Seyfert galaxies over 240 MHz to 5.0 GHz are investigated using 240 MHz, 610 MHz flux densities derived from 
GMRT, and 1.4 GHz and 5.0 GHz flux densities mainly from published VLA data. We test the predictions of Seyfert unification scheme 
by comparing the radio properties of Seyfert type 1s and type 2s.}
{We choose a sample such that the two Seyferts subtypes have matched distributions in parameters that are independent to the orientation of AGN, 
obscuring torus and the host galaxy. 
Our sample selection criteria allow us to assume that the two Seyfert subtypes are intrinsically similar within the framework of the unification scheme.}
{The new observations at 240/610 MHz, together with archival observations at 1.4 GHz, 5.0 GHz show that type 1s and type 2s have statistically similar radio luminosity distributions at 
240 MHz, 610 MHz, 1.4 GHz and 5.0 GHz. 
The spectral indices at selected frequency intervals (${\alpha}_{\rm 240~MHz}^{\rm 610~MHz}$, ${\alpha}_{\rm 610~MHz}^{\rm 1.4~GHz}$ 
and ${\alpha}_{\rm 1.4~GHz}^{\rm 5.0~GHz}$) as well as index measured over 240 MHz to 5.0 GHz (${\alpha}_{\rm int}$) for the two Seyfert 
subtypes have similar distributions with median spectral index ($\alpha$) $\sim$ -0.7 (S$_{\nu}$ $\propto$ ${\nu}^{\alpha}$), 
consistent with the synchrotron emission from optically thin plasma.  
In our snap-shot 240/610 MHz GMRT observations, most of the Seyfert galaxies show primarily an unresolved central radio component, 
except a few sources in which faint kpc-scale extended emission is apparent at 610 MHz. 
Our results on the statistical comparison of the multifrequency radio properties of our sample Seyfert galaxies are in agreement with the predictions of 
the Seyfert unification scheme.}
    {}

   \keywords{galaxies: Seyfert -- galaxies: active -- radio continuum: galaxies}
\authorrunning{Singh et al.}
% \titilerunning{Low-frequency radio observations of Seyfert galaxies}

   \maketitle              

\section{Introduction}
Seyfert galaxies are categorized as low-luminosity (M$_{\rm B-Band}$ $>$ -23; \cite{Schmidt83}), radio-quiet 
($\frac{\rm F_{\rm 5.0~GHz}}{\rm F_{\rm B-Band}}$ $<$ 10; \cite{Kellermann89}) Active Galactic Nuclei (AGN), hosted in spiral or lenticular 
galaxies \citep{Weedman77}. Depending on the presence or absence of the broad permitted emission lines in their nuclear optical spectra, 
Seyfert galaxies are classified as `type~1' and `type~2', respectively. 
The detection of broad permitted emission lines in the spectropolarimetric observations of few Seyfert type 2s laid the foundation of Seyfert 
unification scheme \citep{Ant85}. The unification scheme of Seyfert galaxies hypothesizes that Seyfert type~1s and type~2s constitute the same parent 
population and appear different solely due to the differing orientations of the dusty molecular obscuring material 
having a torus-like geometry around the AGN. 
In Seyfert type 2s, the dusty torus intercepts the observer's line-of-sight and blocks the direct view of the broad line region and 
accreting black hole. While, in type 1s, observer's line-of-sight is away from the obscuring torus and 
therefore, broad line region and accreting black hole are directly visible \citep{Ant85,Ant93,Urry95}.
\\
In the literature, there have been various studies yielding consistent as well inconsistent results to the validity of the Seyfert unification scheme. 
Some of the key results that support the unification scheme include, the presence of broad emission lines in the polarized optical and 
infrared spectra of many Seyfert 2s \citep{Ant85,Moran2000}, the biconical structure of the narrow line region \citep{Mulchaey96}, 
the similar amount of total molecular gas detected by CO measurements in the two Seyfert subtypes \citep{Maiolino97}, the systematic higher X-ray absorption in type 2s 
\citep{Cappi06,Singh11}, and similar nuclear radio properties of both the subtypes \citep{Lal11}. 
However, inconsistencies to the unification scheme remain, such as the absence of hidden Seyfert 1 nuclei in several Seyfert 2s 
\citep{Tran01,Tran03}, higher tendency for Seyfert 1s to be hosted in galaxies of earlier Hubble type \citep{Malkan98}, 
the lack of X-ray absorption in some Seyfert 2s \citep{Panessa02}, and higher propensity of nuclear starbursts in Seyfert 2s 
\citep{Buchanan06}. 
It has been argued that the sample selection is the most crucial issue in testing the predictions of Seyfert unification scheme and the 
samples used in many previous studies suffer from subtle biases \citep{Antonucci02}.
There are suggestions that the optical and UV selected samples are likely to have inherent biases against obscured sources \citep{Ho01}. 
Infrared selected samples can be biased towards unusually dusty sources as well as towards sources with a higher level of nuclear star 
formation \citep{Ho01,Buchanan06}. X-ray selected samples from flux-limited surveys are likely to have obscured type 2 Seyferts that are 
intrinsically more luminous than the selected type 1 counterparts \citep{Heckman05,Wang09}. 
Recent studies on testing the unification scheme have emphasized and attempted to use less-biased samples and reported the results consistent 
with the scheme \citep{Cappi06,Dadina08,Beckmann09,Gallimore10,Lal11}. 
Nonetheless, issues related to sample selection remain and the quest to test the validity and limitations of the Seyfert unification with more improved 
and well-defined samples continues. 
Keeping above sample-selection arguments in mind, we attempt to test the predictions of unification scheme by using a Seyfert sample 
in which types 1s and 2s have matched distributions in parameters that are independent to the orientation of the obscuring torus and AGN axis. 
Our sample selection criteria help in mitigating the biases that are generally inherent in samples derived from flux-limited surveys and it 
also allow us to assume that the two Seyfert subtypes are not intrinsically different within the framework of the unification scheme \citep{Schmitt03a,Lal11}. 
\par
In this paper we attempt to investigate the low-frequency radio properties of the two Seyfert subtypes to test the predictions of unification scheme. 
High resolution radio observations show that Seyferts nuclei produce weak bipolar radio-emitting jets that 
are largely confined within the host galaxy \citep{Thean2000,Lal04}. 
However, at low-frequency significant radio emission may also arise from the extended emission related to either nuclear 
activity \citep{Colbert96,Gallimore06} or to the star-formation \citep{Baum93}. 
Notably, most of the Seyfert samples have been studied at higher frequencies ($\geq$ 1.4 GHz) with 
high resolution ($\sim$ arcsec or less) observations \citep{Ulvestad84a,Ulvestad84b,Ulvestad89,Kukula95,Morganti99,Nagar99,Thean2000,Thean01,Lal11}, 
which effectively filter out emission from low-surface-brightness and extended radio structures. 
In this paper we study the low-frequency radio emission properties of Seyfert galaxies using Giant Meterwave Radio 
Telescope (GMRT) dual frequency 240/610 MHz observations. 
Hitherto, there has been dearth of low-frequency ($<$ 1.0 GHz) radio observations of the sample of Seyfert 
galaxies and our observations are the first attempt, to our knowledge, to make a systematic study of the low-frequency radio properties of 
a well defined sample of Seyfert galaxies. To investigate the nature of multifrequency radio spectra we utilize 1.4 GHz, 5.0 GHz flux density 
measurements derived mainly from Very Large Array (VLA) `D' configuration observations that are sensitive to the low-surface-brightness radio emission.
\par
This paper is structured as follows. Our sample and its selection criteria are described in Section~2. 
The details of our observations and data reductions are described in Section 3. 
The comparison of radio properties of Seyfert type 1s and type 2s are discussed in Section 4. 
Radio properties of individual sources are given in Appendix A. Wherever required we assume cosmological parameters 
H$_{0}$ = 71 km$^{-1}$ Mpc$^{-1}$, ${\Omega}_{\rm m}$ = 0.27, and ${\Omega}_{\rm vac}$ = 0.73. 
\section{The sample}
Our sample consists of 20 Seyfert galaxies with 10 type~1s and 10 type~2s. 
All the sample sources satisfy basic criteria of being Seyfert galaxy {\ie}radio quiet 
($\frac{\rm F_{5.0~GHz}}{\rm F_{B-band}}$ $<$ 10) \citep{Kellermann89}, low optical luminosity (M$_{\rm B}$ $>$ -23) AGN \citep{Schmidt83}, 
nuclear line width (FWHM) of the permitted line H$_{\beta}$ is larger than 1000 km s$^{-1}$ for Seyfert type 1s \citep{Khachikian74}, 
line intensity ratio of [O~III] $\lambda$5007 {\AA} to H$_{\beta}$ is greater that 3.0 for Seyfert type 2s \citep{Dahari88} and sources are 
hosted in spiral or lenticular galaxies \citep{Weedman77}. Thus, we ensure that our Seyfert sample is not contaminated by LINERs, quasars and radio loud 
AGNs. Also, our sample sources fall in `Seyfert region' of [O~III]/O[~II] versus [O~I]/H$_{\alpha}$ diagnostic diagram of 
\cite{Kewley06}.   
Seyfert galaxies which show any broad permitted emission line component in their optical spectra ({\ie}subclasses 1.0, 1.2, 1.5, 1.8, 1.9) 
are considered as type~1 while those which show only narrow permitted emission lines are considered as type~2.
\\
Our sample selection criteria are based on the method proposed by \cite{Lal11}. An elaborate discussion on the sample selection issues and 
adopted methodology is presented in \cite{Lal11}. We choose our sample using isotropic properties that are independent to the orientation 
of the obscuring torus, AGN and host galaxy. 
The criterion of using isotropic orientation-independent parameter mitigates biases that are caused by obscuring torus 
or by the orientation of AGN jet-axis or host galaxy. It is to be noted that the inherent biases in Seyfert samples selected from optical, UV and X-ray 
surveys are interpreted due to the obscuration mainly from the torus \citep{Ho01,Heckman05}.
The criterion of matching the distributions of the two subtypes in the orientation-independent parameters, allows us to assume that two Seyfert 
subtypes in our sample are intrinsically similar within the framework of the unification scheme. 
Essentially, we ascertain that we are not comparing completely intrinsically different sources.
We consider five orientation-independent parameters {\ie}cosmological redshift, 
[O~III] $\lambda$5007$\mbox{\AA}$ luminosity, Hubble type of the host galaxy, 
total absolute stellar luminosity of the host galaxy and absolute bulge magnitude. 
These parameters are independent to the orientation of obscuring torus, AGN or the host galaxy and are also intimately link to 
the evolution of AGN and host galaxy. 
A brief description on the chosen orientation-independent parameters is given below.\\
(i) {\it Cosmological redshift}: it allows us to have control over cosmological evolution effect. 
In our sample the two Seyfert subtypes have similar distributions of cosmological redshift spanning over a narrow interval.  
This implies that the two Seyfert subtypes in our sample belong to similar cosmological epoch.
\\
(ii) {\it [O~III] $\lambda$5007$\mbox{\AA}$ line luminosity}: 
it originates from the narrow line region which is outside the torus and therefore, is not affected by the torus obscuration. 
Also, [O~III] $\lambda$5007$\mbox{\AA}$ luminosity  is expected to be correlated with nuclear ionizing continuum 
as well as nuclear X-ray luminosity and can be considered as the proxy for intrinsic AGN power \citep{Nelson95,Heckman05}. 
In our sample the two subtypes are chosen such that they have similar narrow distributions of [O~III] luminosity. 
This allows us to assume that the AGN powers of the two subtypes are matched.
% In our sample the two subtypes are chosen such that they have similar distributions of [O~III] luminosity which in turn indicates 
% that AGN power of the two subtypes are matched.
\\
(iii) {\it Hubble type of the host galaxy}: it takes into account for the effect of the host galaxy morphology 
and its environment on AGN evolution and vice-versa. Hubble type of a galaxy does not depend on the orientation of the torus or 
AGN-jet axis \citep{Pringle99}. 
In our sample two Seyfert subtypes have similar distributions of Hubble type, 
albeit, matching is less strong for Hubble stage (T) values higher than 2. 
The values of Hubble stage for our sample sources are taken from \cite{Malkan98} and \cite{de-Vaucouleurs91}. 
% We preferred Hubble stage values derived using HST observations in \cite{Malkan98}, over values given in \cite{de-Vaucouleurs91}, whenever available}.
\\
(iv) {\it Total stellar absolute magnitude of the host galaxy}: it can be considered as a characteristic property of the host galaxy. 
The total stellar absolute magnitudes of the host galaxies of our sample sources are taken from \citep{Whittle92} and these are corrected for 
non-stellar and emission line flux, redshift (K) correction, the internal absorption and the Galactic absorption. 
\\
(v) {\it Absolute magnitude of the bulge}: it is independent to the orientation of obscuring torus, AGN axis. 
Also, absolute magnitude of bulge is roughly proportional to the mass of the central supermassive black hole 
\citep{Kormendy01,McConnell13}, arguably a fundamental parameter of the AGN system. 
The values of absolute magnitude of bulge of our sample sources are taken from \cite{Whittle92}. 
\par
We chose our sample of 20 Seyfert galaxies (10 type 1s and 10 type 2s) such that the two subtypes have matched distributions in all 
the five orientation independent parameters. 
Our sample is extracted from \cite{Whittle92} sample of 140 Seyfert galaxies (78 type 1s and 62 type 2s). To ensure the bona fide nature of Seyfert type 
we selected only those sources that host in Spiral or lenticular galaxies {\ie}Hubble type S0 or later \citep{Weedman77}. 
% The values of Hubble types given in \cite{Whittle92} were updated whenever newer data were available ({\eg}\cite{Malkan98}). 
After excluding sources with early type, 
peculiar or uncertain host galaxy morphology we obtained a sample of 92 Seyferts (47 type 1s and 45 type 2s). 
In order to minimize the effects of obscuration by the host galaxy disk on the optical properties we selected Seyferts with relatively face-on 
host galaxies ({\ie}ratio of minor to major isophotal diameter axis greater than 0.5), noting the fact that 
there is no correlation between AGN and host galaxy axis \citep{Pringle99,Nagar99a}. It resulted the sample of 76 (41 type 1s and 35 type 2s) Seyfert galaxies. 
Of these 76 sources, 7 (2 type 1s and 5 type 2s) sources had declination (Dec $\leq$ - 53$^{\circ}$) beyond the GMRT coverage and were excluded. 
In the sample of 69 (39 type 1s and 30 type 2s) sources we further imposed redshift cutoff z $\geq$ 0.031, as per our selection criteria of 
choosing sources from narrow span of cosmological redshift. The redshift cut off rendered the sample of 49 (23 type 1s and 26 type 2s) sources. 
To ensure that source is easily detected above 6$\sigma$ in a ten minute snap-shot scan with GMRT we picked sources which had 1.4 GHz 
NVSS\footnote{http://www.cv.nrao.edu/nvss/}/FIRST\footnote{http://sundog.stsci.edu/index.html} flux density higher than 6.0 mJy {\ie}corresponding extrapolated 610 MHz flux density assuming a less steep radio spectral index of -0.5, 
higher than 9.0 mJy. This yielded a sample of 41 (19 type 1s and 22 type 2s) Seyfert galaxies. 
From these 41 Seyferts we picked 20 Seyferts (10 type 1s and 10 type 2s) that satisfied our selection criteria {\ie}matched distributions of 
the two subtypes in the orientation independent parameters and could be observed in our limited telescope observing time.
All our sample sources comply to the observing feasibility with GMRT {\ie}sources are within the GMRT declination coverage range and  
are detectable with sufficient signal-to-noise ratio in a ten-minute snap shot with GMRT, there is no strong radio source in the neighborhood that 
can impact the dynamical range {\ie}flux density of target source.
Table 1 lists our sample sources and the values of their orientation-independent parameters. 
Redshift values are taken from NASA Extragalactic Database (NED) and are rounded off to the fourth place of decimal. 
The values of [O~III] luminosity, absolute stellar magnitude and absolute bulge magnitude are taken from \cite{Whittle92}. 
A quality factor has been assigned to all the parameters in \cite{Whittle92} and it reflects the level of reliability. 
We have used only those parameter values that have reliable quality rating {\ie}`a' to `c' in \cite{Whittle92} catalog. 
We opted for updated values of parameters, whenever available in the literature, for example, [O~III] luminosity and Hubble stage values 
obtained from HST observations \citep{Schmitt03a,Malkan98} were preferred over values given in \cite{Whittle92}.
HST observations of [O~III] emission reported in \cite{Schmitt03a} yield spatially resolved [O~III] emission associated with NLR regions. 
A comparison between [O~III] fluxes obtained from HST and ground based observations shows that both kinds of observations give similar 
[O~III] flux measurements (Figure 2 in \cite{Schmitt03a}) and therefore, in general, contamination to [O~III] emission by HII regions is likely to be 
not significant in Seyfert galaxies.
The presence of circumnuclear starburst may affect the measurements of [O~III] flux, bulge and stellar magnitudes. 
However, most of our sample sources are free from circumnuclear starburst contamination, except NGC 3227 and NGC 7469 
that are known to posses circumnuclear starburst \citep{Gonz97,Genzel95}. 
Thus, our sample selection is not much affected by the circmnuclear starburst contamination.

% The bulge magnitude has been derived from total galaxy magnitude and bulge-to-disk ratio that depends on Hubble type of the host galaxy. 
% The observational error associated with bulge-to-disk ratio can impact the estimate of bulge magnitude. 
% 
Figure 1 shows the matched distributions of the orientation-independent parameters for the two Seyfert subtypes of our sample.
The matched distributions allow us to assume that we are not comparing entirely intrinsically different
sources selected from different parts of the (luminosity, bulge mass, Hubble type, redshift) evolution function \citep{Schmitt03b}.
Indeed, following the same sample selection criteria there is possibility to obtain a larger sample, however, 
we would like to emphasize that the more important is the sample selection and not the sample size to rigorously test the predictions of 
the unification scheme. Larger but heterogeneous and biased sample is likely to result incorrect conclusions. 
The relatively small size of our sample is the result of the combined effect of restrictive selection criteria and observational constraints. 
And, in future, we plan to extend our analysis to a larger sample.
\begin{table*} 
%\small
% \tiny
\centering
\begin{minipage}{140mm}
\caption{Our Seyfert sample}
%\vspace {0.2cm} ,
\resizebox{14.0cm}{!}{\begin{tabular} {@{}cccccccc@{}}
% \resizebox{14.0cm}{6.8cm}{\begin{tabular} {@{}cccccccccc@{}}
\hline
Source &  RA   &  Dec    & Redshift & log L$_{\rm [O~III]}$ &   Hubble  &  M$_{\rm B_{(Total)}}$    & M$_{\rm B_{(Bulge)}}$  \\ 
Name   & (h m s) & (d m s)& (z)    & (erg s$^{-1}$) & Stage (T) &         &          \\ \hline
\multicolumn{8}{c}{\it Seyfert 1s} \\
MRK 6      & 06 52 12.2 & +74 25 37    & 0.0188$^{7}$ & 41.79$^{2}$ & 0$^{5}$ &  -20.30$^{1}$ &  -19.44$^{1}$    \\
NGC 3227   & 10 23 30.6 & +19 51 54    & 0.0039$^{7}$ & 40.31$^{1}$ & 1$^{6}$ &  -21.47$^{1}$ &  -20.46$^{1}$    \\
NGC 3516   & 11 06 47.5 & +72 34 07    & 0.0088$^{7}$ & 41.07$^{2}$ & 0$^{5}$ &  -21.61$^{1}$ &  -20.88$^{1}$   \\
NGC 4151   & 12 10 32.6 & +39 24 21    & 0.0033$^{7}$ & 41.35$^{1}$ & 2$^{6}$ &  -21.22$^{1}$ &  -19.98$^{1}$     \\ 
MRK 766    & 12 18 26.5 & +29 48 46    & 0.0129$^{7}$ & 41.61$^{2}$ & 3$^{5}$ &  -21.03$^{1}$ &  -20.10$^{1}$     \\ 
MRK 279    & 13 53 03.4 & +69 18 30    & 0.0305$^{7}$ & 41.46$^{1}$ & 1$^{5}$ &  -21.59$^{1}$ &  -20.98$^{1}$   \\
NGC 5548   & 14 17 59.5 & +25 08 12    & 0.0172$^{7}$ & 41.42$^{2}$ & 1$^{5}$ &  -21.82$^{1}$ &  -20.89$^{1}$  \\
ARK 564     & 22 42 39.3 & +29 43 31   & 0.0247$^{7}$ & 41.38$^{1}$ & 1$^{6}$ &  -21.65$^{1}$ &  -20.11$^{1}$     \\ 
NGC 7469    & 23 03 15.6 & +08 52 26   & 0.0163$^{7}$ & 41.51$^{1}$ & 4$^{5}$ &  -22.01$^{1}$ &  -20.90$^{1}$     \\ 
MRK 530     & 23 18 56.6 & +00 14 38   & 0.0295$^{7}$ & 40.98$^{1}$ & 1$^{5}$ &  -22.74$^{1}$ &  -21.19$^{1}$     \\ 
\multicolumn{8}{c}{\it Seyfert 2s} \\
MRK 348     & 00 48 47.1 & +31 57 25   & 0.0150$^{7}$  & 41.31$^{2}$ & 0$^{5}$ & -21.13$^{1}$ &  -20.20$^{1}$     \\ 
MRK 1       & 01 16 07.2 & +33 05 22   & 0.0159$^{7}$  & 41.52$^{1}$ & 5$^{5}$ & -20.32$^{1}$ &  -19.46$^{1}$     \\ 
MRK 1066    & 02 59 58.6 & +36 49 14   & 0.0120$^{7}$  & 40.88$^{1}$ & 5$^{5}$ & -21.06$^{1}$ &  -20.45$^{1}$   \\
NGC 2110    & 05 52 11.4 & -07 27 22   & 0.0078$^{7}$  & 40.35$^{1}$ & 1$^{5}$ & -21.57$^{1}$ &  -20.72$^{1}$     \\
NGC 2273    & 06 50 08.6 & +60 50 45   & 0.0061$^{7}$  & 40.09$^{3}$ & 1$^{6}$ & -20.99$^{1}$ &  -19.97$^{1}$     \\ 
NGC 5252    & 13 38 15.9 & +04 32 33   & 0.0230$^{7}$  & 41.41$^{4}$ & 0$^{5}$ & -21.96$^{1}$ &  -21.35$^{1}$  \\
NGC 5728    & 14 42 23.9 & -17 15 11   & 0.0094$^{7}$  & 41.11$^{1}$ & 1$^{6}$ & -22.35$^{1}$ &  -21.12$^{1}$  \\
NGC 7212    & 22 07 01.3 & +10 13 52   & 0.0266$^{7}$  & 42.15$^{2}$ & 1$^{6}$ & -21.24$^{1}$ &  -20.22$^{1}$     \\ 
NGC 7682    & 23 29 03.9 & +03 32 00   & 0.0171$^{7}$  & 41.16$^{1}$ & 0$^{5}$ & -21.11$^{1}$ &  -19.88$^{1}$     \\ 
MRK 533     & 23 27 56.7 & +08 46 45   & 0.0289$^{7}$  & 41.99$^{2}$ & 5$^{5}$ & -22.65$^{1}$ &  -20.69$^{1}$    \\     \hline
\end{tabular}} 
%\label{table:nolin} (is used to refer this table in the text "table:nolin")
%\vspace {0.2cm} 
% \justify
% \\
\footnotesize{{\bf Notes.} Column 1: source name; Columns 2 and 3: right ascension (hours, minutes, and seconds) and declination 
(degrees, arcminutes, and arcseconds) in J2000; Column 4: cosmological redshift; Column 5: [O~III] $\lambda$5007$\mbox{\AA}$ luminosity in log; 
Column 6: Hubble Stage (T); Column 7: total stellar absolute magnitude (M$_{\rm B_{Total}}$) corrected for the nuclear non-stellar continuum 
and emission line flux, redshift correction (K), the internal absorption and the Galactic absorption; Column 8: absolute bulge magnitude in `B' band 
(M$_{\rm B_{Bulge}}$).\\
{\bf References.} (1) \cite{Whittle92}; (2) \cite{Schmitt03a}; (3) \cite{Ferruit2000}; (4) \cite{Polletta96}; (5) \cite{Malkan98}; 
(6) \cite{de-Vaucouleurs91}; (7) NASA Extragalactic Database (NED).}
\end{minipage}
\end{table*}
\begin{figure*}
\centering
\includegraphics[angle=0,width=6.0cm,height=0.18\textheight]{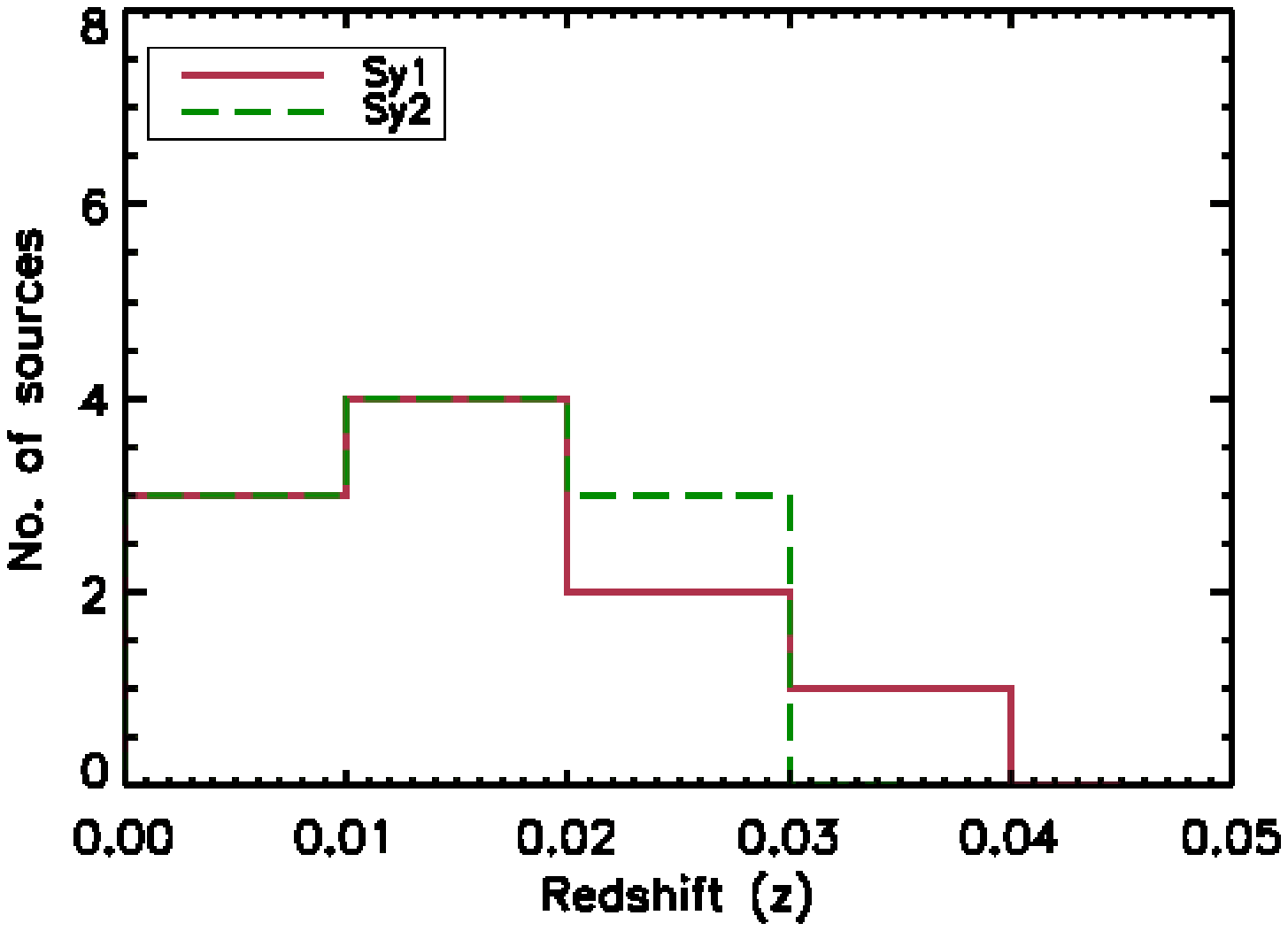}{\includegraphics[angle=0,width=6.0cm,height=0.18\textheight]{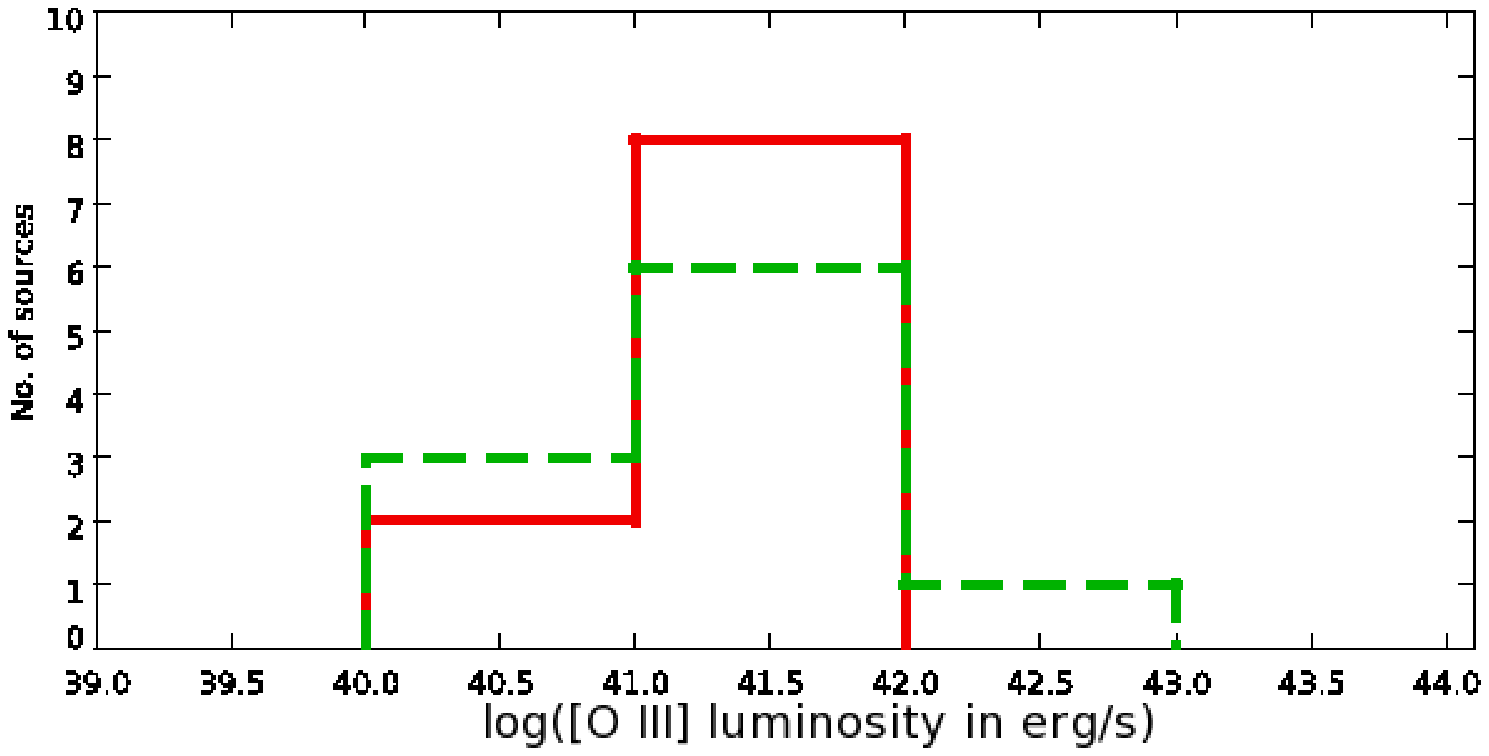}}
\includegraphics[angle=0,width=6.0cm,height=0.18\textheight]{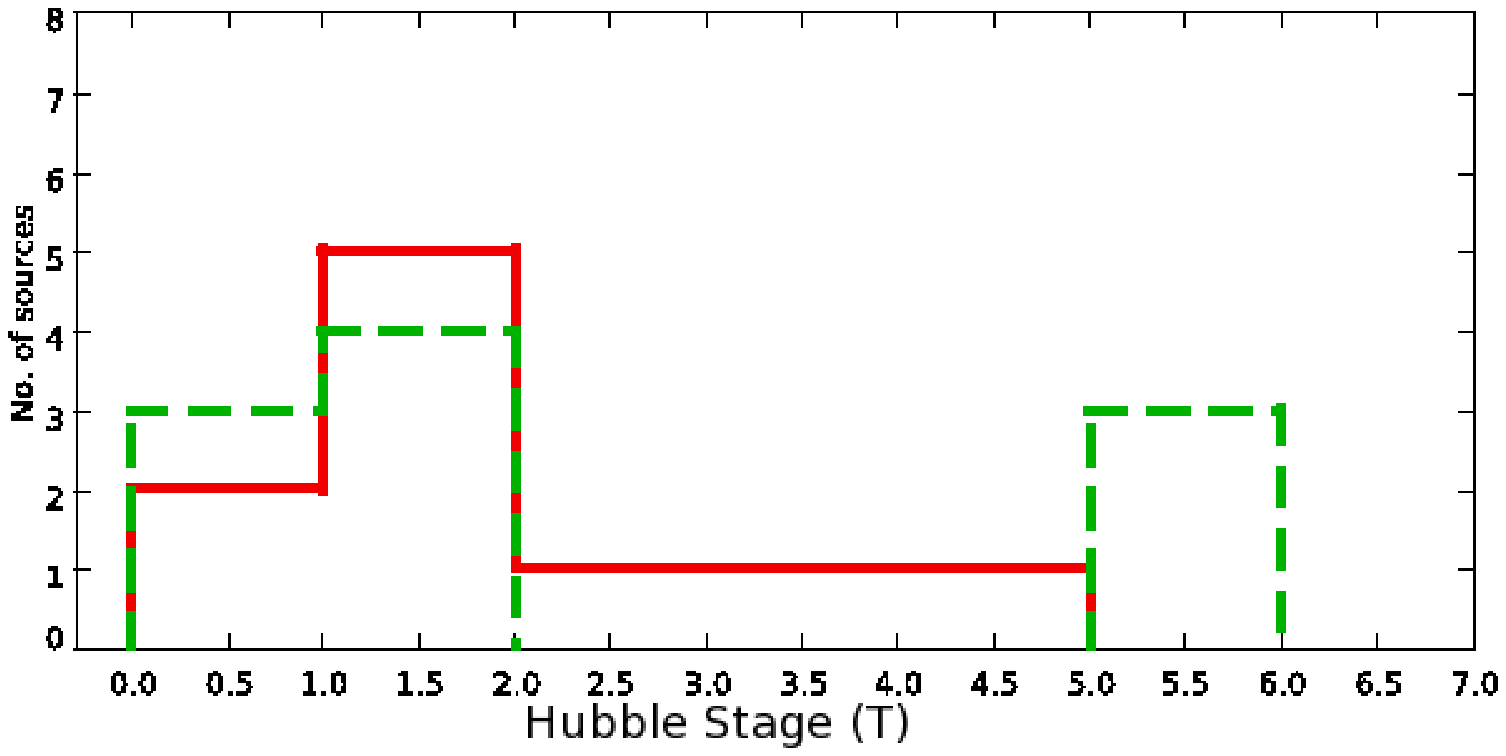}{\includegraphics[angle=0,width=6.0cm,height=0.18\textheight]{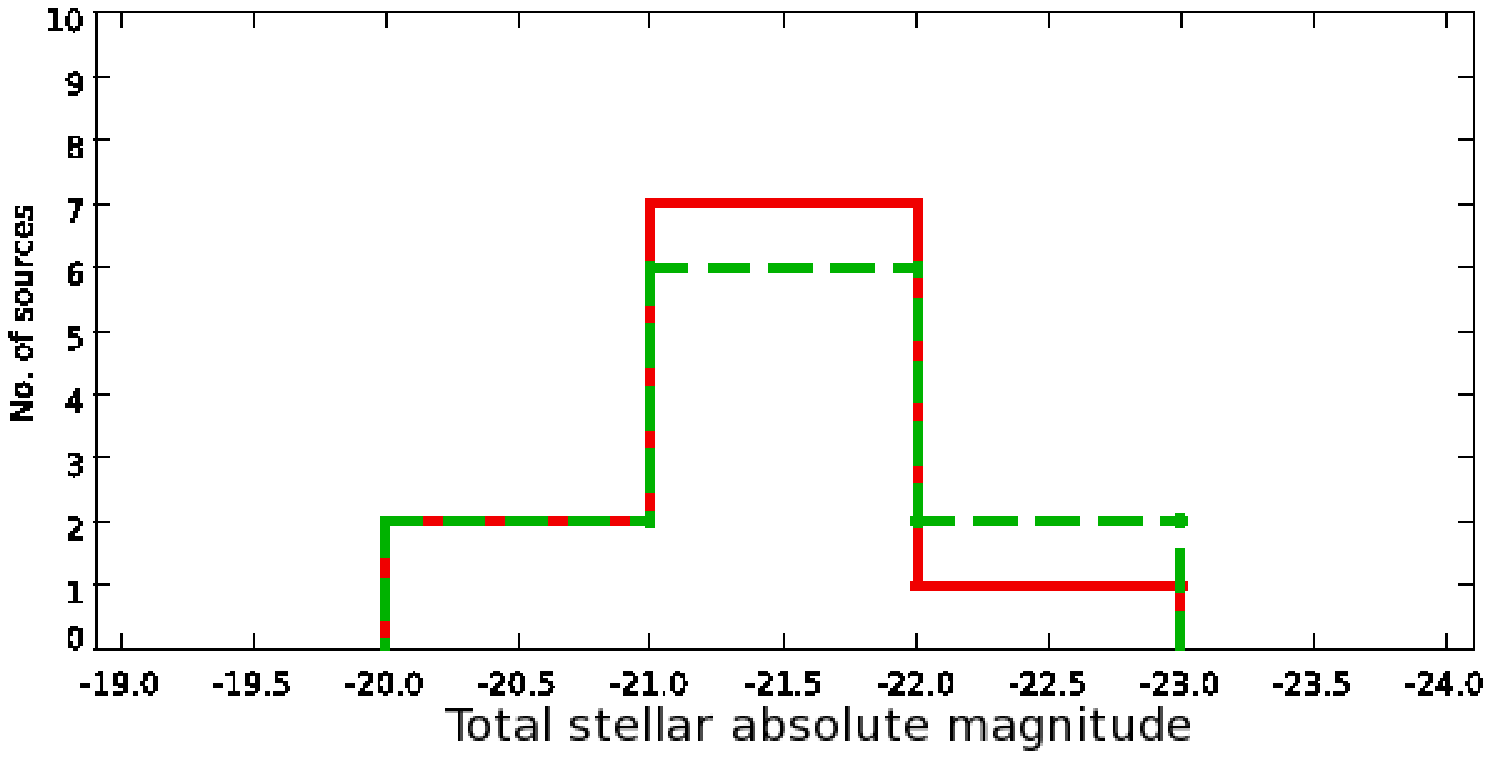}}
{\includegraphics[angle=0,width=6.0cm,height=0.18\textheight]{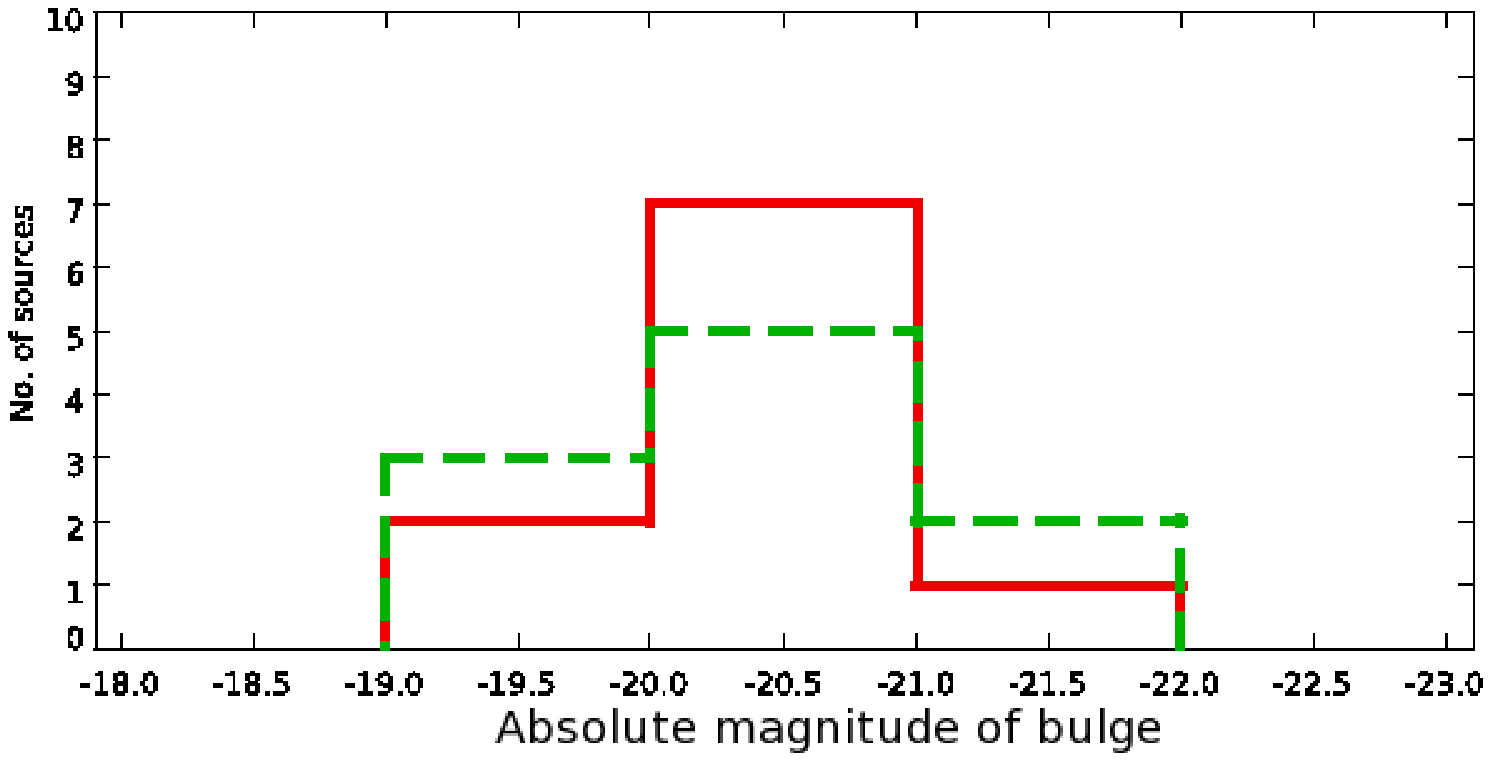}}
\caption{Histograms showing matched distributions of Seyfert type 1s and type 2s in redshift, [O~III] $\lambda$5007$\mbox{\AA}$ luminosity, Hubble type of the host galaxy, total stellar absolute magnitude 
of the host galaxy and absolute magnitude of the bulge. The histograms for type 1s and type 2s are plotted with `Red colored solid lines' 
and `Green colored dashed lines', respectively.}
\end{figure*}
\section{Observations and Data Reduction}
We carried out full array GMRT \citep{Swarup91} snapshot observations at dual frequency 610/240 MHz 
using 4 second integration time and bandwidth of 32 MHz. 
Our observing log is given in Table 2.
All the 20 Seyfert galaxies are observed with 2 - 4 scans (except NGC 7469 with only one scan and NGC 5548 with 5 scans) 
with each scan spanning to 10 minutes. We gave more observing scans to weaker radio sources and less to stronger ones by 
using 1.4 GHz NVSS flux densities as the indicator of source radio strength.
Absolute flux and bandpass calibration is done by observing standard 
flux calibrators 3C~147 and 3C~286, at the start and end of the observing run. 
The phase calibration is done by observing a nearby phase calibrator source for $\sim$ 4 minutes, before and 
after every scan of the target source. 
Our data are reduced in a standard way using `Astronomical Image Processing System' (AIPS)\footnote{http://www.aips.nrao.edu} package. 
For each run, bad visibility points are edited out, after which the data are calibrated. 
The edited and calibrated visibilities are Fourier transformed into radio maps using `IMAGR' task in AIPS. 
We performed wide field imaging as the primary beam size of full array GMRT at 610/240 MHz is rather 
large $\sim$ 43$^{\prime}$/114$^{\prime}$. 
For all our sample sources the signal-to-noise ratio is sufficiently high enough to apply self-calibration 
which removes antenna-based phase and amplitude errors. 
\begin{table}
\centering
\begin{minipage}{140mm}
\caption{GMRT observational log}
\small
% \tiny
% \hspace{2cm}
\begin{tabular}{@{}cccc@{}}
\hline
Source     &  No.   & Phase  & Obs.  \\ 
name       &  of scans     & Cal & date              \\ \hline
\multicolumn{4}{c}{\it Seyfert 1s} \\
MRK 6      &   3    &  0614+607   &  2008 August 09, 10  \\
NGC 3227   &   3    &  1111+199   &  2008 August 09    \\
NGC 3516   &   4    &  1313+675   &  2008 August 09, 10   \\
NGC 4151   &   2    &  1331+305   &  2008 August 09    \\
MRK 766    &   3    &  1331+305   &  2008 August 09, 10  \\
MRK 279    &   4    &  1313+675   &  2008 August 09, 10    \\
NGC 5548   &   5    &  1331+305   &  2008 August 09, 10 \\
ARK 564    &   3    &  2236+284   &  2008 August 10     \\
NGC 7469   &   1    &  2212+018   &  2008 August 10   \\
MRK 530    &   4    &  2212+018   &  2008 August 10  \\
\multicolumn{4}{c}{\it Seyfert 2s} \\
MRK 348    &   3    &  0137+331   &  2008 August 10   \\
MRK 1      &   3    &  0137+331   &  2008 August 10\\
MRK 1066   &   3    &  0137+331   &  2008 August 10 \\ 
NGC 2110   &   2    &  0607-085   &  2008 August 10 \\
NGC 2273   &   3    &  0614+607   &  2008 August 09, 10     \\
NGC 5252   &   4    &  1351-148   &  2008 August 09, 10  \\
NGC 5728   &   2    &  1351-148   &  2008 August 09 \\
NGC 7212   &   2    &  2212+018   &  2008 August 10 \\
NGC 7682   &   2    &  2212+018   &  2008 August 10\\
MRK 533    &   2    &  2212+018   &  2008 August 10  \\ \hline
\end{tabular}
\end{minipage}
\end{table}
\begin{table*}
\centering
\begin{minipage}{140mm}
\caption{610 MHz radio image parameters}
\small
% \tiny
\begin{tabular}{@{}cccccccccc@{}}
\hline
            &  \multicolumn{4}{c}{Map parameter} & \multicolumn{5}{c}{Source parameter}  \\
Sources     &  Scale & Beam size & P.A. &  rms noise   & S$_{\rm Peak}$ & S$_{\rm Int.}$ & \multicolumn{3}{c}{Fitted size} \\ 
name       &  (kpc/$^{\prime\prime}$) & (arcsec$^2$) & (deg)  & (mJy/b) &   (mJy/b)    &   (mJy)   & Maj.  & Min.  & P.A.  \\ 
           &             &                   &        &        &               &       &  (arcsec)    &  (arcsec)   &  (deg) \\ \hline
\multicolumn{10}{c}{\it Seyfert 1s} \\
MRK 6      &     0.377   & 10.20$\times$4.95 & -28.38 &  1.65  &  476.2        & 502   & 10.42& 5.06& 150.8          \\
NGC 3227   &     0.079   & 8.03$\times$6.44  & -65.60 &  2.10  &  130.1        & 148   & 8.85 & 6.70& 119.4            \\
NGC 3516   &     0.179   & 7.51$\times$4.91  & -12.96 &  1.20  &  13.5         & 25    & 10.44 & 7.38& 4.0             \\
NGC 4151   &     0.068   & 7.79$\times$5.04  & -25.02 &  3.50  & 312.9 & 341.5 & 8.12 & 5.28& 147.7            \\
           &             &                   &        &        & 218.1 & 264.5 & 8.30 & 5.74& 162.1      \\
MRK 766    &     0.261   & 7.52$\times$4.63  & -25.68 &  1.15  &  53.3         &  58.0  & 7.92 & 4.81& 150.0             \\
MRK 279    &     0.601   & 7.47$\times$5.27  &  19.68 &  0.90  &  39.6         &  44    & 7.86 & 5.54& 19.0              \\
NGC 5548   &     0.344   & 6.31$\times$5.19  & -10.90 &  0.95  &  19.7         &  56    & 14.33& 7.11& 158.5             \\
ARK 564    &     0.491   & 7.68$\times$5.16  & -45.09 &  1.00  &  61.0         &  62.5  & 7.62 & 4.91& 132.3          \\
NGC 7469   &     0.328   & 9.67$\times$5.36  & -24.77 &  1.15  &  218.5        &  292   & 10.43& 6.70 &154.7           \\
MRK 530    &     0.584   & 7.18$\times$6.85  &  35.11 &  0.60  &  37.1         &  40    & 7.28 & 7.18& 97.0             \\
\multicolumn{10}{c}{\it Seyfert 2s} \\
MRK 348    &     0.302   & 6.28$\times$5.69  & -14.78 &  2.00  &  480.9        &  499   & 6.34 & 5.81 & 157.0             \\
MRK 1      &     0.320   & 6.27$\times$5.19  &  0.87  &  0.75  &  109.8        &  115   & 6.31 & 5.32& 177.3            \\
MRK 1066   &     0.243   & 6.56$\times$5.21  &  19.39 &  1.20  &  166.8        &  190   & 6.67 & 5.63 & 18.9             \\ 
NGC 2110   &     0.158   & 6.88$\times$5.09  &  5.93  &  1.20  &  494.4        &  554   & 7.68 & 5.11 & 9.0         \\
NGC 2273   &     0.125   & 7.93$\times$4.77  & -29.88 &  0.80  &  85.9         &  102   & 8.32 & 5.34 & 149.6           \\
NGC 5252   &     0.458   & 10.68$\times$5.12 & -22.24 &  0.70  &  18.9         & 21     & 10.56 & 5.66 & 159.6         \\
NGC 5728   &     0.189   & 11.59$\times$5.98 & -22.67 &  0.75  & 25.6  & 34.3  & 12.60 & 7.46 & 154.7             \\
           &             &                   &        &        & 17.5  & 39.4  & 16.10 & 9.39 & 159.9      \\
NGC 7212   &     0.528   & 8.32$\times$4.84  & -21.85 &  0.45  & 189.4         & 192   & 8.40 & 4.85 & 158.5         \\
NGC 7682   &     0.344   & 10.20$\times$6.41 & -23.64  & 0.55  & 92.3         & 98     & 10.21 & 6.82 & 150.8       \\
MRK 533    &     0.572   & 6.80$\times$6.15  &  46.30 &  1.20  & 340.7         & 364   & 6.98 & 6.26 & 44.1            \\
\hline
\end{tabular}
\end{minipage}
\end{table*}
\begin{table*}
\centering
\begin{minipage}{140mm}
\caption{240 MHz radio image parameters}
\small
%\tiny
\begin{tabular}{@{}cccccccccc@{}}
\hline
           &  \multicolumn{4}{c}{Map parameter} & \multicolumn{5}{c}{Source parameter} \\
Source     &  Scale & Beam size &  P.A.  &  rms noise & S$_{\rm Peak}$ & S$_{\rm Int.}$   & \multicolumn{3}{c}{Fitted size}    \\ 
name       &  (kpc/$^{\prime\prime}$) & (arcsec$^2$) & (deg)    & (mJy/b) &   (mJy/b)    &  (mJy)   & Maj.  & Min.  & P.A.   \\ 
          &             &                   &        &        &               &       &  (arcsec)    &  (arcsec)   &  (deg) \\ \hline
\multicolumn{10}{c}{\it Seyfert 1s}        \\
MRK 6      &   0.377   & 23.04$\times$11.36 & -22.04 & 3.0   &   853.2        & 944.6  & 24.14 & 11.76& 158.8   \\
NGC 3227   &   0.079   & 27.46$\times$14.27 & -50.31 & 5.6   &   117.0        & 170.0  & 32.61 & 18.02 & 125.8 \\
NGC 3516   &   0.179   & 16.26$\times$11.44 & -15.03 & 3.8   & 36.9  & 81     & 37.09 & 14.31 & 25.82         \\
NGC 4151   &   0.068   & 17.27$\times$11.68 & -14.00 &  12.0 &  1324.4       &  1348  & 17.10 & 11.64& 165.9  \\
MRK 766    &   0.261   & 18.13$\times$12.75 & -18.95 &  7.5  &   74.8        &  85    & 16.20 & 9.90 & 152.5 \\
MRK 279    &   0.601   & 16.09$\times$10.75 &  28.19 &  3.5  &  80.3         &  96    & 17.69 & 11.52& 31.7  \\
NGC 5548   &   0.344   & 21.65$\times$15.17 & -30.72 &  6.5  &    77.6       &  83    & 23.54 & 14.83 & 154.0 \\
ARK 564    &   0.491   & 25.99$\times$13.81 & -45.7  &  4.4  & 164.4         & 166    & 24.29 & 12.92 & 131.97 \\
NGC 7469   &   0.328   & 38.01$\times$22.06 & -1.01  &  9.5  &   793.5       &  807   & 37.86 & 20.68 & 177.1 \\
MRK 530    &   0.584   & 38.13$\times$18.73 & -17.42 &  9.0  &   107.3       &  130   & 46.92 & 21.46 & 156.4 \\
\multicolumn{10}{c}{\it Seyfert 2s}        \\
MRK 348    &   0.302   & 18.52$\times$12.88 & -35.86 &  5.6  &  763.4        &  810   & 18.96& 12.91 & 148.2   \\
MRK 1      &   0.320   & 18.14$\times$12.64 & -22.12 &  5.5  &  181.9        &  186   & 17.57& 12.36& 163.6    \\
MRK 1066   &   0.243   & 16.89$\times$13.25 & -1.87  &  6.0  &  288.6        &  300   & 17.25& 13.53& 177.0    \\
NGC 2110   &   0.158   & 43.63$\times$34.50 & -0.18  &  25.0  &  1498.5       &  1499 & 38.85 & 32.47 & 1.9    \\
NGC 2273   &   0.125   & 19.04$\times$11.06 &  -26.39 & 4.0   &  113.8        &  127  & 20.84 & 11.53 & 152.3 \\
NGC 5252   &   0.458   & 38.21$\times$12.37 &  -25.16 & 3.0  & 19.4         &  26.4   & 46.20 & 19.31 & 152.5  \\
NGC 5728   &   0.189   & 27.42$\times$12.94 &  -19.36&  3.0 &  22.8         &  40     & 37.62 & 19.20 & 131.6  \\
NGC 7212   &   0.528   & 40.16$\times$29.61 & -1.95  &  7.5  &    523.1      &    551 & 32.85 & 25.79 & 175.4 \\
NGC 7682   &   0.344   & 46.13$\times$34.28 & 0.87   &  7.5  &    227.9      &    228 & 38.68 & 29.23 & 17.14  \\
MRK 533    &   0.572   & 17.87$\times$16.98 &-33.83  &  9.0  &  639.2        &   842  & 20.01 & 19.04 & 66.3 \\
\hline
\end{tabular}
\end{minipage}
\end{table*}
% 
% \newpage
\begin{table*}
% \centering
\begin{minipage}{140mm}
\caption{Radio luminosities and spectral indices}
% \small
\tiny
\begin{tabular}{@{}ccccccccccc@{}}
\hline
Source     &  L$_{\rm 240 MHz}$ & L$_{\rm 610 MHz}$ &  S$_{\rm 1.4 GHz}$ & L$_{\rm 1.4 GHz}$ & S$_{\rm 5.0 GHz}$ & L$_{\rm 5.0 GHz}$ 
& ${\alpha}^{\rm 610 MHz}_{\rm 240 MHz}$ & ${\alpha}^{\rm 1.4 GHz}_{\rm 610 MHz}$ & ${\alpha}^{\rm 5.0 GHz}_{\rm 1.4 GHz}$ & ${\alpha}_{\rm int.}$ \\ 
name       &  (erg s$^{-1}$ Hz$^{-1}$) & (erg s$^{-1}$ Hz$^{-1}$) & (mJy) & (erg s$^{-1}$ Hz$^{-1}$) & (mJy) & (erg s$^{-1}$ Hz$^{-1}$) &     &     &    \\ \hline
\multicolumn{11}{c}{\it Seyfert 1s}        \\
MRK 6      & 7.87$\times$10$^{30}$ & 4.18$\times$10$^{30}$ & 269.5  & 2.24$\times$10$^{30}$ & 100.5$^{\rm K}$ & 8.37$\times$10$^{29}$ & -0.68 & -0.75 & -0.77 & -0.74  \\
NGC 3227   & 6.09$\times$10$^{28}$ & 5.30$\times$10$^{28}$ & 97.5   & 3.49$\times$10$^{28}$ & 35.0$^{\rm Ga}$ & 1.25$\times$10$^{28}$ & -0.15 & -0.50 & -0.80 & -0.53  \\
NGC 3516   & 1.48$\times$10$^{29}$ & 4.58$\times$10$^{28}$ & 31.3   & 5.74$\times$10$^{28}$ & 7.4$^{\rm Ga}$  & 1.36$\times$10$^{28}$ & -1.26 & +0.27& -1.13 &  -0.71  \\
NGC 4151   & 2.71$\times$10$^{29}$ & 1.22$\times$10$^{29}$ & 359.6  & 7.24$\times$10$^{28}$ & 128.0$^{\rm Ga}$& 2.58$\times$10$^{28}$ & -0.86 & -0.63 & -0.81 & -0.76  \\
MRK 766    & 3.28$\times$10$^{29}$ & 2.24$\times$10$^{29}$ & 38.1   & 1.47$\times$10$^{29}$ & 20.4$^{\rm Ga}$ & 7.86$\times$10$^{28}$ & -0.41 & -0.51 & -0.49 & -0.47  \\
MRK 279    & 2.03$\times$10$^{30}$ & 9.28$\times$10$^{29}$ & 23.2   & 4.89$\times$10$^{29}$ & 7.4$^{\rm E}$   & 1.56$\times$10$^{29}$ & -0.84 & -0.77 & -0.90 & -0.84  \\
NGC 5548   & 5.51$\times$10$^{29}$ & 3.72$\times$10$^{29}$ & 28.2   & 1.87$\times$10$^{29}$ & 11.2$^{\rm Ga}$ & 7.44$\times$10$^{28}$ & -0.42 & -0.83 & -0.73 & -0.68  \\
ARK 564    & 2.41$\times$10$^{30}$ & 9.09$\times$10$^{29}$ & 28.6   & 4.16$\times$10$^{29}$ & 11.4$^{\rm L}$ &  1.66$\times$10$^{29}$ & -1.05 & -0.94 & -0.72 & -0.88  \\
NGC 7469   & 4.74$\times$10$^{30}$ & 1.71$\times$10$^{30}$ & 180.5  & 1.06$\times$10$^{30}$ & 61.6$^{\rm Ga}$ & 3.62$\times$10$^{29}$ & -1.09 & -0.58 & -0.84 & -0.82  \\
MRK 530    & 2.56$\times$10$^{30}$ & 7.88$\times$10$^{29}$ & 24.4   & 4.80$\times$10$^{29}$ & 11.5$^{\rm E}$  & 2.26$\times$10$^{29}$ & -1.26 & -0.59 & -0.59 & -0.77  \\ 
\multicolumn{11}{c}{\it Seyfert 2s}        \\
MRK 348    & 4.17$\times$10$^{30}$ & 2.57$\times$10$^{30}$ & 292.2  & 1.51$\times$10$^{30}$ & 801.7$^{\rm Ga}$ & 4.13$\times$10$^{29}$ & -0.52 & -0.64 & +0.79 & -0.58  \\
MRK 1      & 1.09$\times$10$^{30}$ & 6.75$\times$10$^{29}$ & 75.4   & 4.43$\times$10$^{29}$ & 32.0$^{\rm GC}$  & 1.88$\times$10$^{29}$ & -0.52 & -0.51 & -0.67  & -0.58 \\
MRK 1066   & 9.83$\times$10$^{29}$ & 6.23$\times$10$^{29}$ & 100.4  & 3.29$\times$10$^{29}$ & 35.0$^{\rm GC}$  & 1.15$\times$10$^{29}$ & -0.49 & -0.77 & -0.83  & -0.72 \\
NGC 2110   & 2.17$\times$10$^{30}$ & 8.01$\times$10$^{29}$ & 298.8  & 4.32$\times$10$^{29}$ & 165.0$^{\rm Gr}$ & 2.38$\times$10$^{29}$ & -1.07 & -0.74 & -0.47  & -0.72 \\
NGC 2273   & 1.03$\times$10$^{29}$ & 8.26$\times$10$^{28}$ & 62.6   & 5.07$\times$10$^{28}$ & 44.0$^{\rm GC}$  & 3.56$\times$10$^{28}$& -0.24 & -0.59 & -0.28   & -0.37 \\
NGC 5252   & 3.31$\times$10$^{29}$ & 2.58$\times$10$^{29}$ & 16.3   & 2.00$\times$10$^{29}$ & 18.1$^{\rm E}$   & 2.22$\times$10$^{29}$& -0.27 & -0.30 & +0.08   &  -0.14 \\
NGC 5728   & 7.33$\times$10$^{28}$ & 1.36$\times$10$^{29}$ & 70.0   & 1.28$\times$10$^{29}$ & 17.5$^{\rm S}$   & 3.09$\times$10$^{28}$& +0.66 & -0.07 & -1.09   &  .... \\
NGC 7212   & 9.37$\times$10$^{30}$ & 3.27$\times$10$^{30}$ & 128.0$^{\rm W}$ & 2.18$\times$10$^{30}$ & 46.0$^{\rm Gr}$  & 7.83$\times$10$^{29}$& -1.13 & -0.49 & -0.80   & -0.79 \\
NGC 7682   & 1.51$\times$10$^{30}$ & 6.51$\times$10$^{29}$ & 59.8   & 3.97$\times$10$^{29}$ & 24.6$^{\rm E}$   & 1.63$\times$10$^{29}$& -0.91 & -0.59 & -0.70   & -0.72 \\
MRK 533    & 1.66$\times$10$^{31}$ & 7.17$\times$10$^{30}$ & 220.9  & 4.35$\times$10$^{30}$ & 75.1$^{\rm E}$   & 1.48$\times$10$^{30}$& -0.90 & -0.60 & -0.85   & -0.78 \\ \hline
\end{tabular}
% \vspace{0.05cm}\\
\footnotesize{{\bf Notes.} 1.4 GHz flux densities are from NVSS catalog \citep{Condon98} except for NGC 7212 for which NVSS data are unavailable. \\
{\bf References.} (K) \cite{Kharb06}; (E) \cite{Edelson87}; (Ga) \cite{Gallimore06}; (GC) \cite{Gregory91}; (Gr) \cite{Griffith95}; (L) \cite{Lal11}; (S) \cite{Schommer88}, 
(W) \cite{White92} .}
\end{minipage}
\end{table*}
\begin{figure*}[!htbp]
%\epsscale{0.8}
\centering
\includegraphics[angle=0,width=6.8cm,height=0.26\textheight]{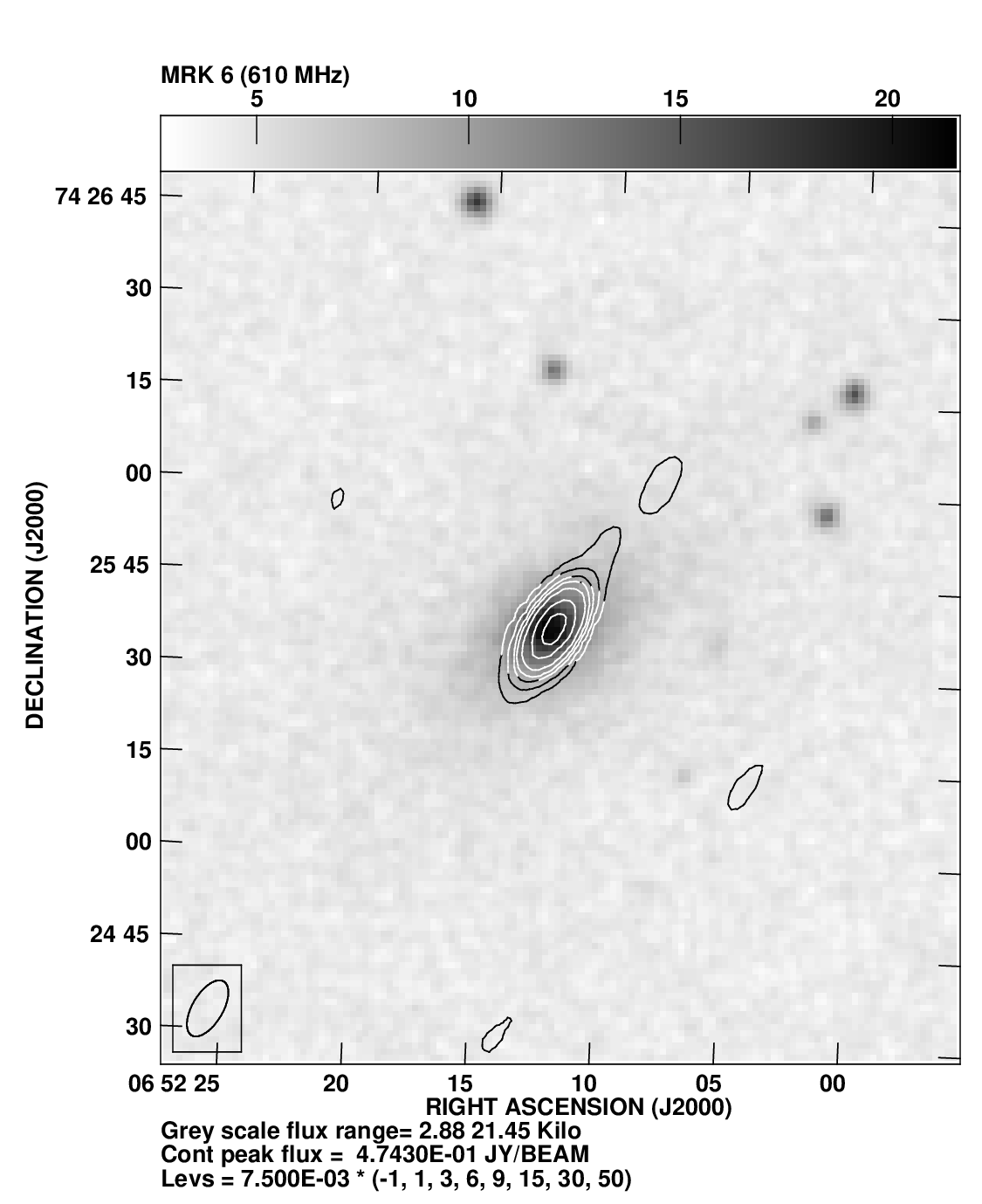}{\includegraphics[angle=0,width=6.8cm,height=0.26\textheight]{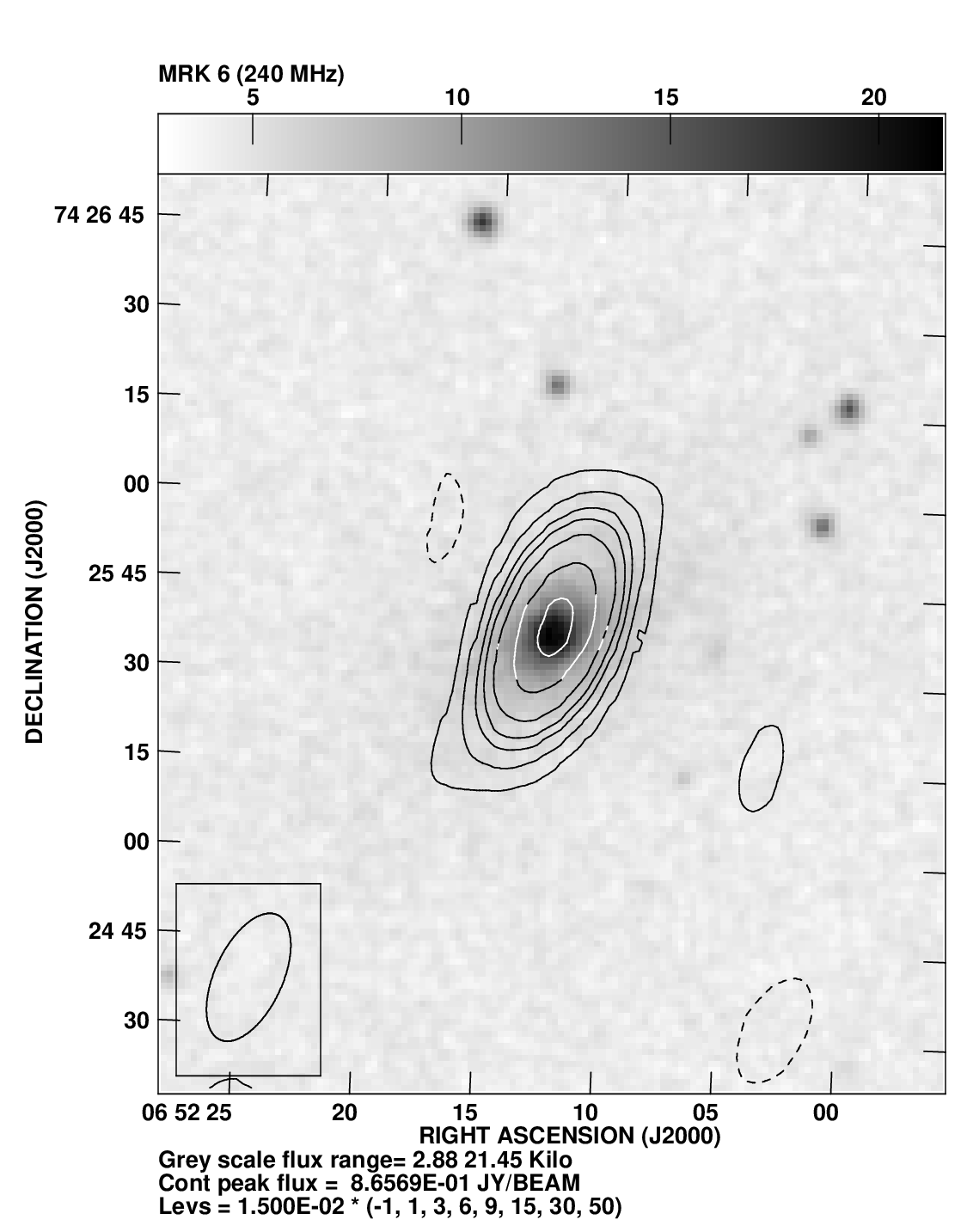}}
\includegraphics[angle=0,width=6.8cm,height=0.26\textheight]{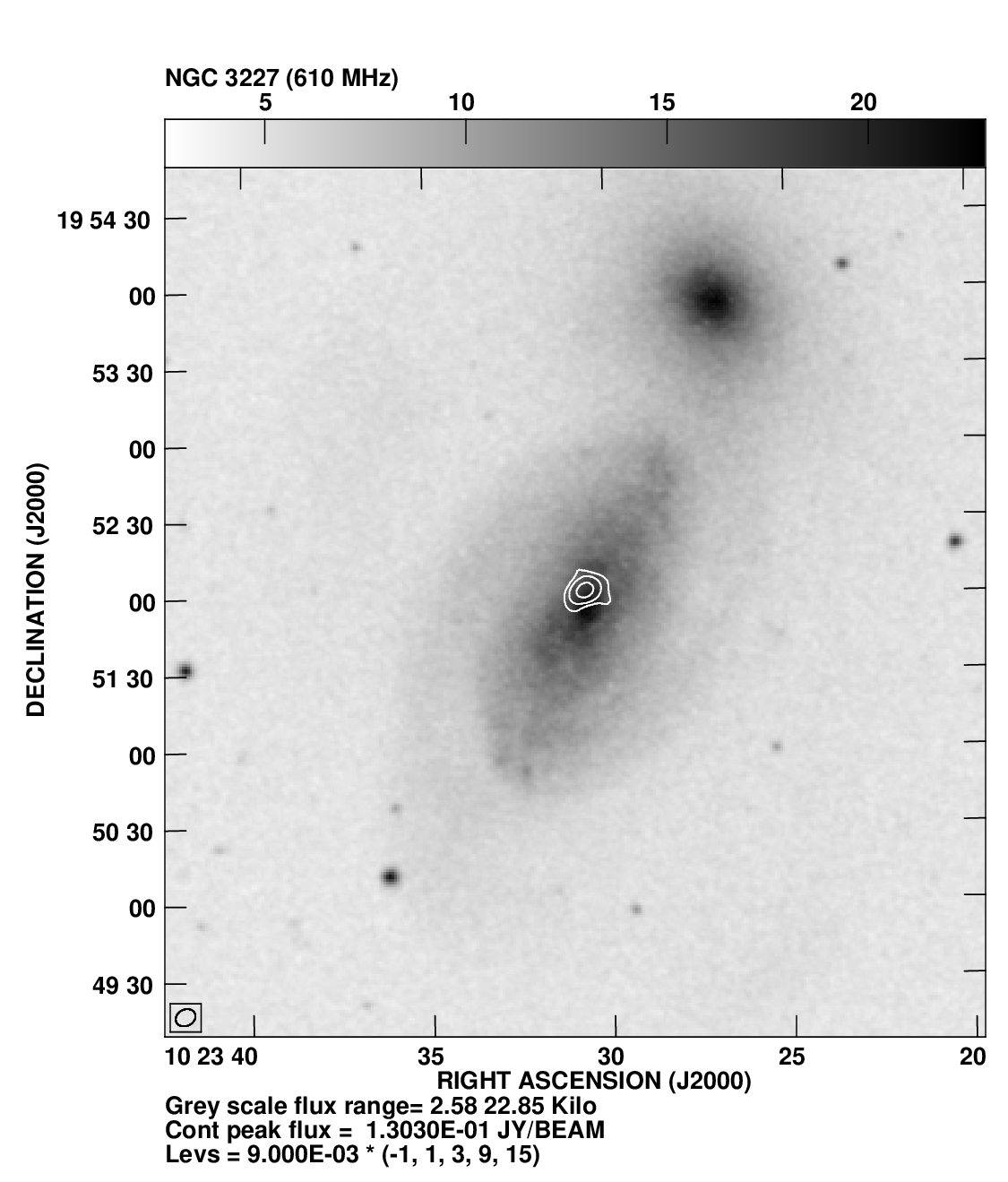}{\includegraphics[angle=0,width=6.8cm,height=0.26\textheight]{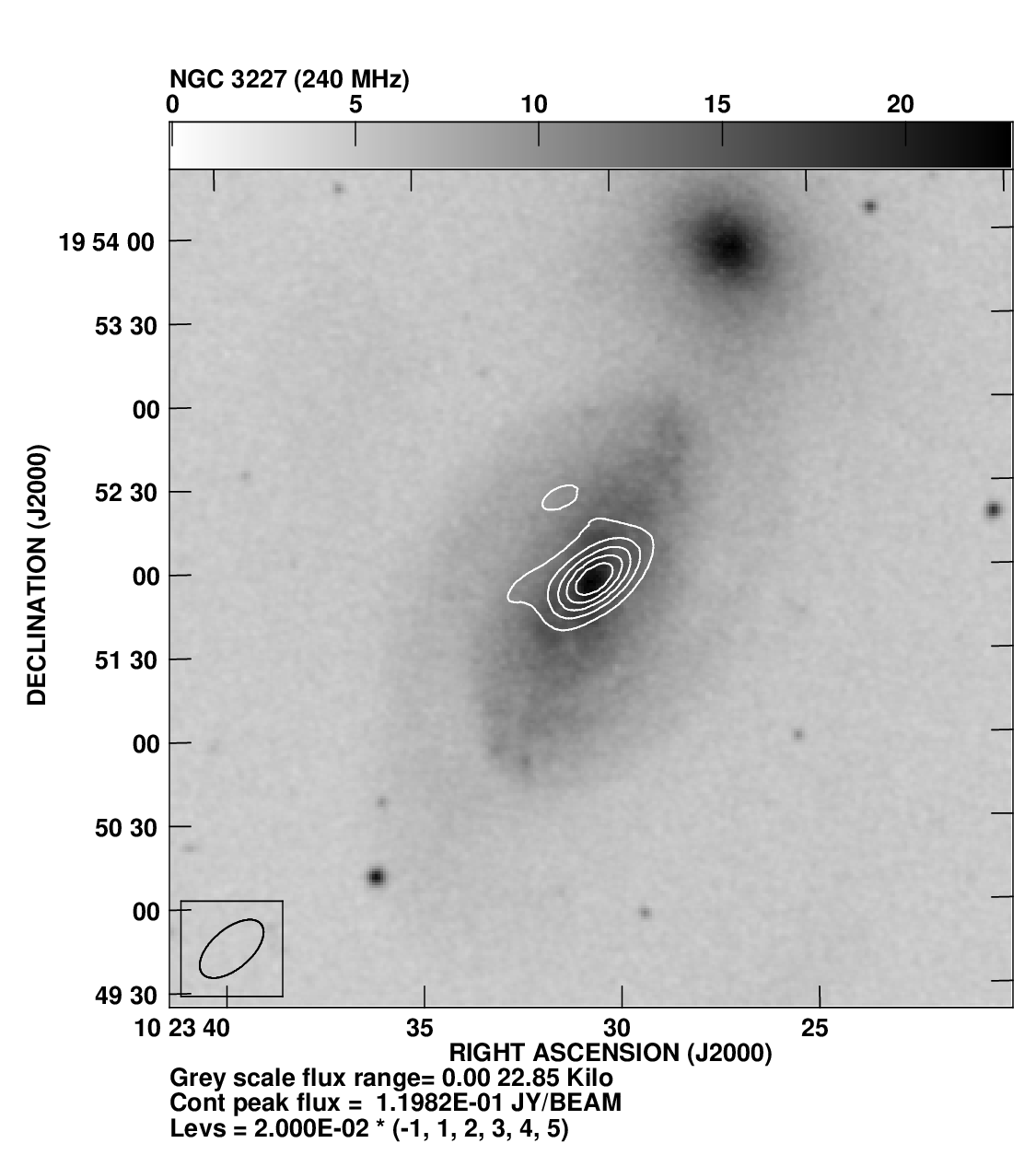}}
\includegraphics[angle=0,width=6.8cm,height=0.26\textheight]{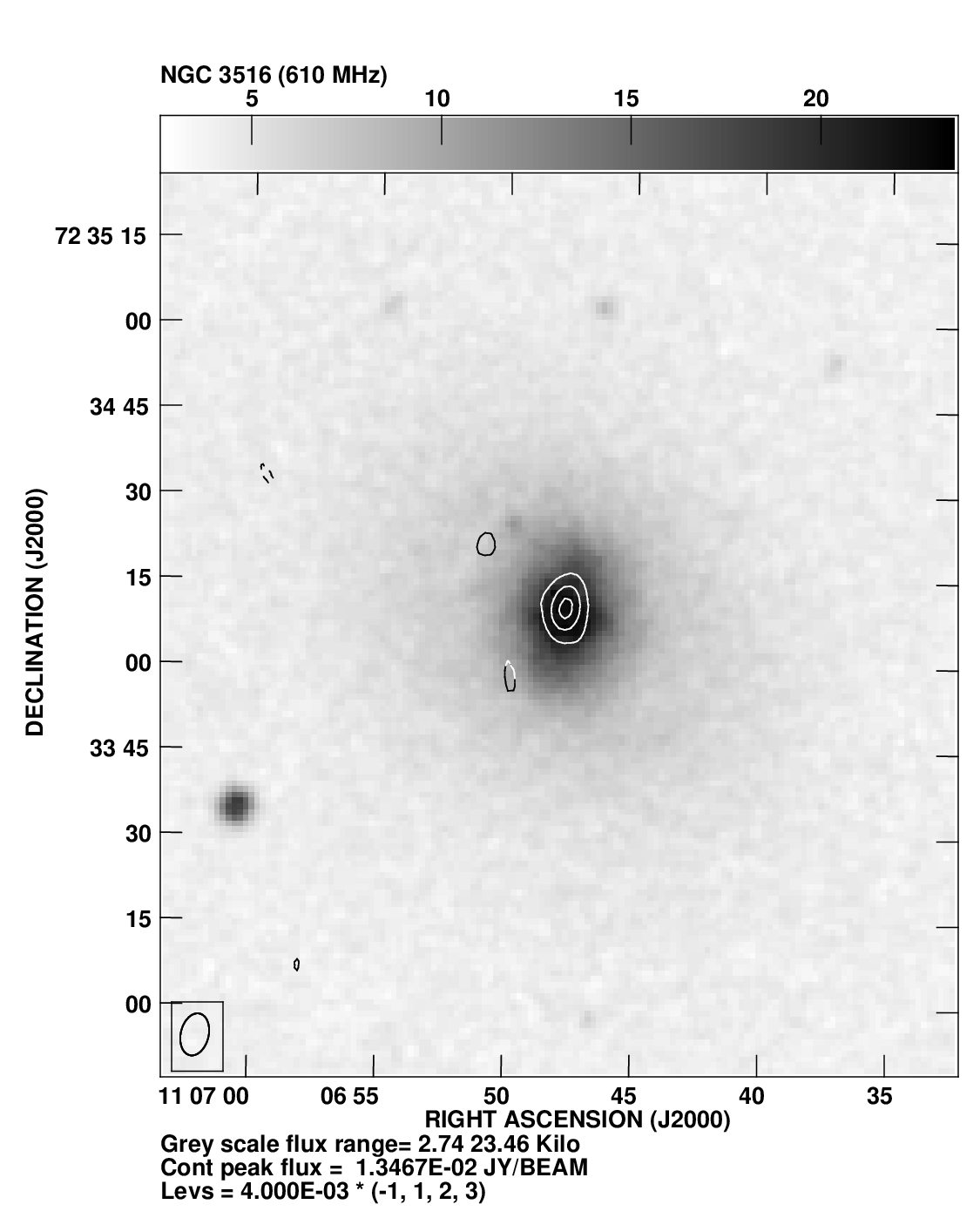}{\includegraphics[angle=0,width=6.8cm,height=0.26\textheight]{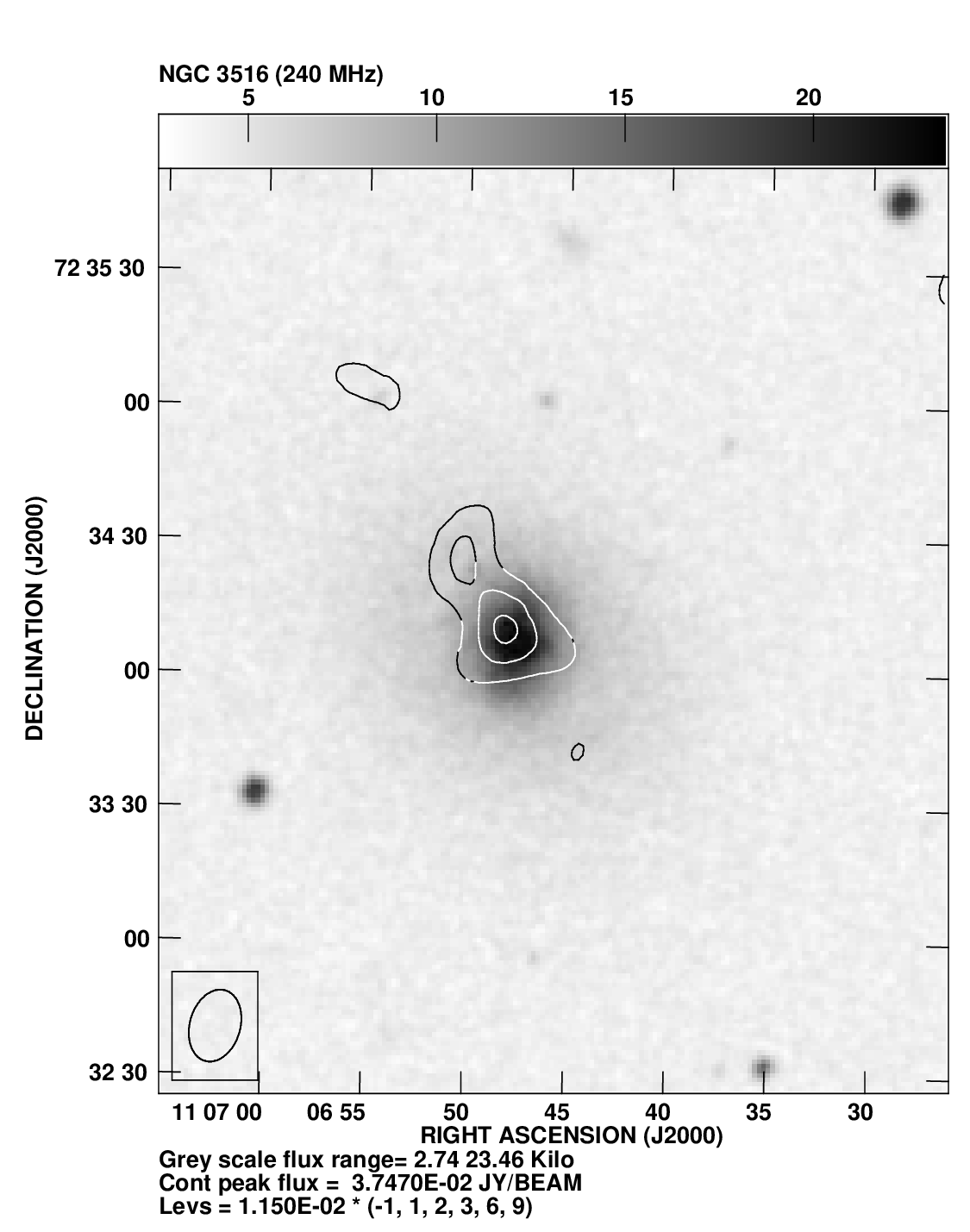}}
\includegraphics[angle=0,width=6.8cm,height=0.26\textheight]{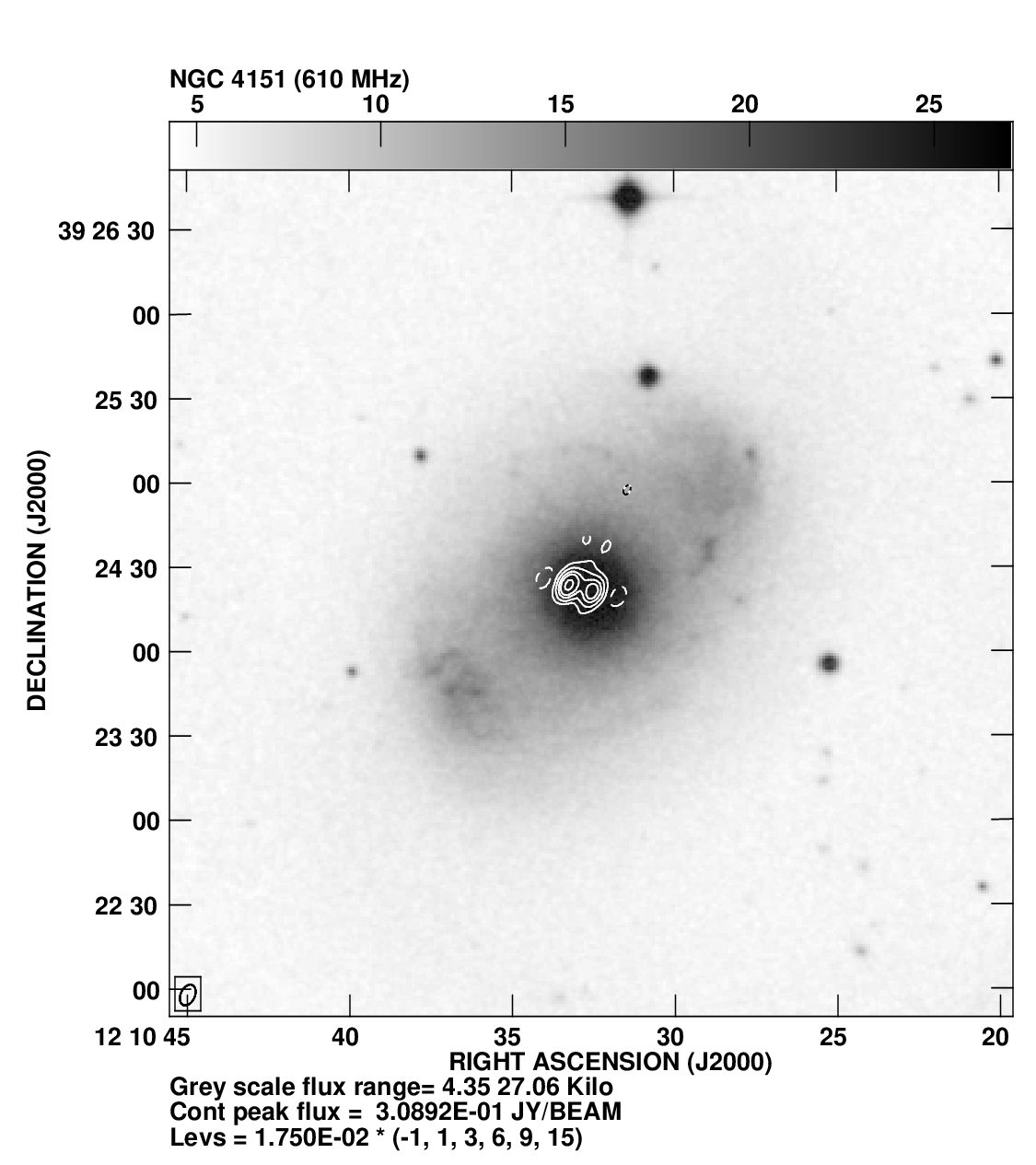}{\includegraphics[angle=0,width=6.8cm,height=0.26\textheight]{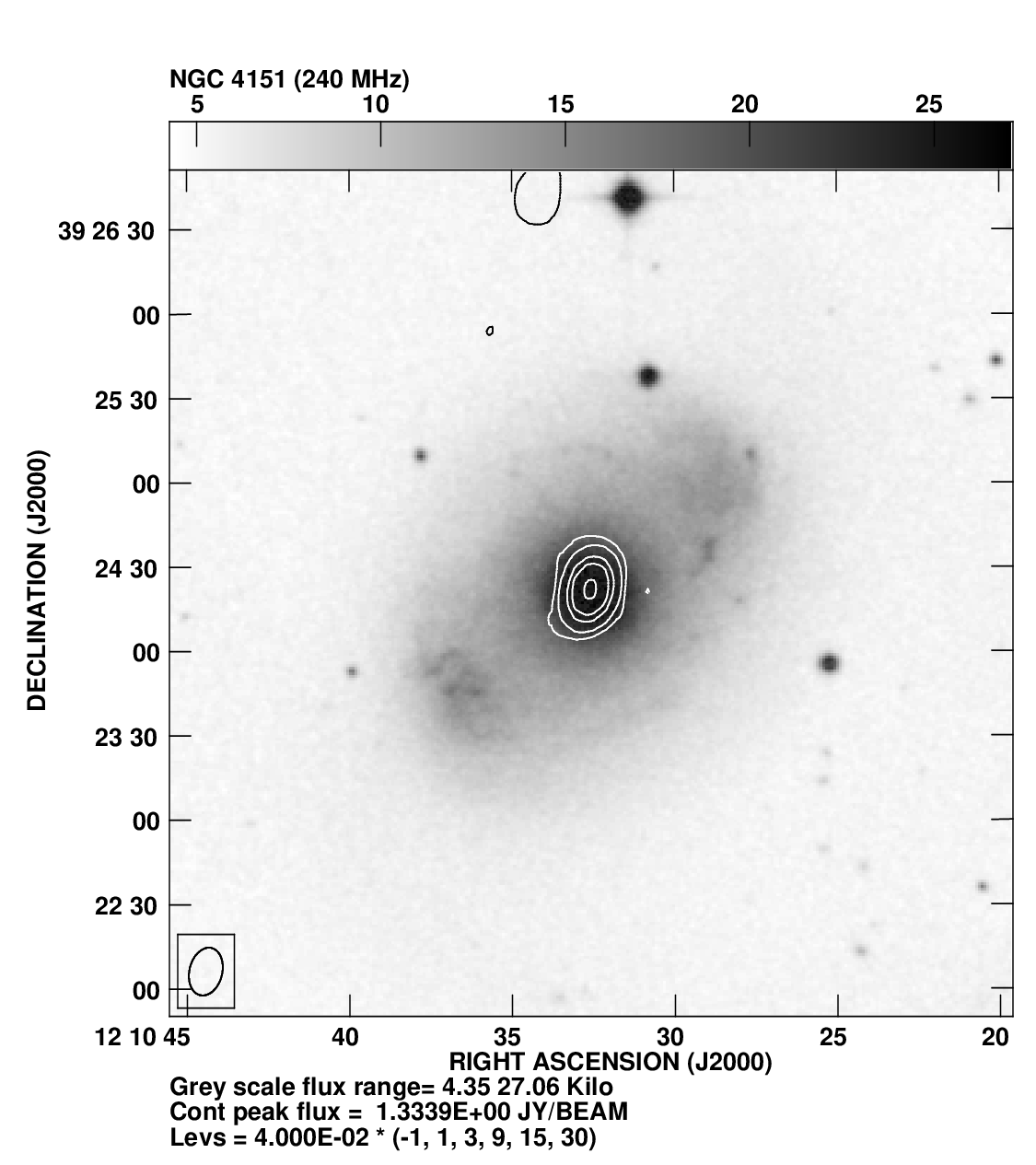}}
\caption{610 MHz (left panel) and 240 MHz (right panel) radio contours overlaid on their DSS optical images. 
The restoring beam is shown in lower left corner of each map. The contour levels are shown at the bottom of each map. 
The first lowest radio contour is about 4$\sigma$ $-$ 5$\sigma$ of the rms noise value in each map. 
The source name and radio frequency is mentioned at the top left of each map. The same plotting convention is followed for other sources.}
\end{figure*}
\addtocounter{figure}{-1}
\begin{figure*}[!htbp]
\centering
\includegraphics[angle=0,width=6.8cm,height=0.26\textheight]{M766+610RADIOCONTDSS.PS}{\includegraphics[angle=0,width=6.8cm,height=0.26\textheight]{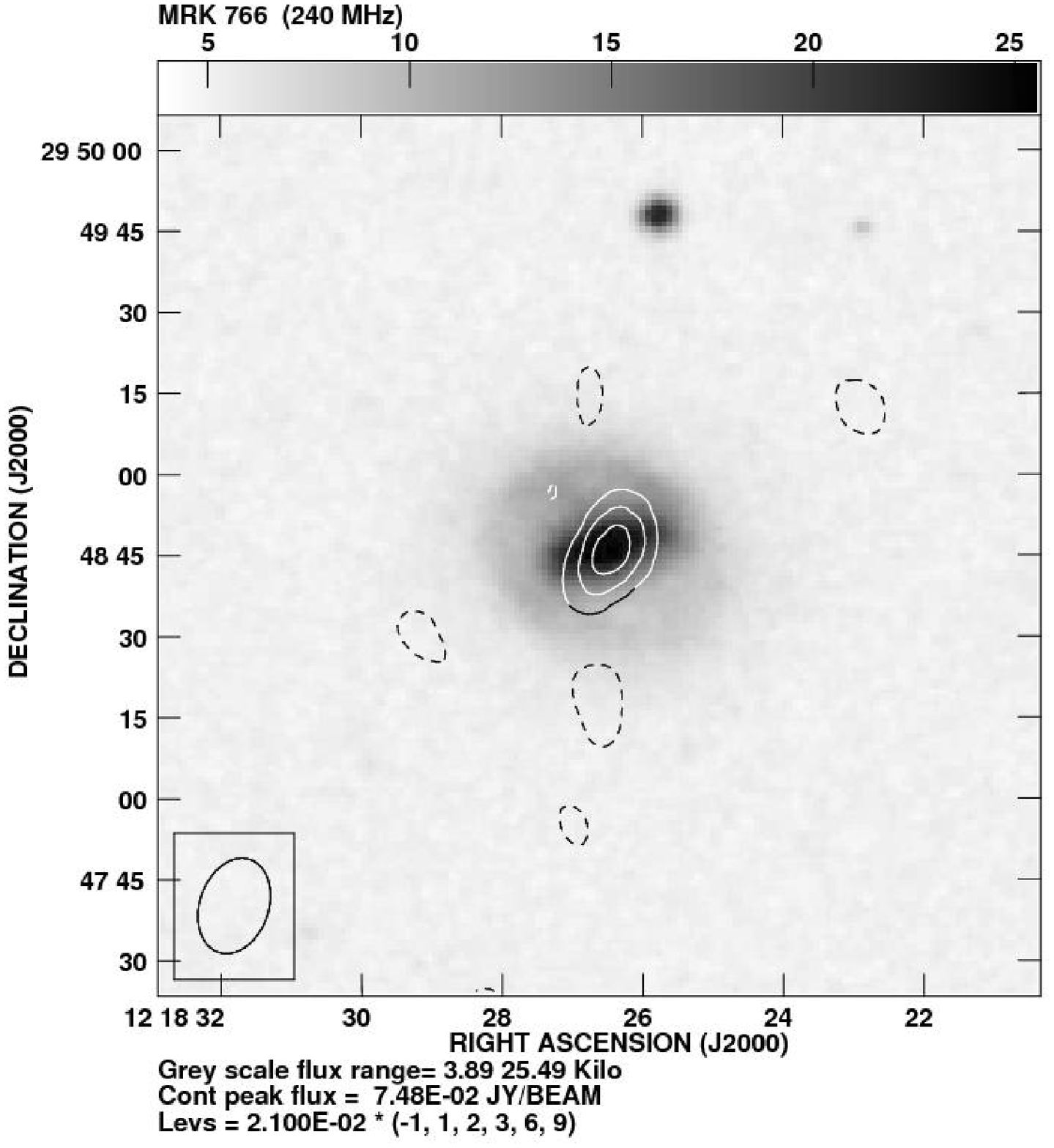}}
\includegraphics[angle=0,width=6.8cm,height=0.26\textheight]{M279+610RADIOCONTDSS.PS}{\includegraphics[angle=0,width=6.8cm,height=0.26\textheight]{M279+240RADIOCONTDSSSHIFT.PS}}
\includegraphics[angle=0,width=6.8cm,height=0.26\textheight]{N5548+610RADIOCONTDSS.PS}{\includegraphics[angle=0,width=6.8cm,height=0.26\textheight]{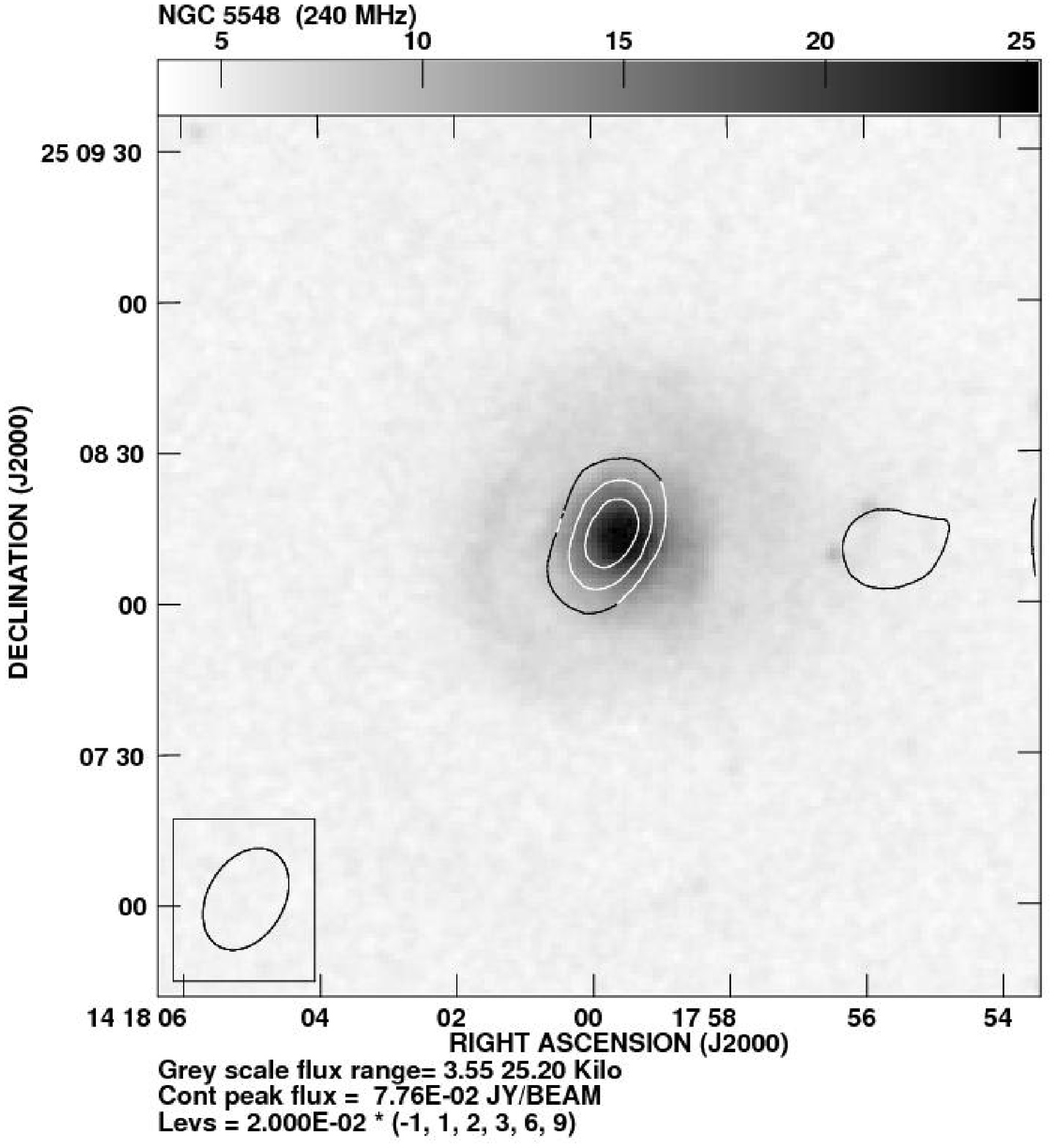}}
\includegraphics[angle=0,width=6.8cm,height=0.26\textheight]{A564+610RADIOCONTDSS.PS}{\includegraphics[angle=0,width=6.8cm,height=0.26\textheight]{A564+240RADIOCONTDSS.PS}}
\caption{{\it continued}: 610 MHz (left panel) and 240 MHz (right panel) radio contours overlaid on their DSS optical images.}
\end{figure*}
\addtocounter{figure}{-1}
\begin{figure*}[!htbp]
\centering
\includegraphics[angle=0,width=6.8cm,height=0.26\textheight]{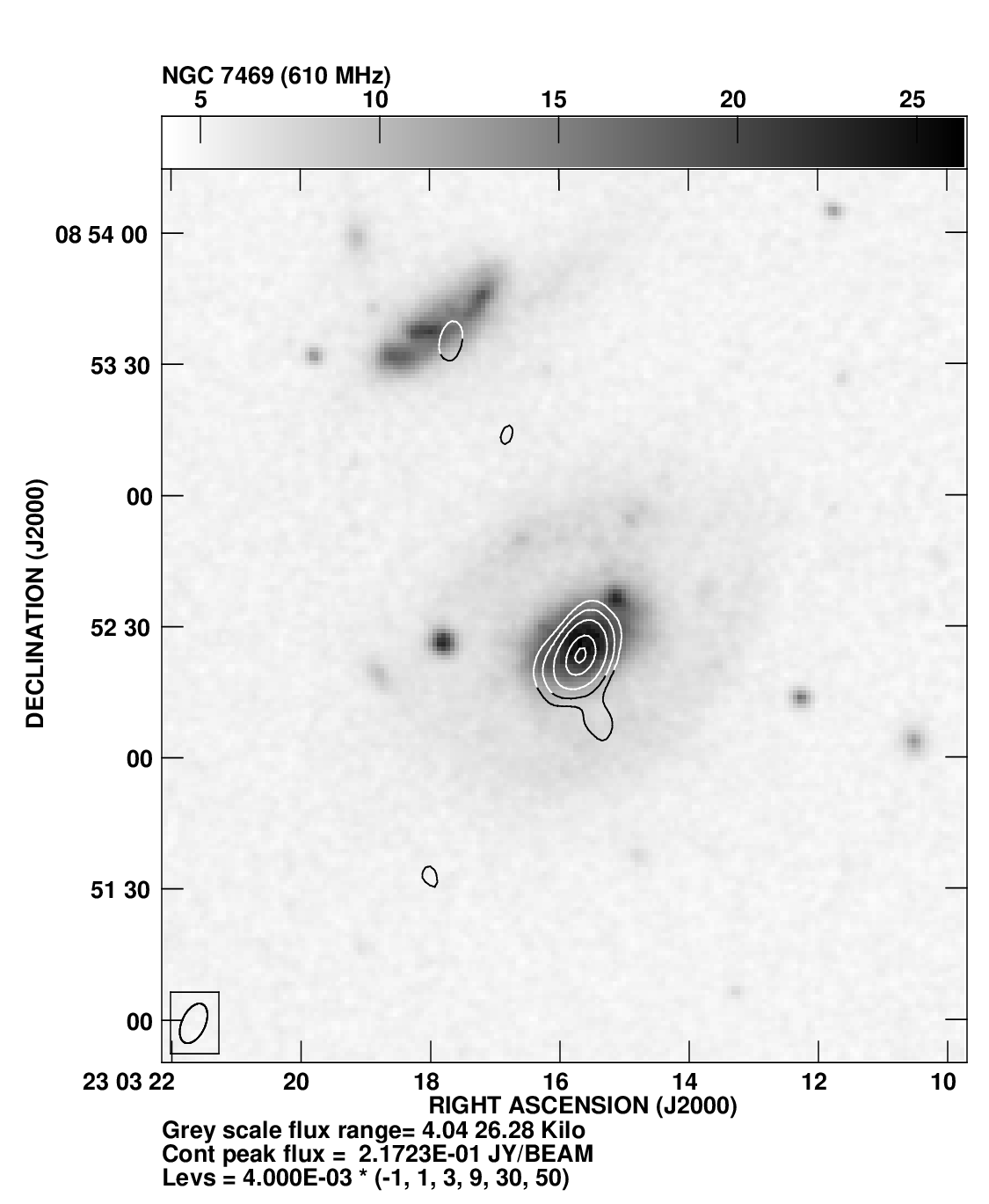}{\includegraphics[angle=0,width=6.8cm,height=0.26\textheight]{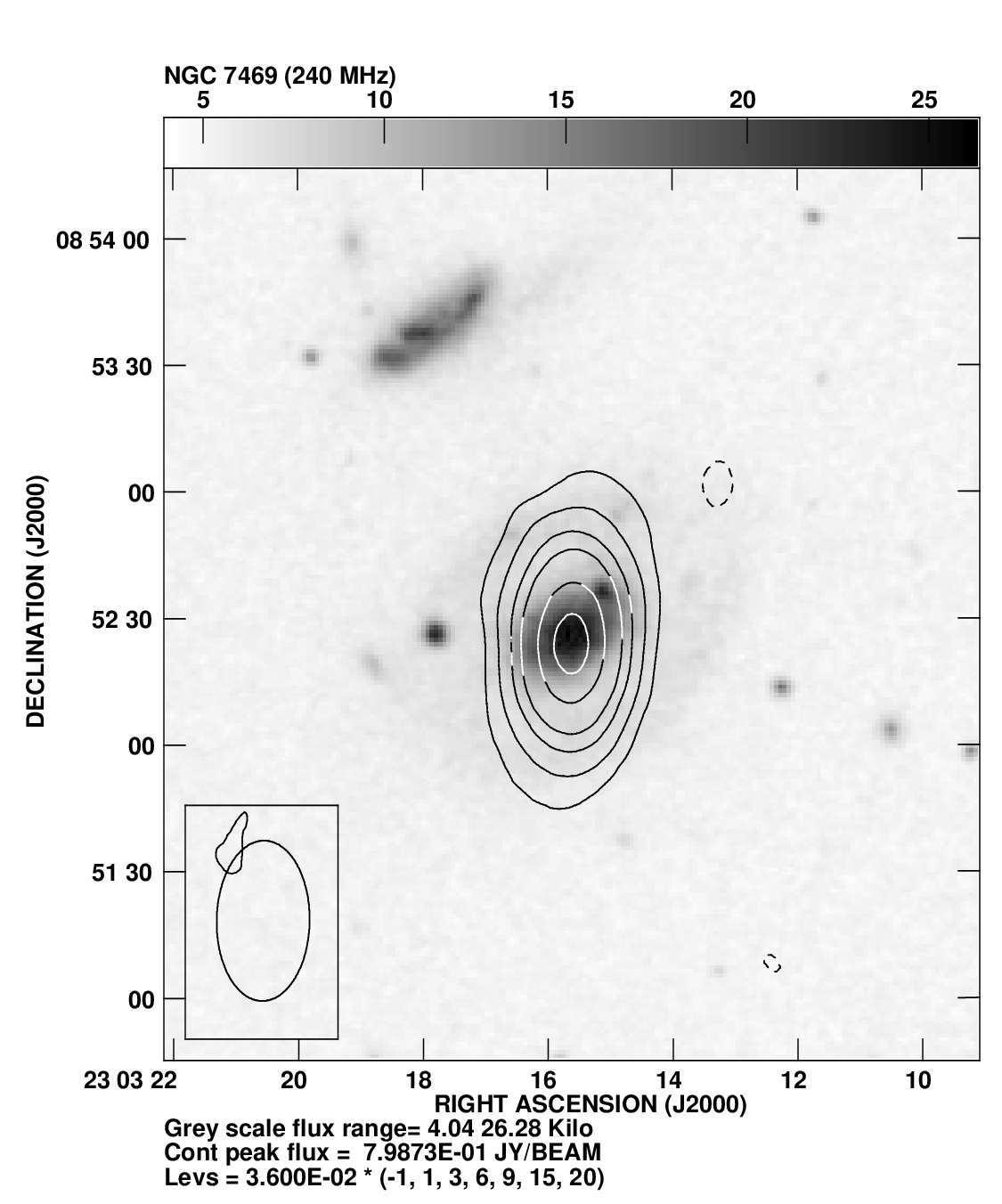}}
\includegraphics[angle=0,width=6.8cm,height=0.26\textheight]{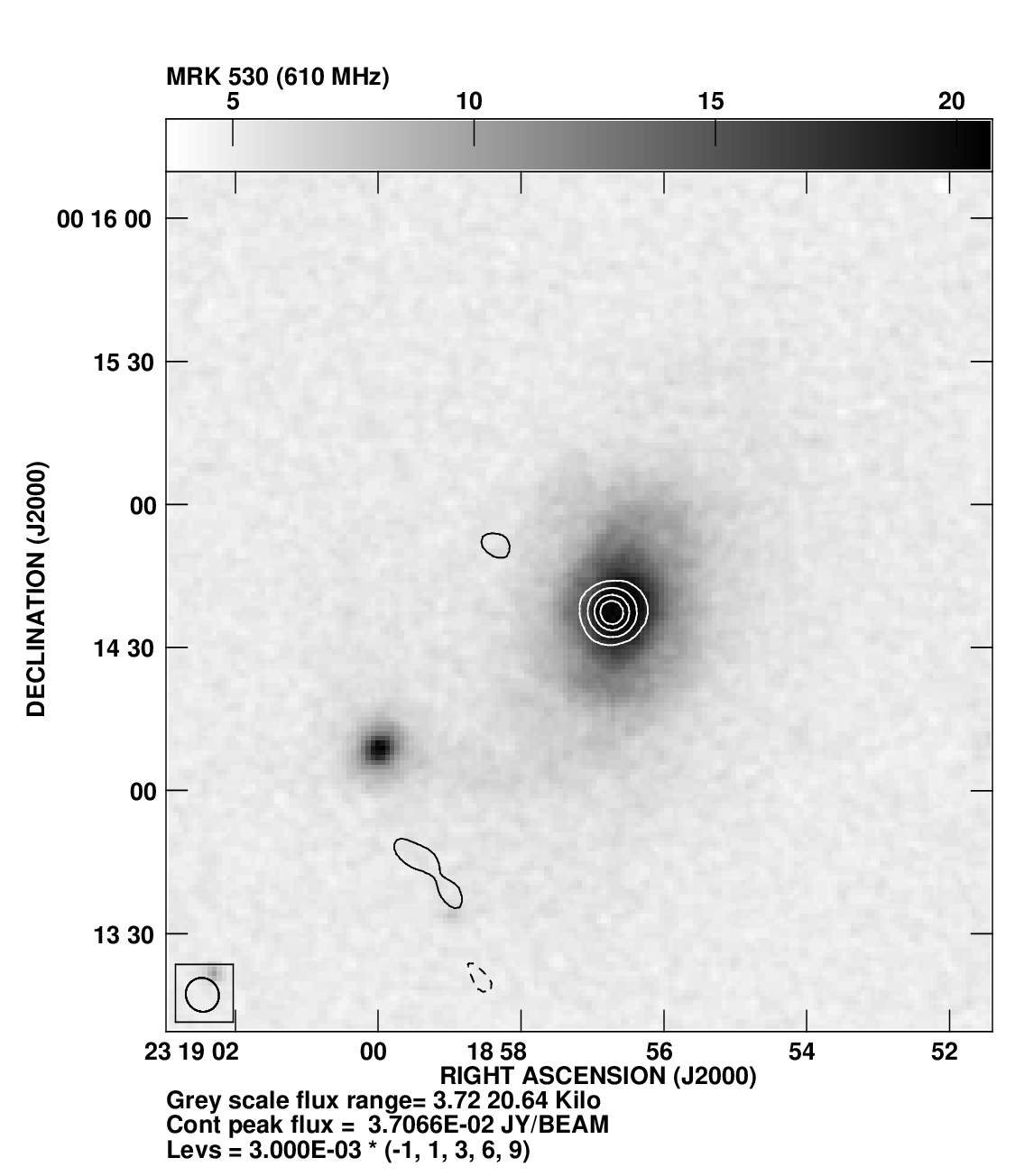}{{\includegraphics[angle=0,width=6.8cm,height=0.26\textheight]{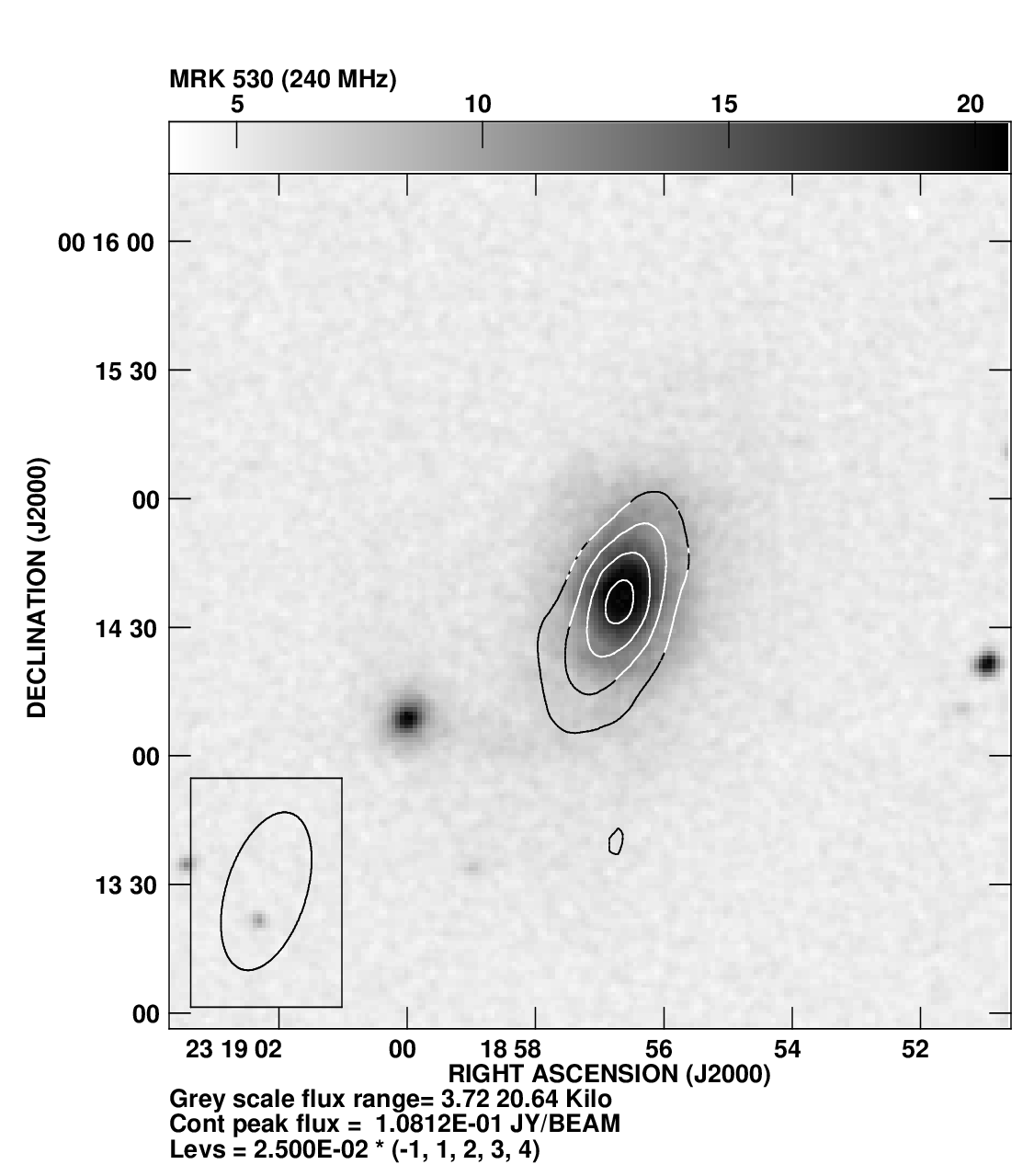}}}
\includegraphics[angle=0,width=6.8cm,height=0.26\textheight]{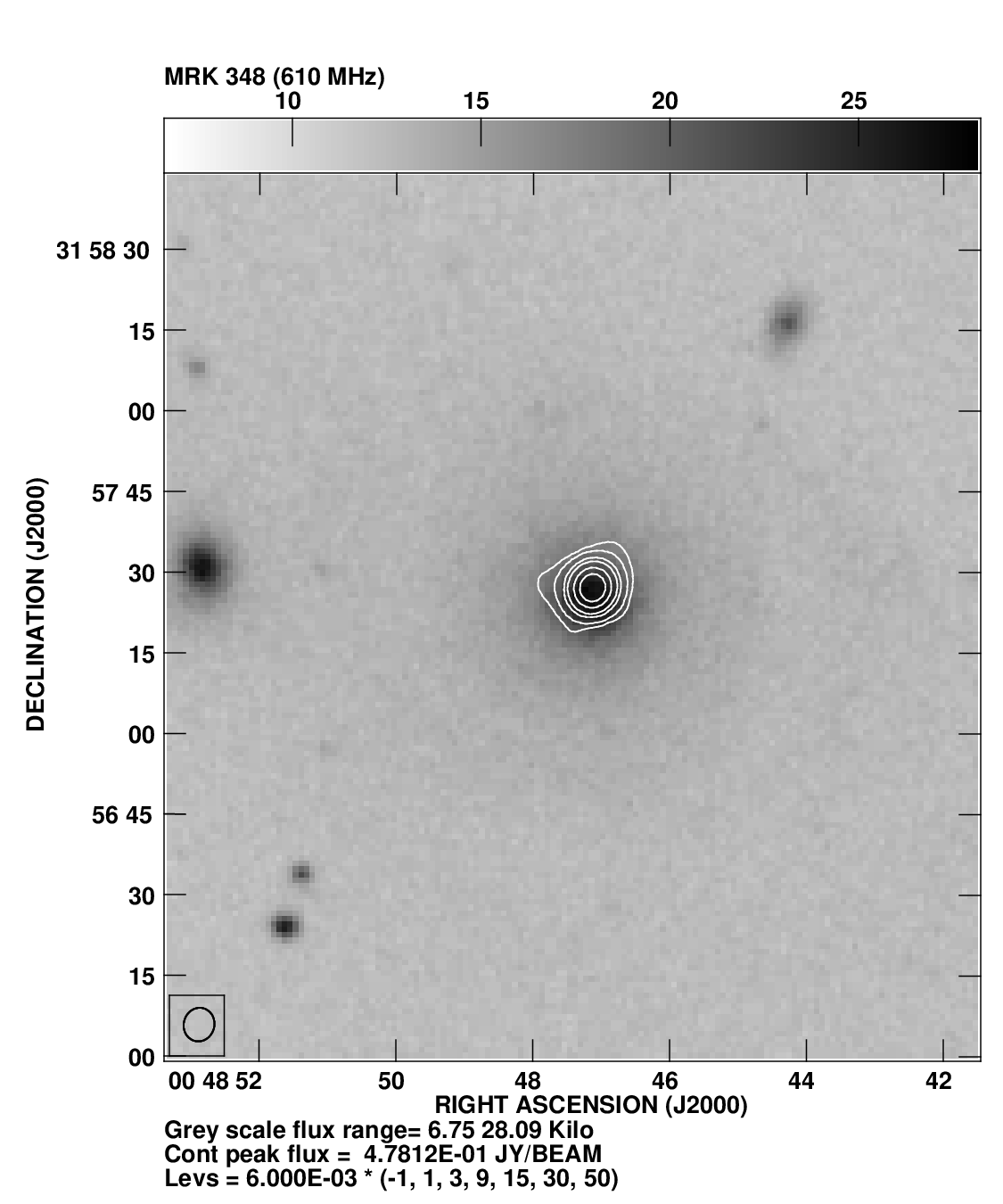}{\includegraphics[angle=0,width=6.8cm,height=0.26\textheight]{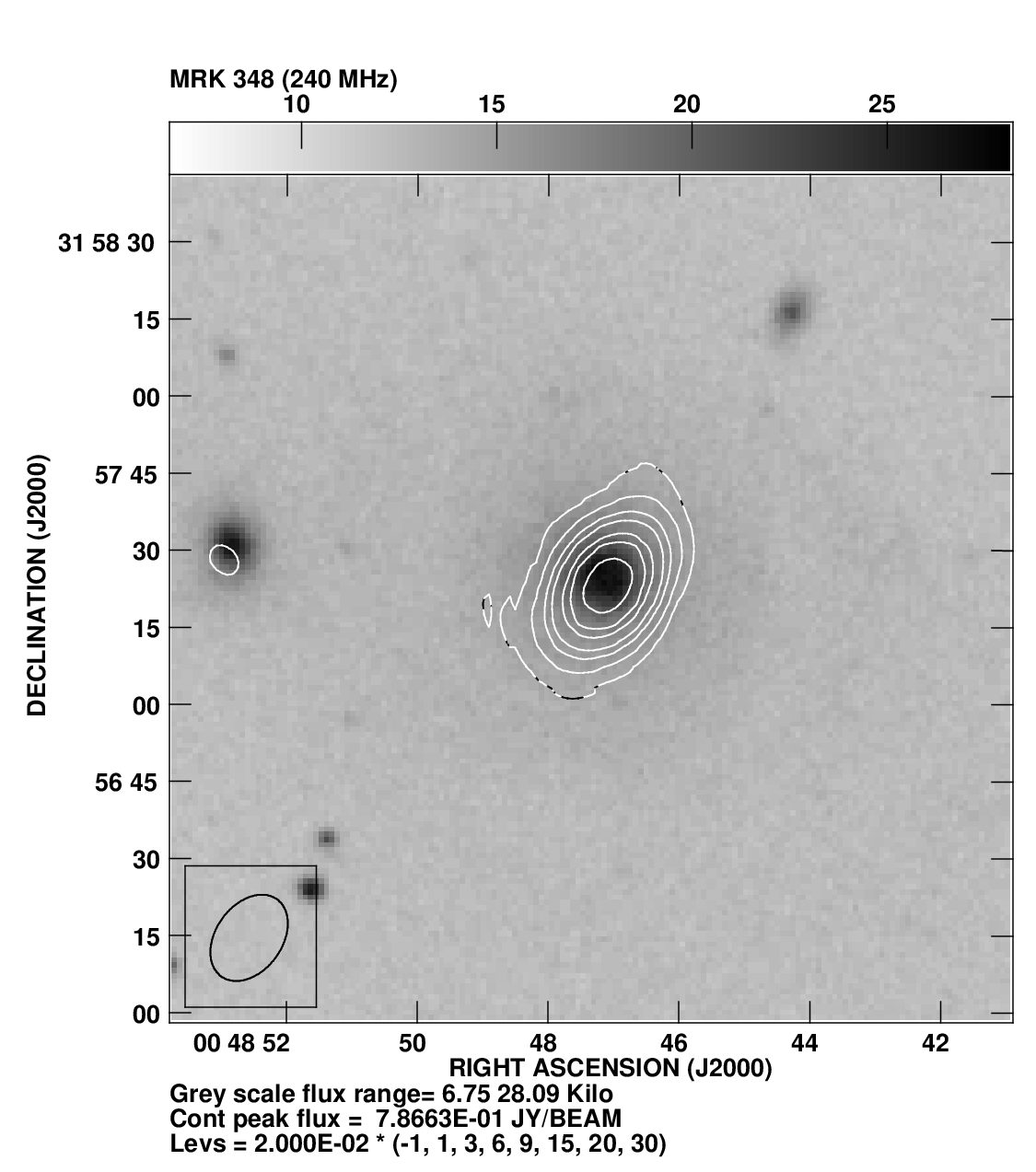}}
\includegraphics[angle=0,width=6.8cm,height=0.26\textheight]{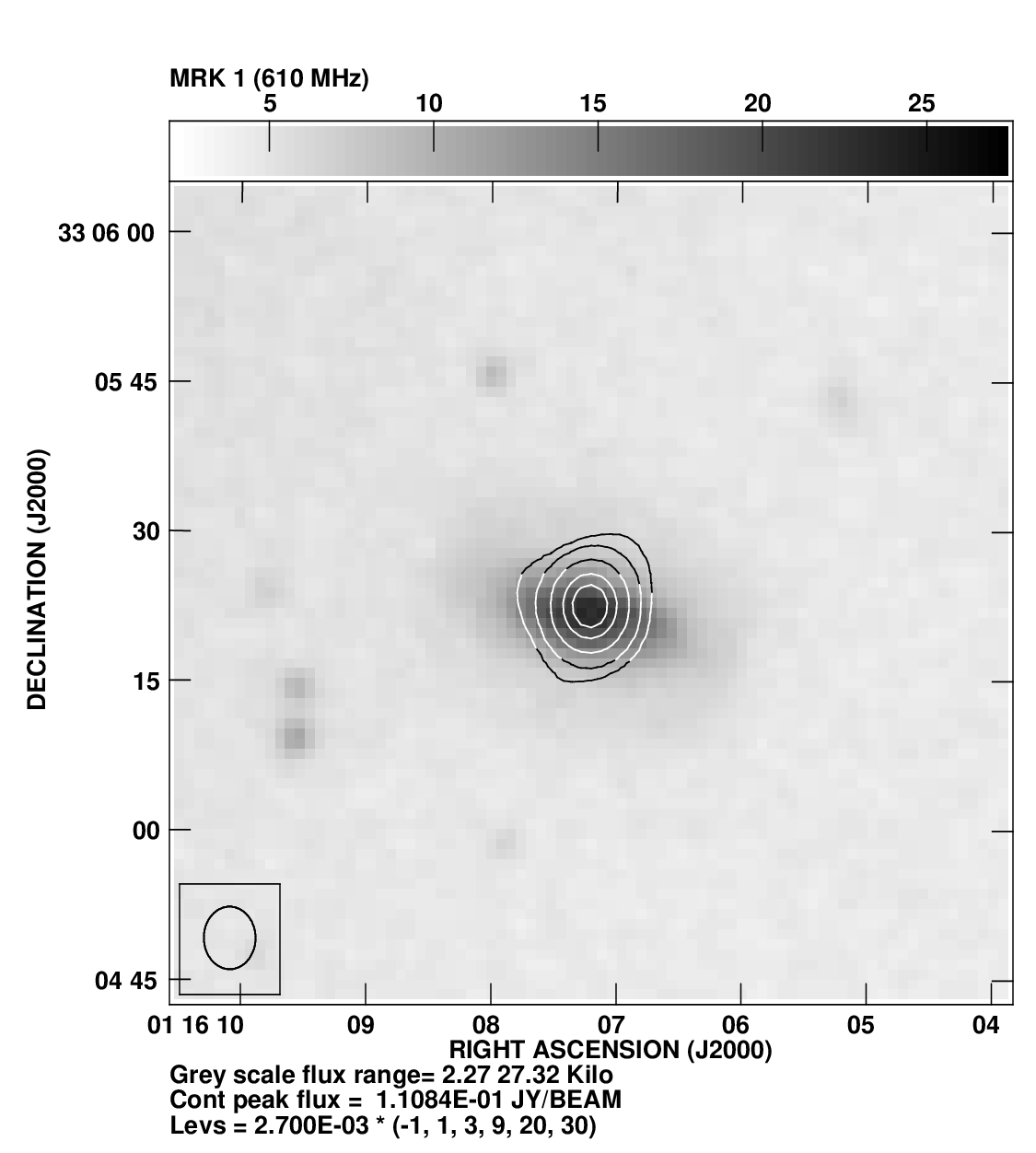}{{\includegraphics[angle=0,width=6.8cm,height=0.26\textheight]{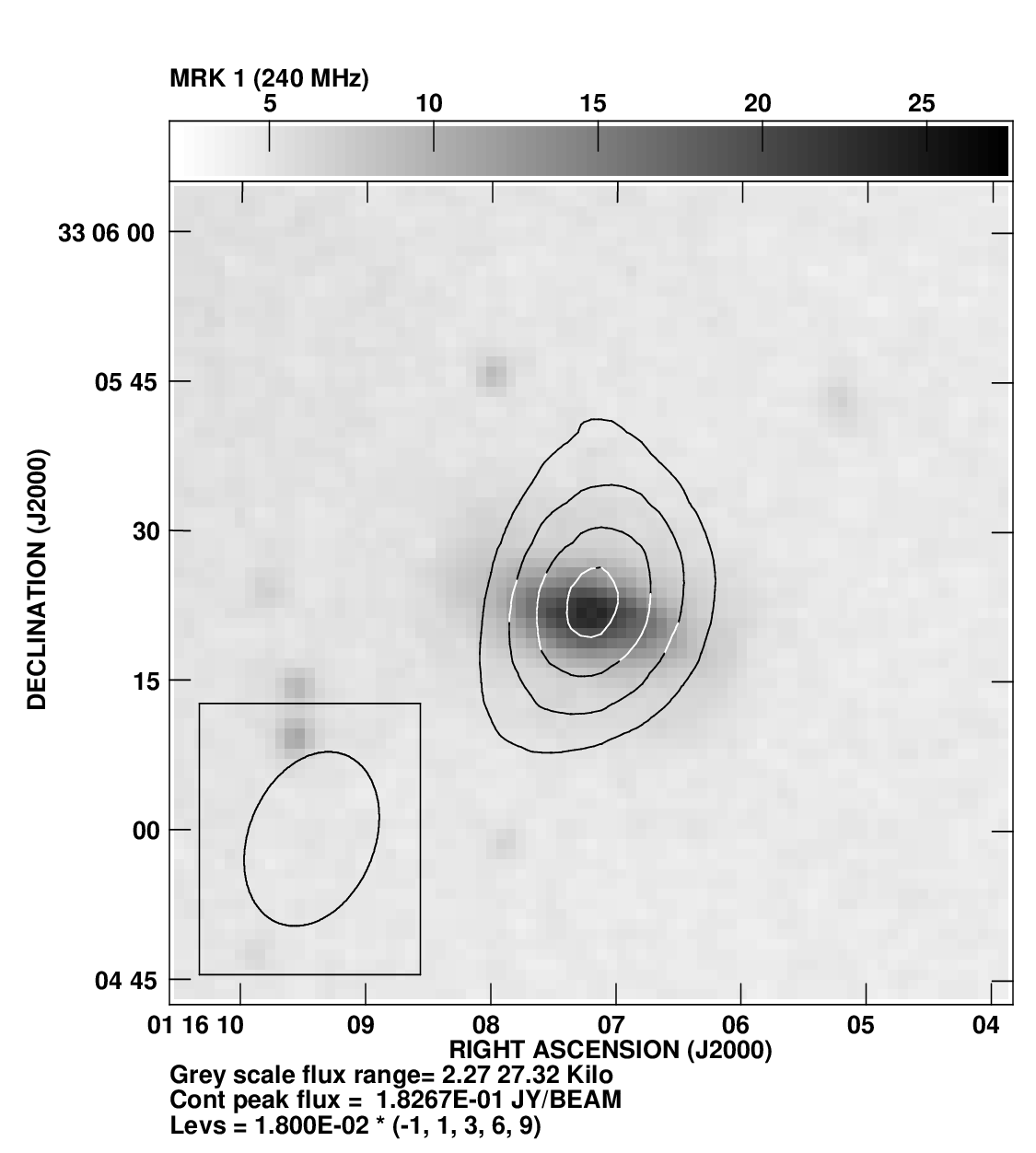}}}
\caption{{\it continued}: 610 MHz (left panel) and 240 MHz (right panel) radio contours overlaid on their DSS optical images.}
\end{figure*}
\addtocounter{figure}{-1}
\begin{figure*}[!htbp]
\centering
\includegraphics[angle=0,width=6.8cm,height=0.26\textheight]{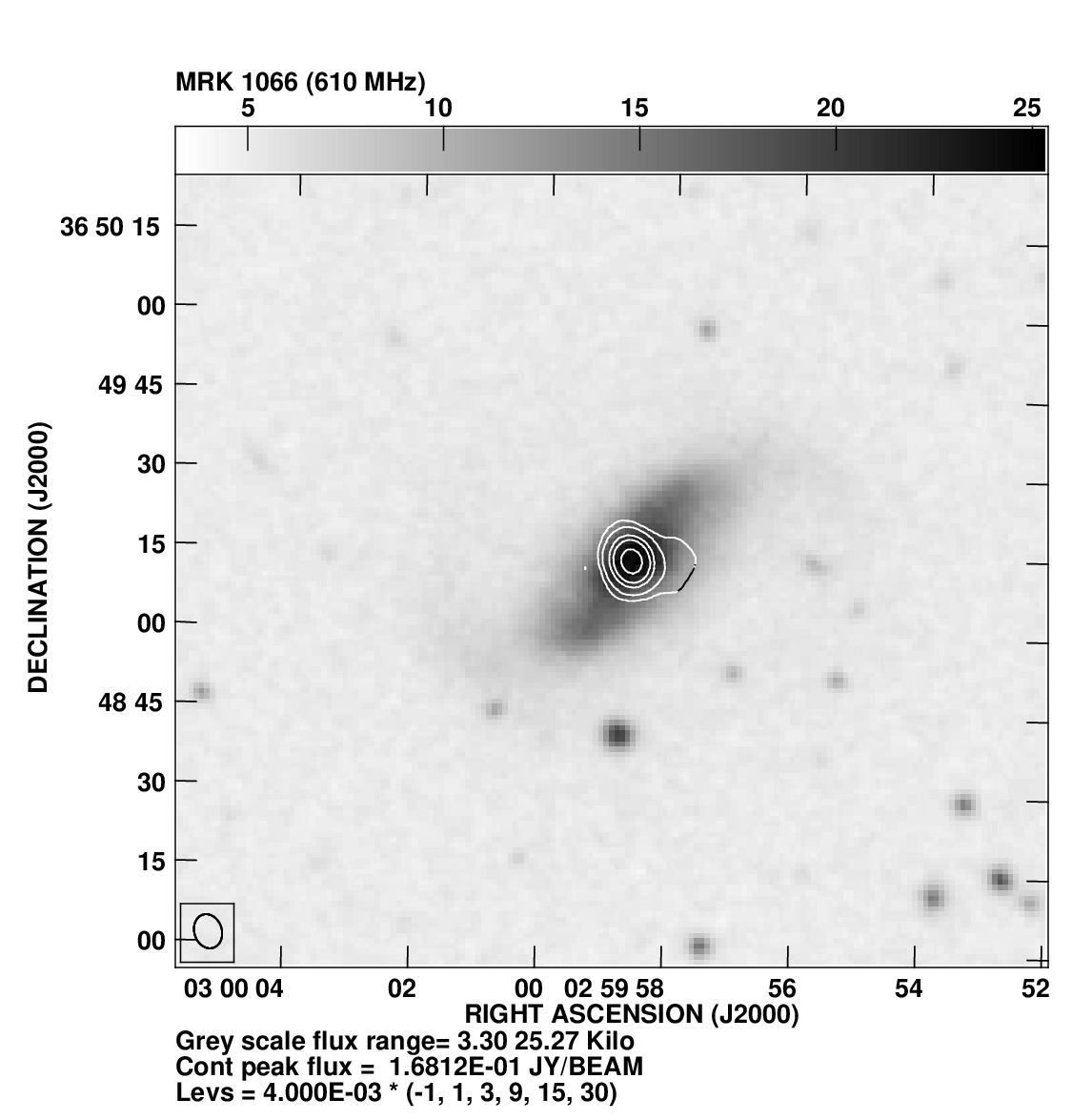}{\includegraphics[angle=0,width=6.8cm,height=0.26\textheight]{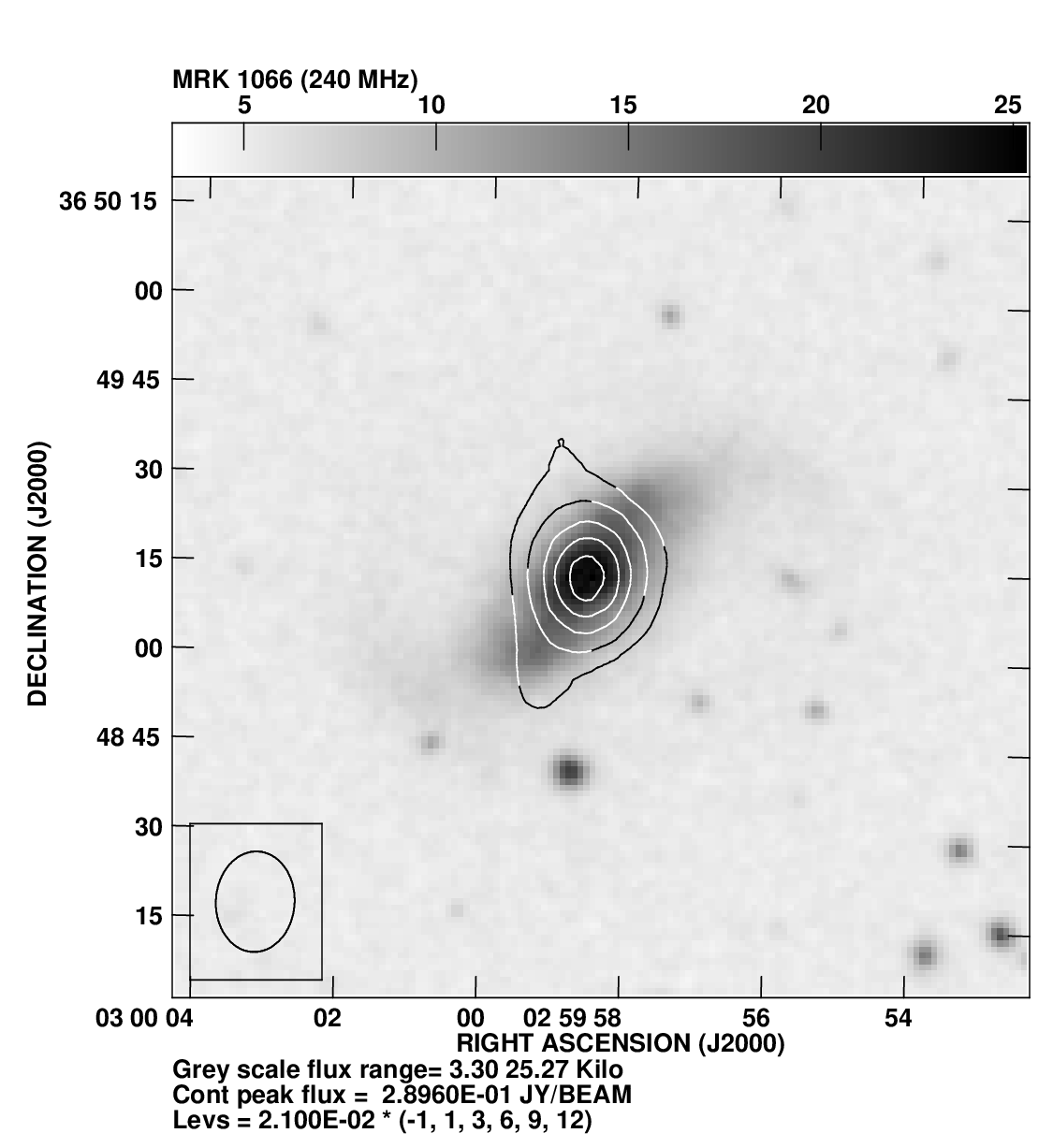}}
\includegraphics[angle=0,width=6.8cm,height=0.26\textheight]{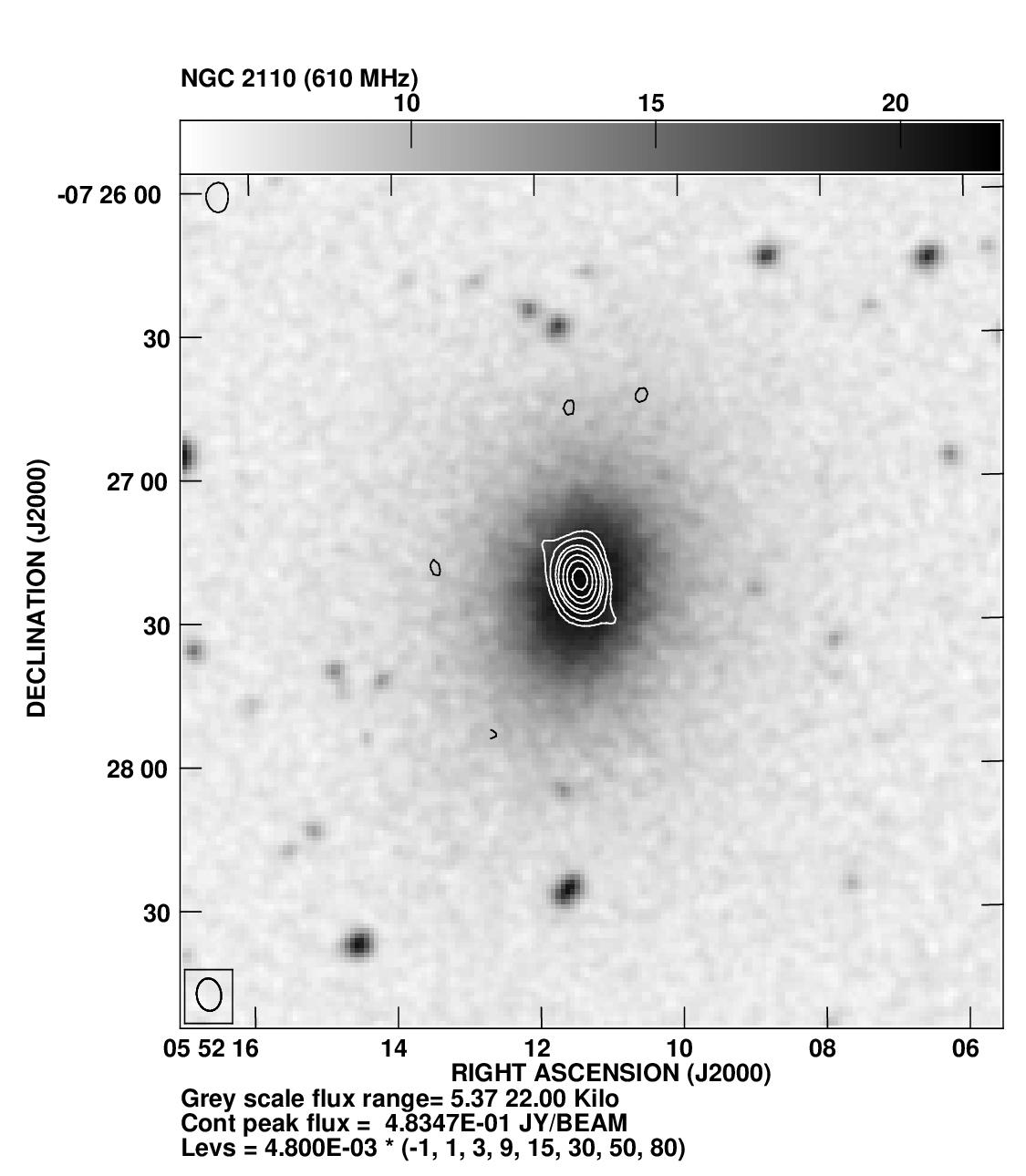}{\includegraphics[angle=0,width=6.8cm,height=0.26\textheight]{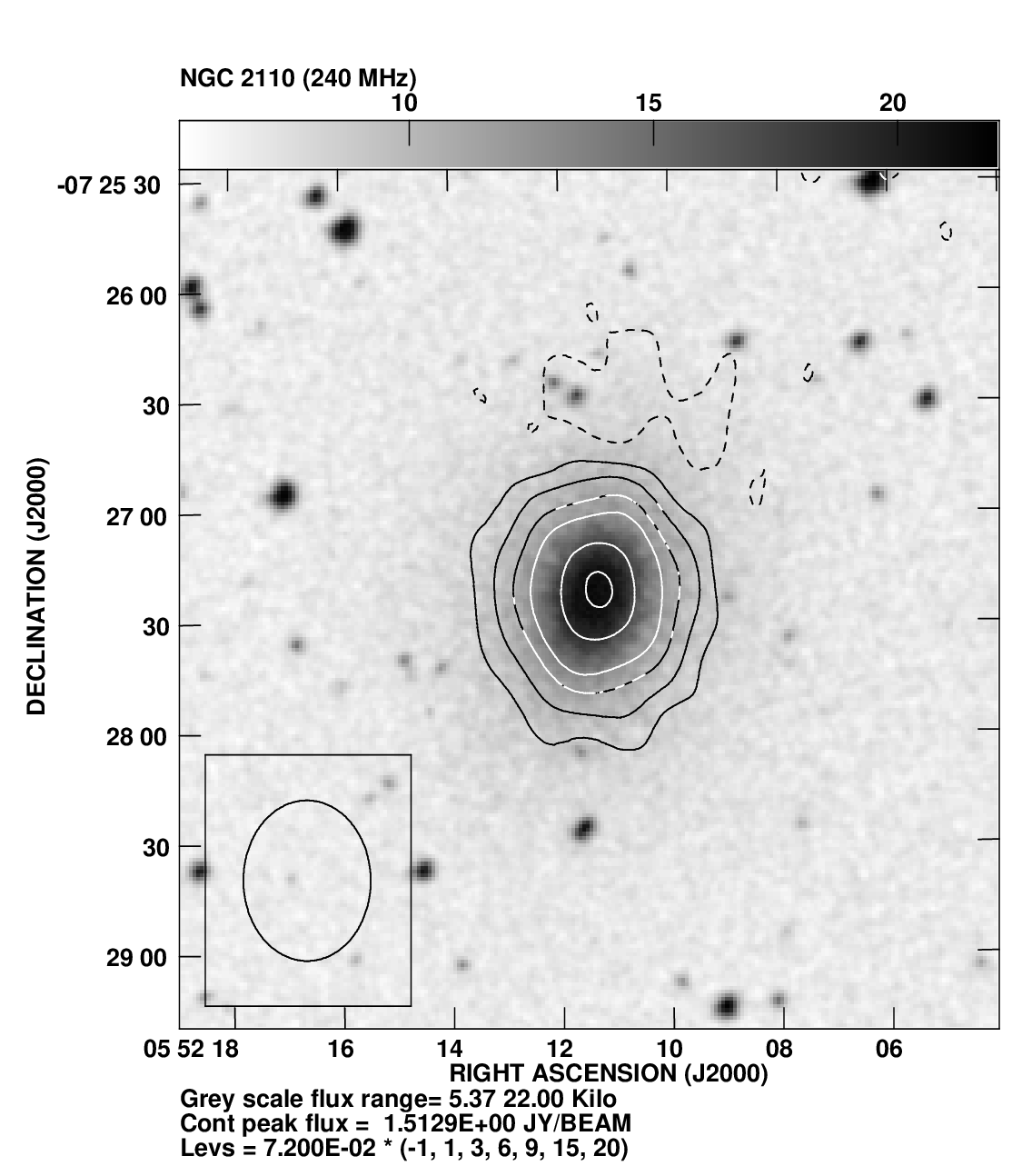}}
\includegraphics[angle=0,width=6.8cm,height=0.26\textheight]{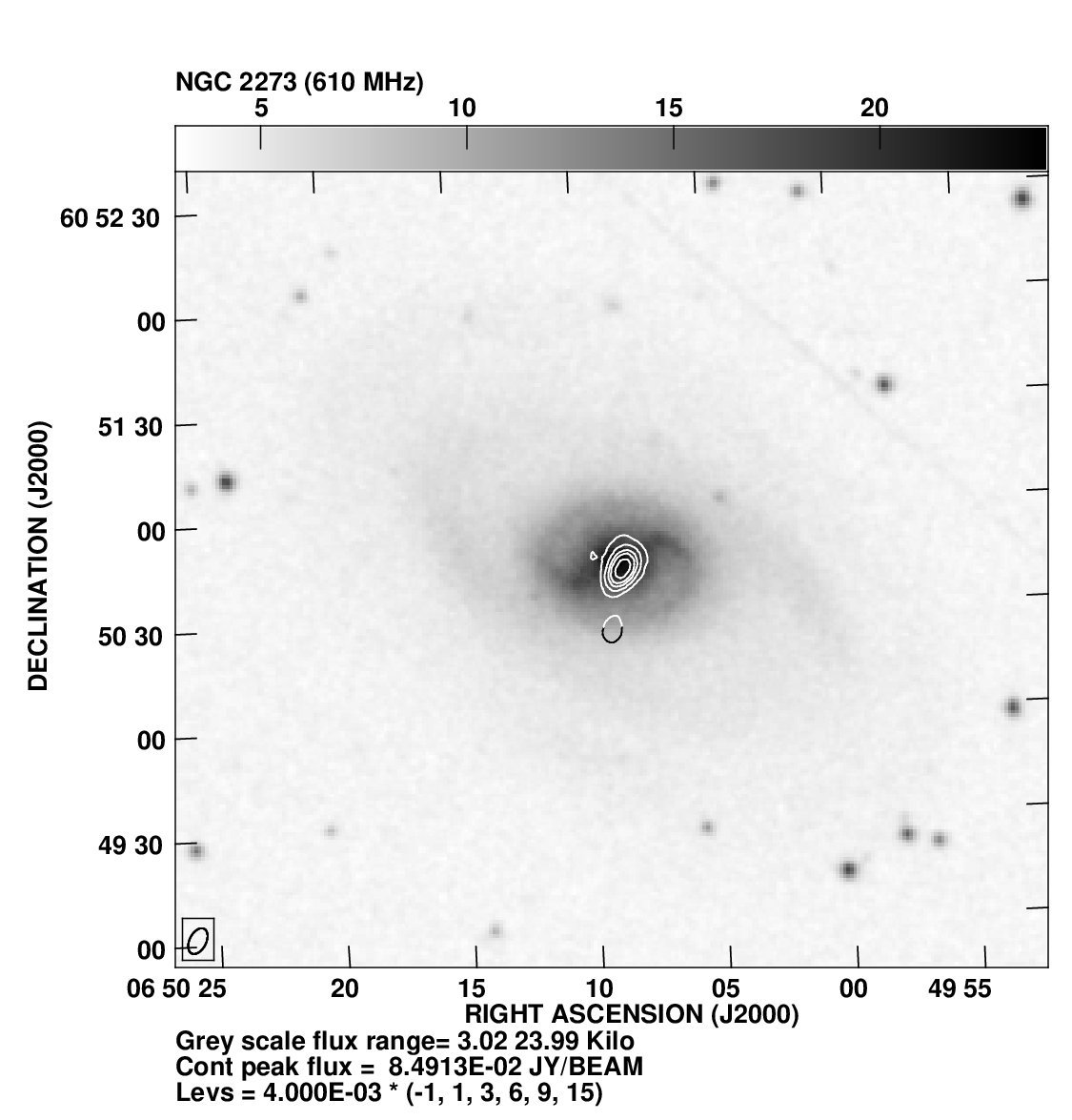}{\includegraphics[angle=0,width=6.8cm,height=0.26\textheight]{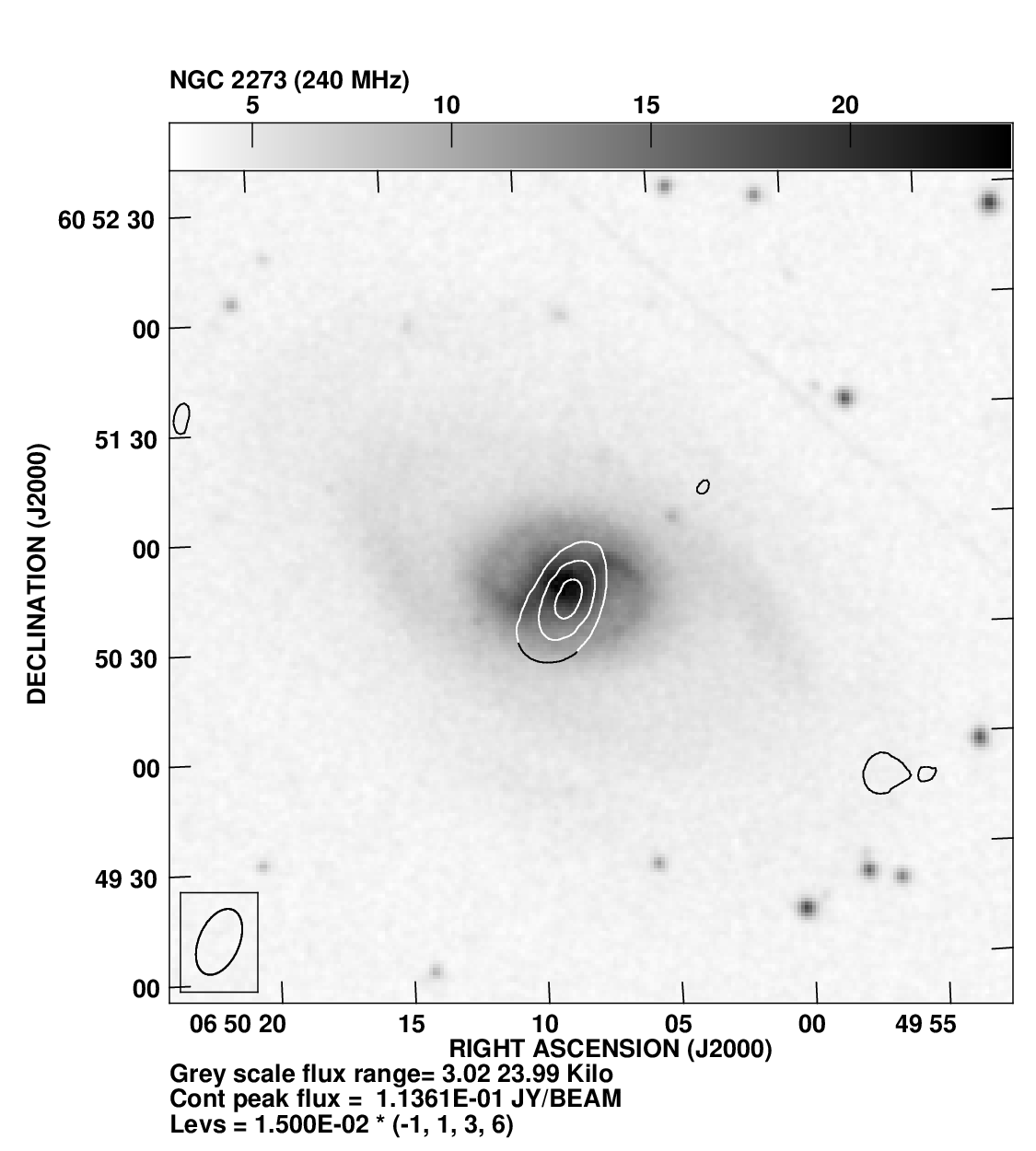}}
\includegraphics[angle=0,width=6.8cm,height=0.26\textheight]{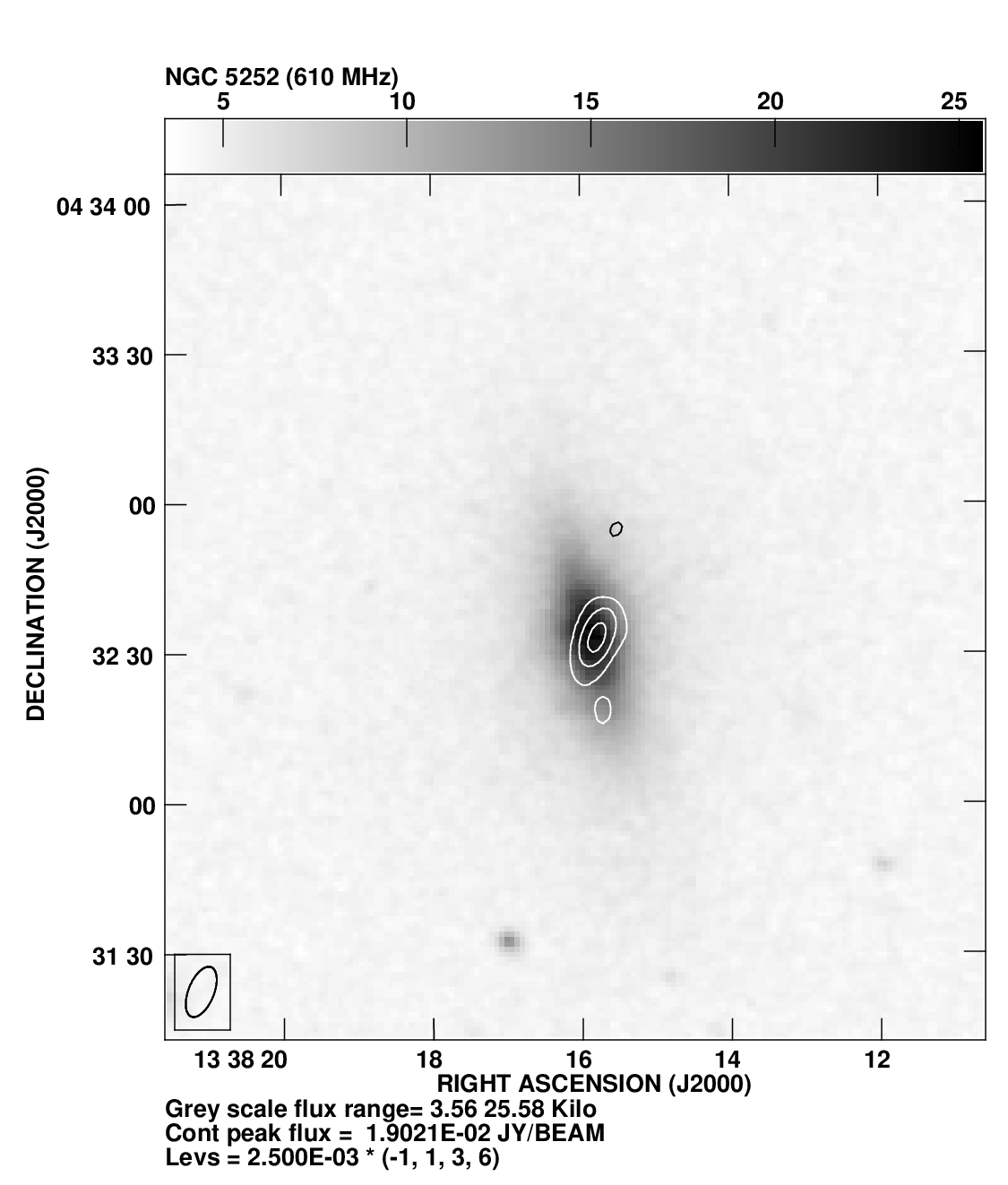}{\includegraphics[angle=0,width=6.8cm,height=0.26\textheight]{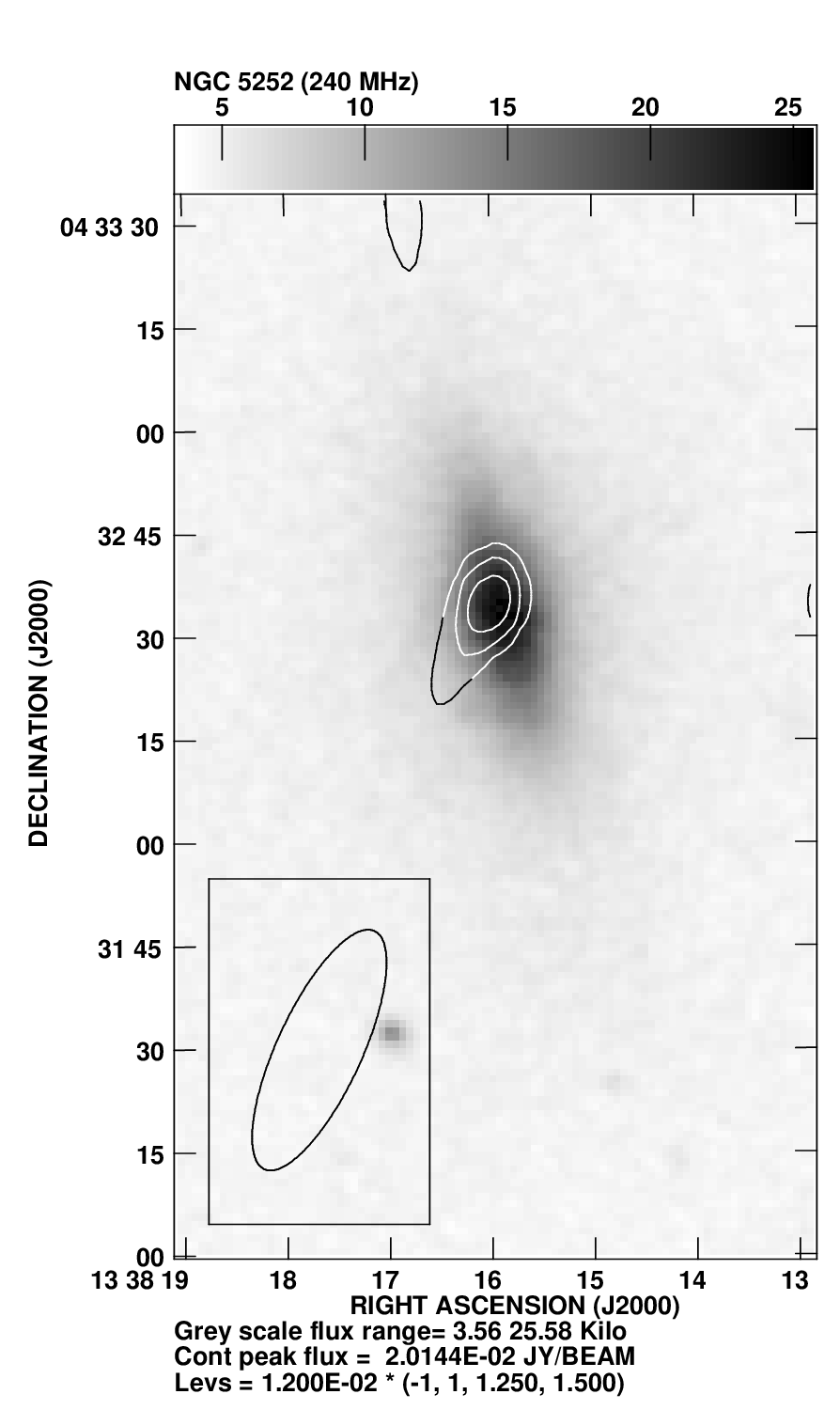}}
\caption{{\it continued}: 610 MHz (left panel) and 240 MHz (right panel) radio contours overlaid on their DSS optical images.}
\end{figure*}
\addtocounter{figure}{-1}
\begin{figure*}[!htbp]
\centering
\includegraphics[angle=0,width=6.8cm,height=0.26\textheight]{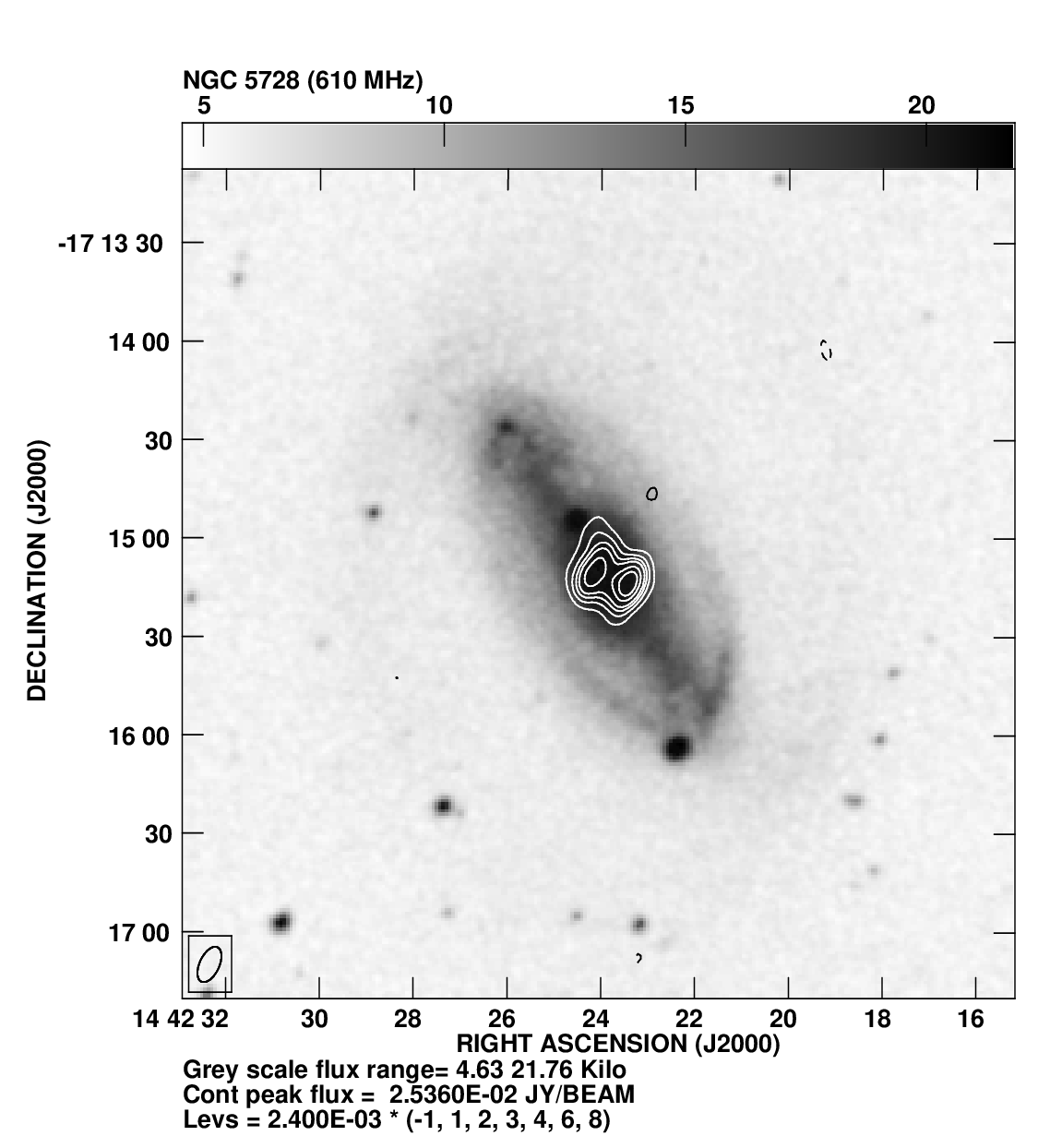}{\includegraphics[angle=0,width=6.8cm,height=0.26\textheight]{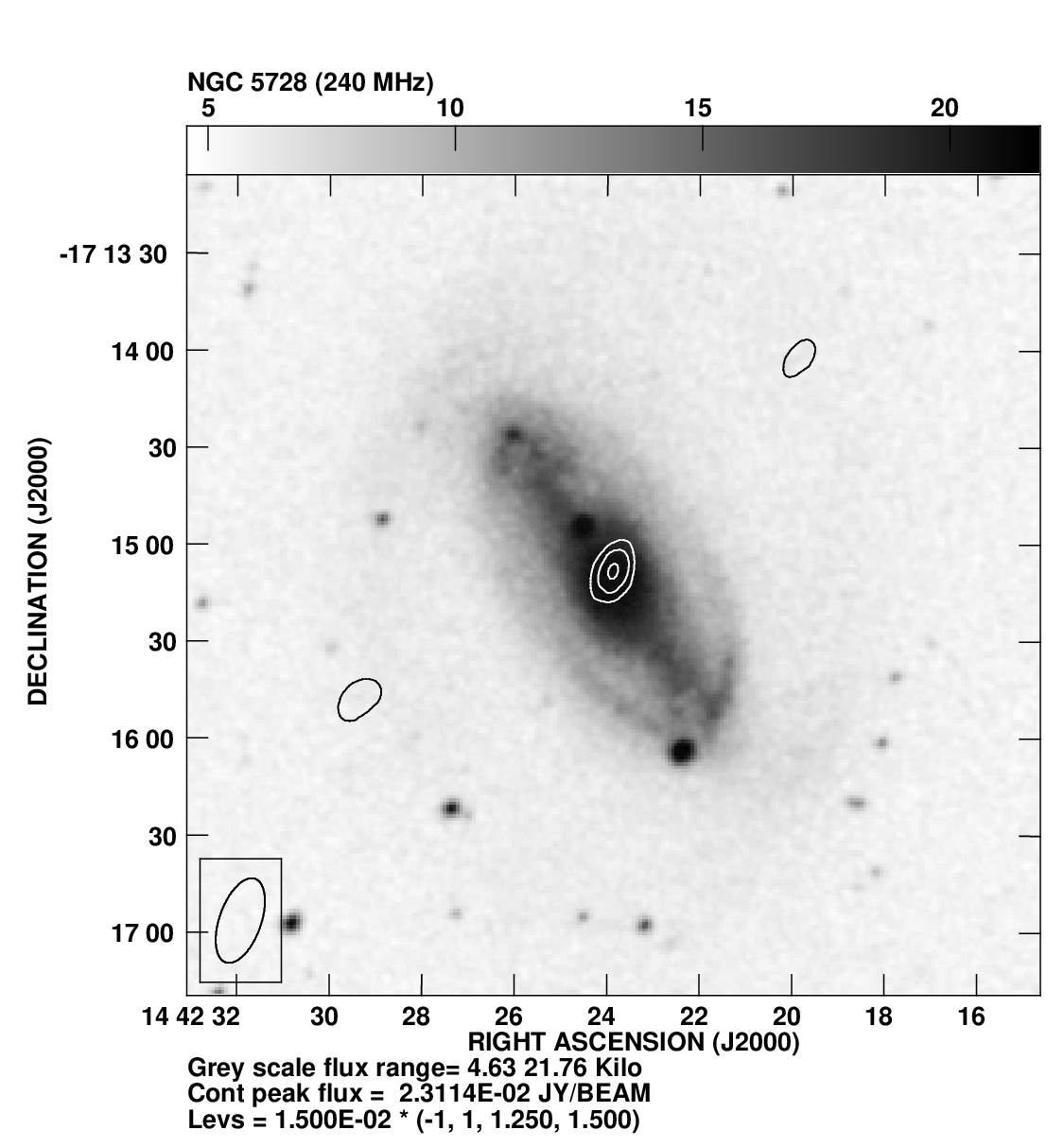}}
\includegraphics[angle=0,width=6.8cm,height=0.26\textheight]{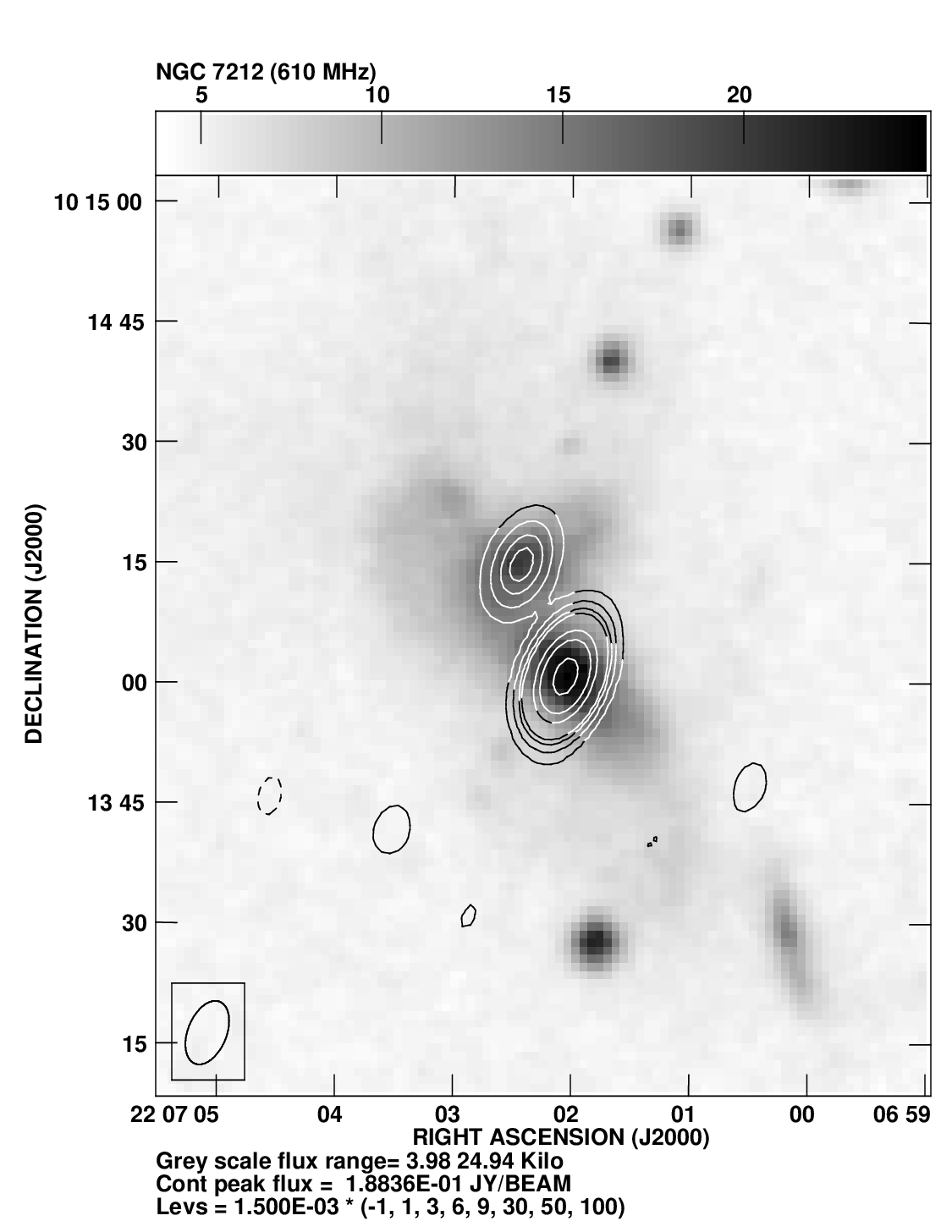}{\includegraphics[angle=0,width=6.8cm,height=0.26\textheight]{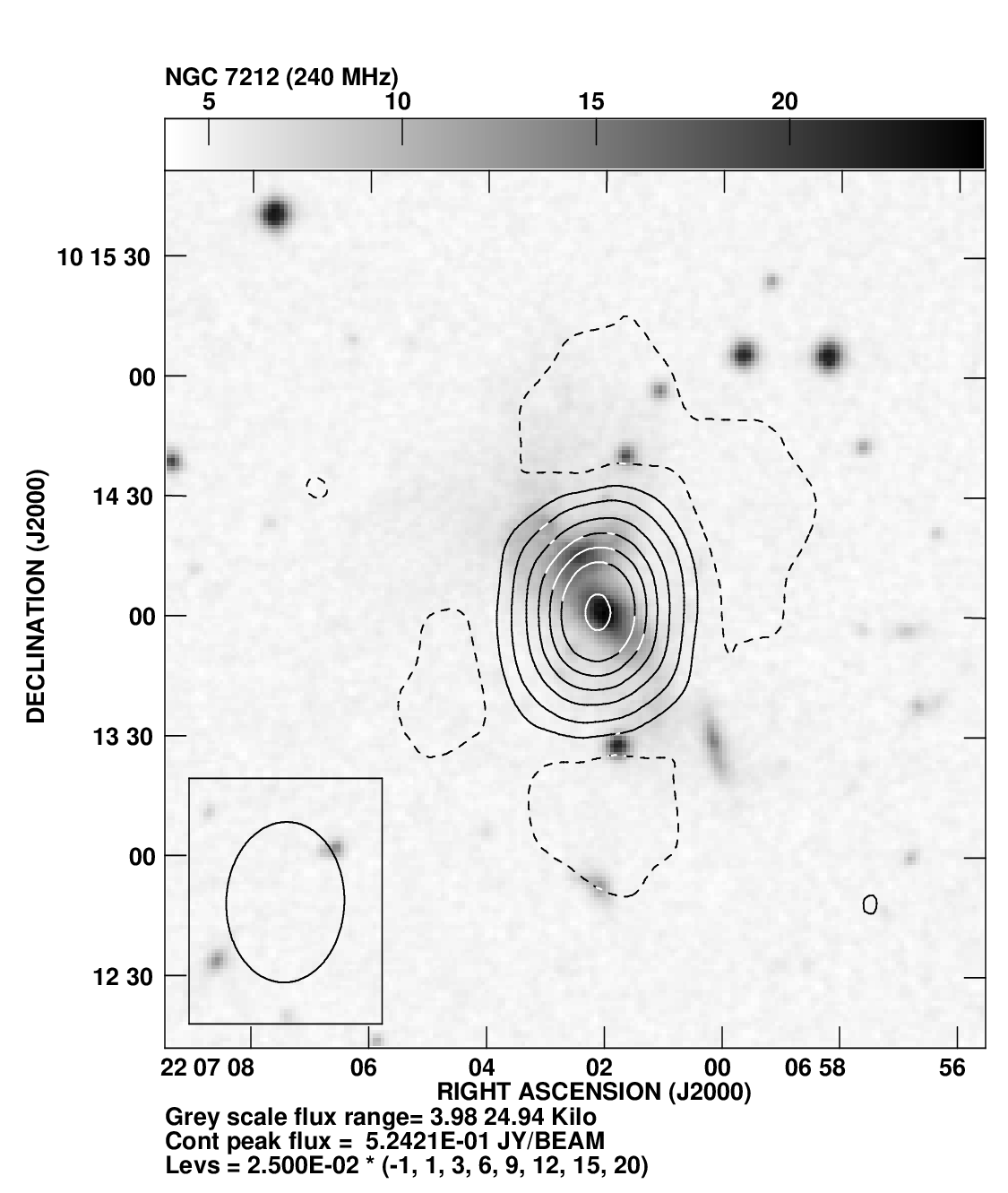}}
\includegraphics[angle=0,width=6.8cm,height=0.26\textheight]{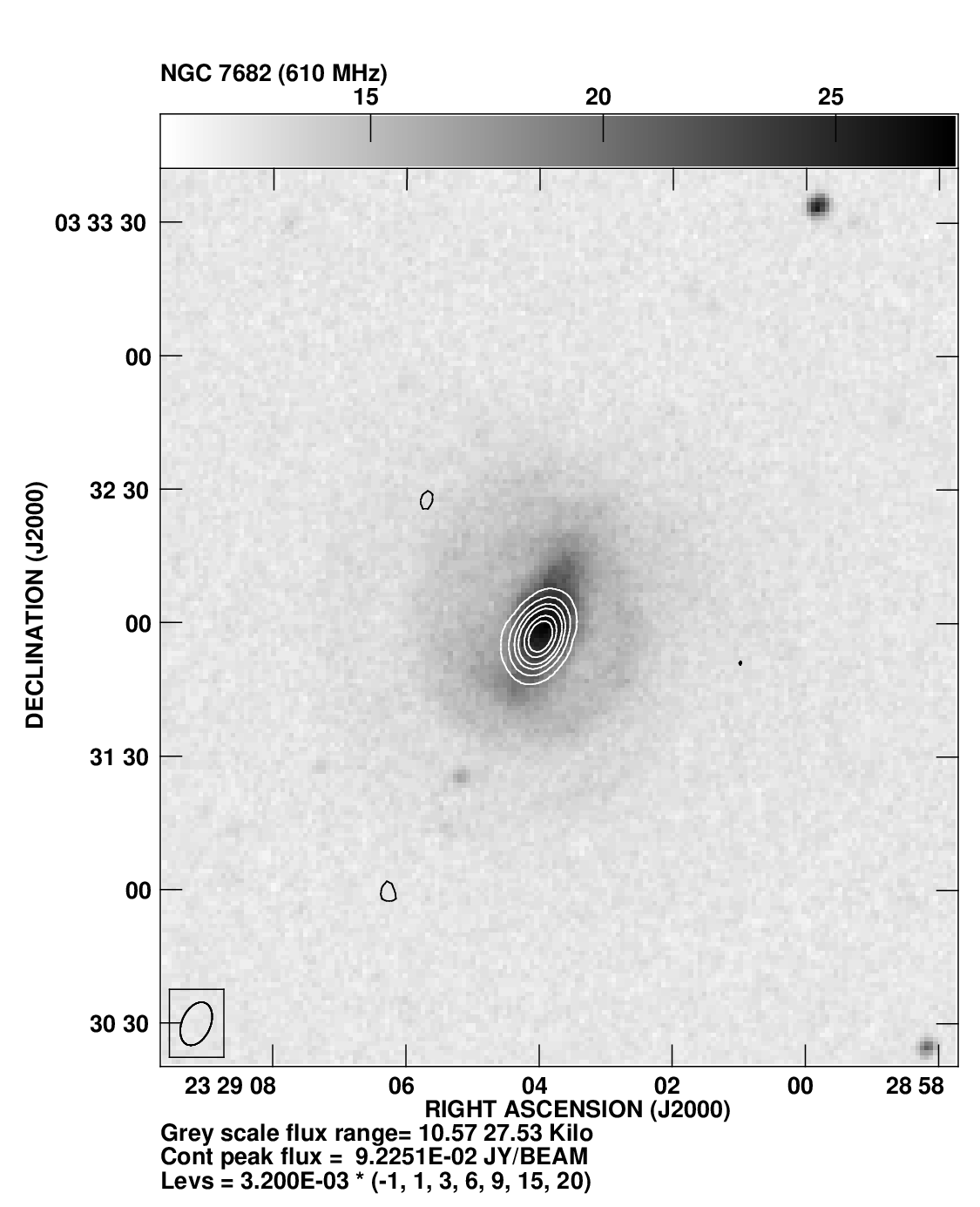}{\includegraphics[angle=0,width=6.8cm,height=0.26\textheight]{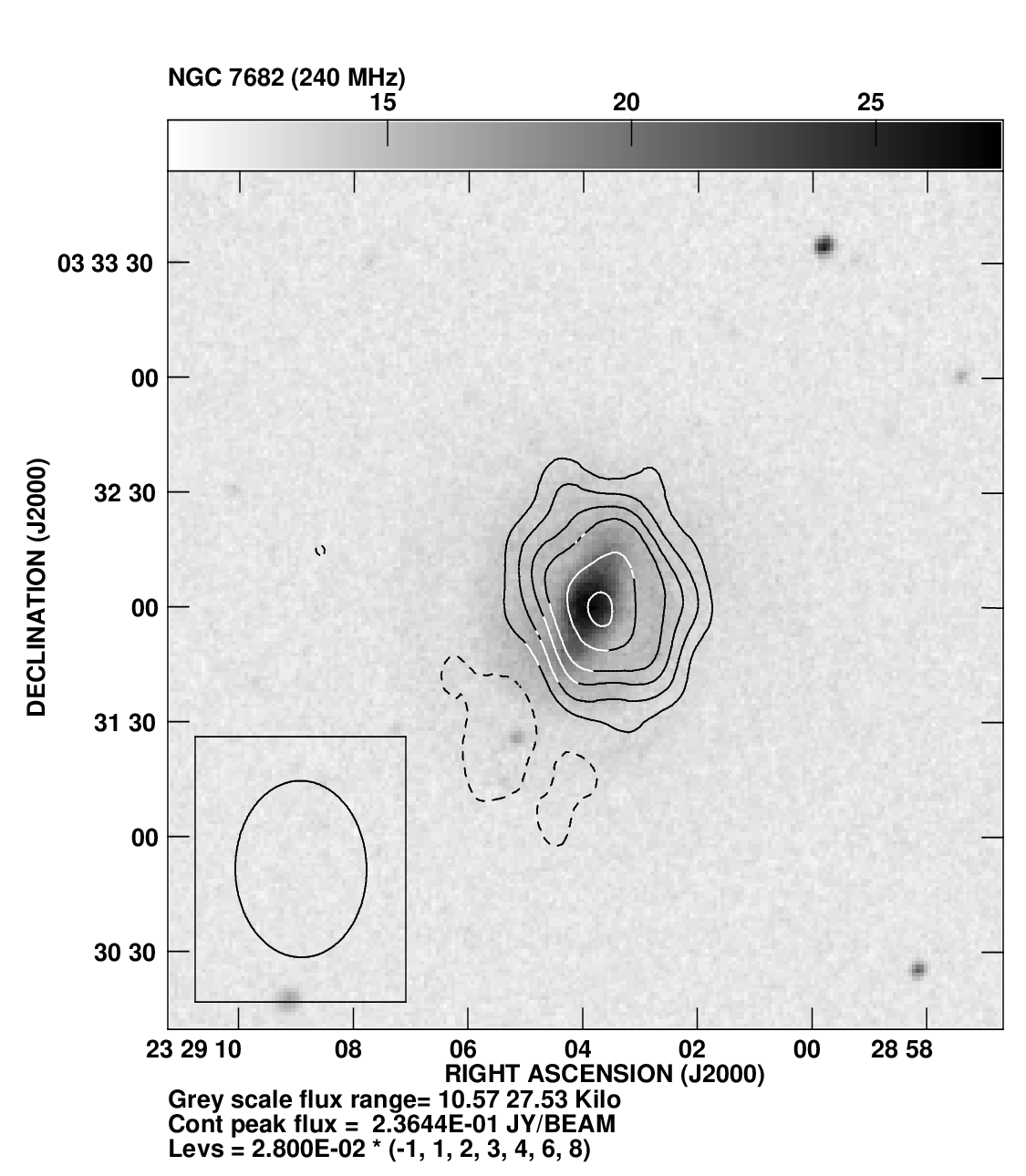}}
\includegraphics[angle=0,width=6.8cm,height=0.26\textheight]{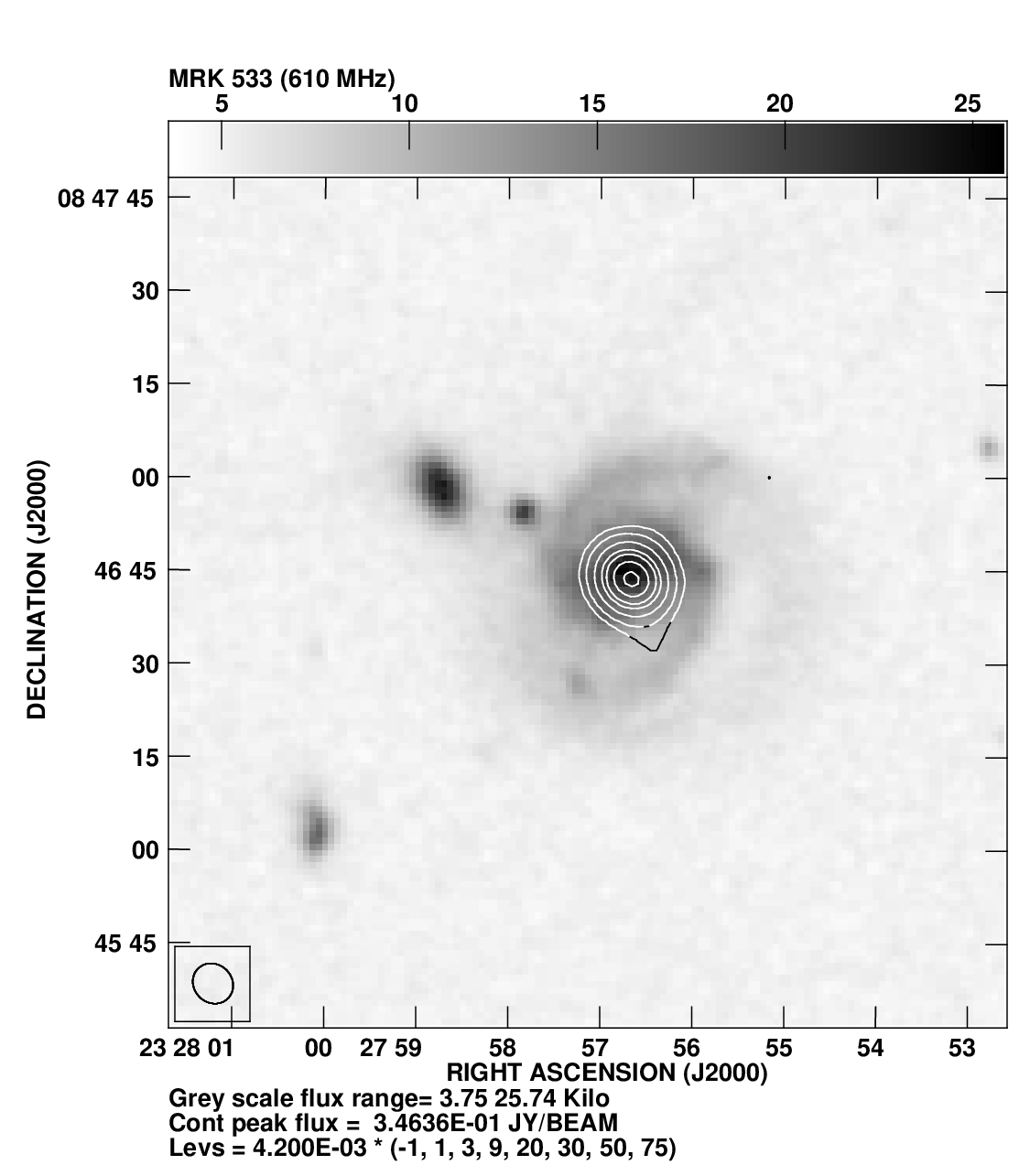}{\includegraphics[angle=0,width=6.8cm,height=0.26\textheight]{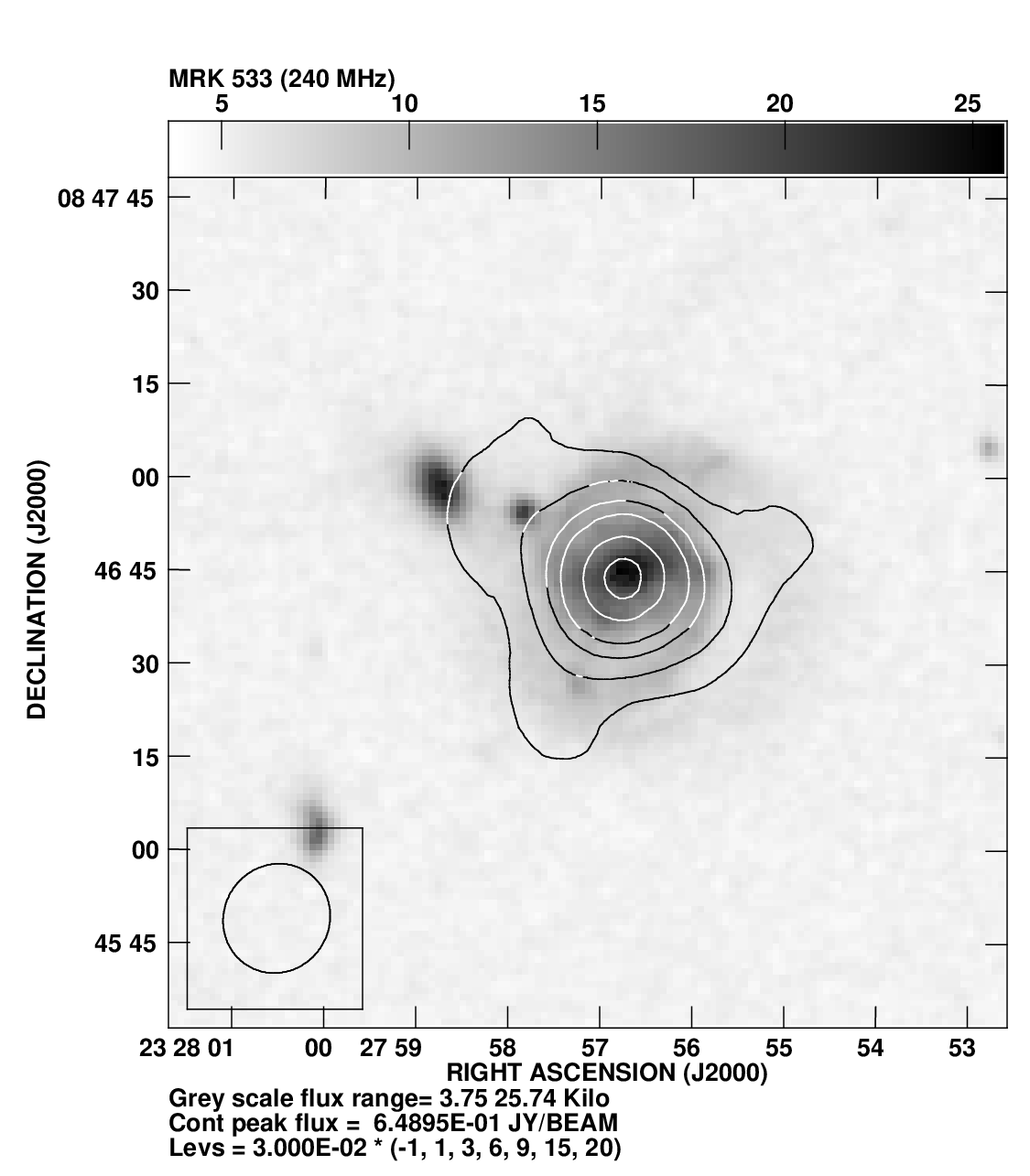}}
\caption{{\it continued}: 610 MHz (left panel) and 240 MHz (right panel) radio contours overlaid on their DSS optical images.}
\end{figure*}
% 
% \addtocounter{figure}{-1}
% % 
% \begin{figure*}[!htbp]
% \centering
% \caption{{\it continued}: 610 MHz ({\it right panel}) and 240 MHz ({\it left panel}) radio contours overlaid on their DSS optical images.}
% \end{figure*}
% % 
% \addtocounter{figure}{-1}
% % 
% \begin{figure*}[!htbp]
% \centering
% \caption{{\it continued}: 610 MHz ({\it right panel}) and 240 MHz ({\it left panel}) radio contours overlaid on their DSS optical images.}
% \end{figure*}
% 
% 
\section{Radio properties of Seyfert galaxies} 
From our GMRT observations we obtained 240 MHz and 610 MHz radio images of all our 20 Seyfert galaxies (\cf Figure 2.0, Table 3.0 and Table 4.0). 
In the following sections we discuss the radio properties ({\ie}luminosities, spectra and morphologies) of Seyfert type 1s and 2s in the 
framework of the Seyfert unification scheme. 
\subsection{Radio luminosities of Seyfert type 1s and type 2s}
The obscuring torus around the AGN is optically thin at centimeter radio wavelengths and 
there is no relativistic beaming effect in Seyfert galaxies \citep{Shastri03,Middelberg04,Ulvestad05}. Therefore, 
the measured radio luminosity is expected to be independent to the orientation of the obscuring torus and the radio jet axis.
According to the unification scheme both the Seyfert subtypes are intrinsically similar and therefore, both type 1s and type 2s are 
expected to show similar radio luminosities. 
This prediction of the unification scheme has been tested in some previous studies yielding different results. 
For example, early studies reported in \cite{deBruyn78,Ulvestad84a} suggest that Seyfert type 2s are more radio powerful than Seyfert type 1s 
at centimeter wavelengths, giving results inconsistent with the unification scheme. 
However, later studies \citep{Ho01} argued that the samples used in earlier studies were biased towards radio powerful Seyfert type 2s.
High resolution 3.6 cm VLA observations showed that the two Seyfert subtypes have similar radio luminosities in samples 
based on 12 $\mu$m \citep{Thean01} and 60 $\mu$m \citep{Schmitt01b}.
We make statistical comparison of the radio luminosities of the two subtypes using 240 MHz, 610 MHz GMRT observations and 
1.4 GHz (NVSS), 5.0 GHz observations of relatively low resolution ($\sim$ 20$\arcsec$ - 45$\arcsec$). 
Our observations are likely to pick up extended kpc-scale emission along with the nuclear AGN emission.
In the unification scheme both Seyfert subtypes expected to have similar likelihood of the presence of kpc-scale radio emission. 
Therefore, type 1s and type 2s are expected to show similar luminosity distributions obtained with low as well as high frequency/resolution 
radio observations.
Figure 3.0 shows the radio luminosity distributions at 240 MHz, 610 MHz, 1.4 GHz and 5.0 GHz of the type 1s and type 2s of our sample Seyferts.      
Seyfert type 1s have 240 MHz, 610 MHz, 1.4 GHz and 5.0 GHz luminosities in the range of 
L$_{\rm 240~MHz}$ $\sim$ 6.09$\times$10$^{28}$ - 7.87$\times$10$^{30}$ erg s$^{-1}$ Hz$^{-1}$, 
L$_{\rm 610~MHz}$ $\sim$ 5.30$\times$10$^{28}$ - 4.58$\times$10$^{30}$ erg s$^{-1}$ Hz$^{-1}$,
L$_{\rm 1.4~GHz}$ $\sim$ 3.49$\times$10$^{28}$ - 2.24$\times$10$^{30}$ erg s$^{-1}$ Hz$^{-1}$, 
and L$_{\rm 5.0~GHz}$ $\sim$ 1.25$\times$10$^{28}$ - 8.37$\times$10$^{29}$ erg s$^{-1}$ Hz$^{-1}$, respectively, 
with the median values L$_{\rm 240~MHz, median}$ $\sim$ 2.03$\times$10$^{30}$ erg s$^{-1}$ Hz$^{-1}$, 
L$_{\rm 610~MHz, median}$ $\sim$ 7.88$\times$10$^{29}$ erg s$^{-1}$ Hz$^{-1}$, 
L$_{\rm 1.4~GHz, median}$ $\sim$ 4.16$\times$10$^{29}$ erg s$^{-1}$ Hz$^{-1}$, and 
L$_{\rm 5.0~GHz, median}$ $\sim$ 1.56$\times$10$^{29}$ erg s$^{-1}$ Hz$^{-1}$, respectively.
While type 2s have 240 MHz, 610 MHz, 1.4 GHz and 5.0 GHz luminosities in the range of 
L$_{\rm 240~MHz}$ $\sim$ 7.33$\times$10$^{28}$ - 9.37$\times$10$^{30}$ erg s$^{-1}$ Hz$^{-1}$, 
L$_{\rm 610~MHz}$ $\sim$ 8.26$\times$10$^{28}$ - 7.17$\times$10$^{30}$ erg s$^{-1}$ Hz$^{-1}$,
L$_{\rm 1.4~GHz}$ $\sim$ 5.07$\times$10$^{28}$ - 4.35$\times$10$^{30}$ erg s$^{-1}$ Hz$^{-1}$, 
and L$_{\rm 5.0~GHz}$ $\sim$ 3.09$\times$10$^{28}$ - 1.48$\times$10$^{30}$ erg s$^{-1}$ Hz$^{-1}$, respectively, 
with the median values L$_{\rm 240~MHz, median}$ $\sim$ 1.51$\times$10$^{30}$ erg s$^{-1}$ Hz$^{-1}$, 
L$_{\rm 610~MHz, median}$ $\sim$ 6.75$\times$10$^{29}$ erg s$^{-1}$ Hz$^{-1}$, 
L$_{\rm 1.4~GHz, median}$ $\sim$ 4.32$\times$10$^{29}$ erg s$^{-1}$ Hz$^{-1}$, and 
L$_{\rm 5.0~GHz, median}$ $\sim$ 2.22$\times$10$^{29}$ erg s$^{-1}$ Hz$^{-1}$, respectively. 
We note that the radio luminosity distributions for the two Seyfert subtypes at 240 MHz, 610 MHz, 1.4 GHz 
and 5.0 GHz span over similar range with similar median values at the respective frequencies. 
The two sample Kolmogorov-Smirnov statistical test shows that there is 99$\%$ probability that 
the L$_{\rm 240~MHz}$ and L$_{\rm 610~MHz}$ distributions of the two Seyfert subtypes are drawn from the same parent population.
The 1.4 GHz and 5.0 GHz luminosity distributions of the two Seyfert subtypes are also not statistically different (\cf Table 6.0).
Our results on the comparison of radio luminosity are in complement with the previous studies which reported that the pc-scale nuclear radio 
luminosities at higher frequencies are similar for the two Seyfert subtypes (\eg\cite{Lal11}). 
\begin{figure*}
\centering
\includegraphics[angle=0,width=7.5cm,height=0.26\textheight]{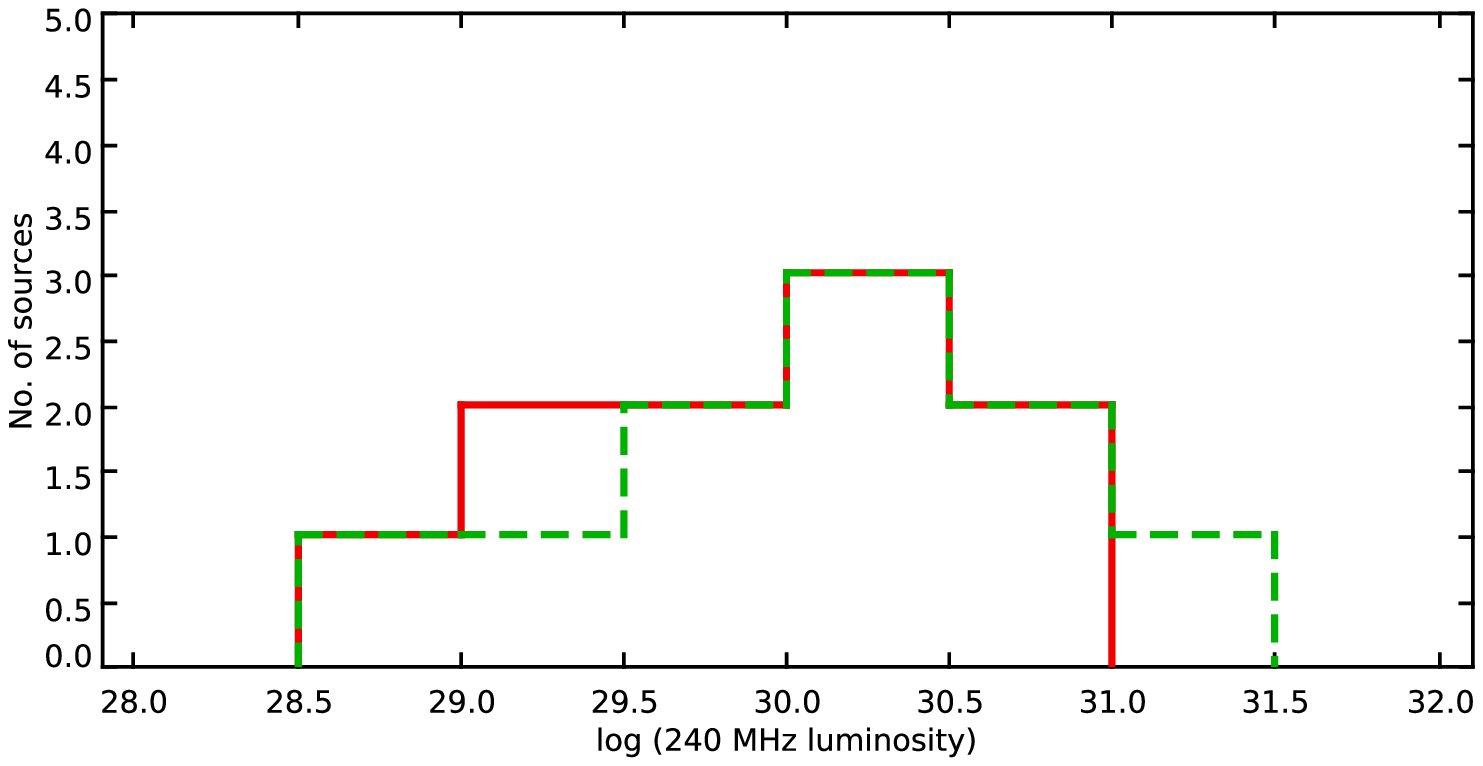}{\includegraphics[angle=0,width=7.5cm,height=0.26\textheight]{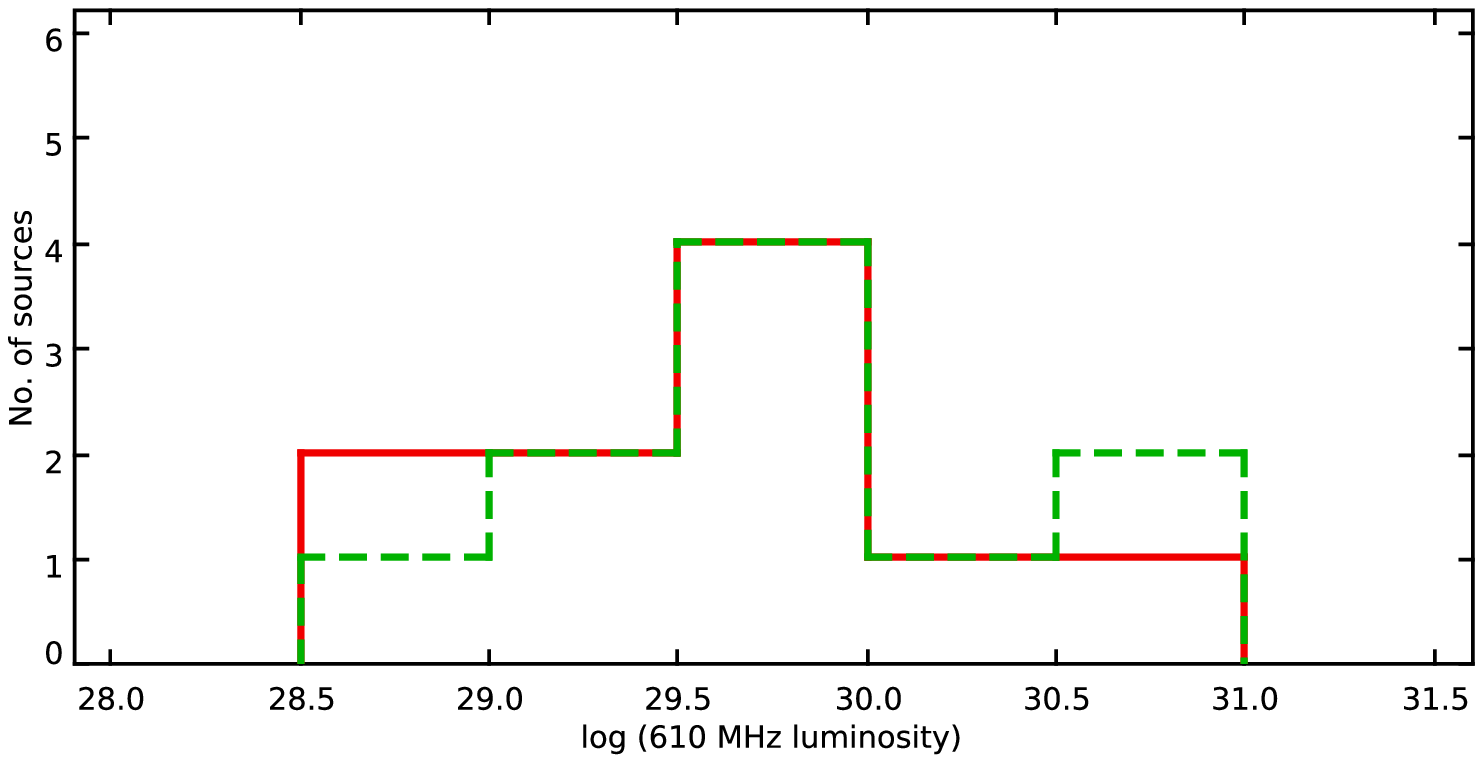}}
\includegraphics[angle=0,width=7.5cm,height=0.26\textheight]{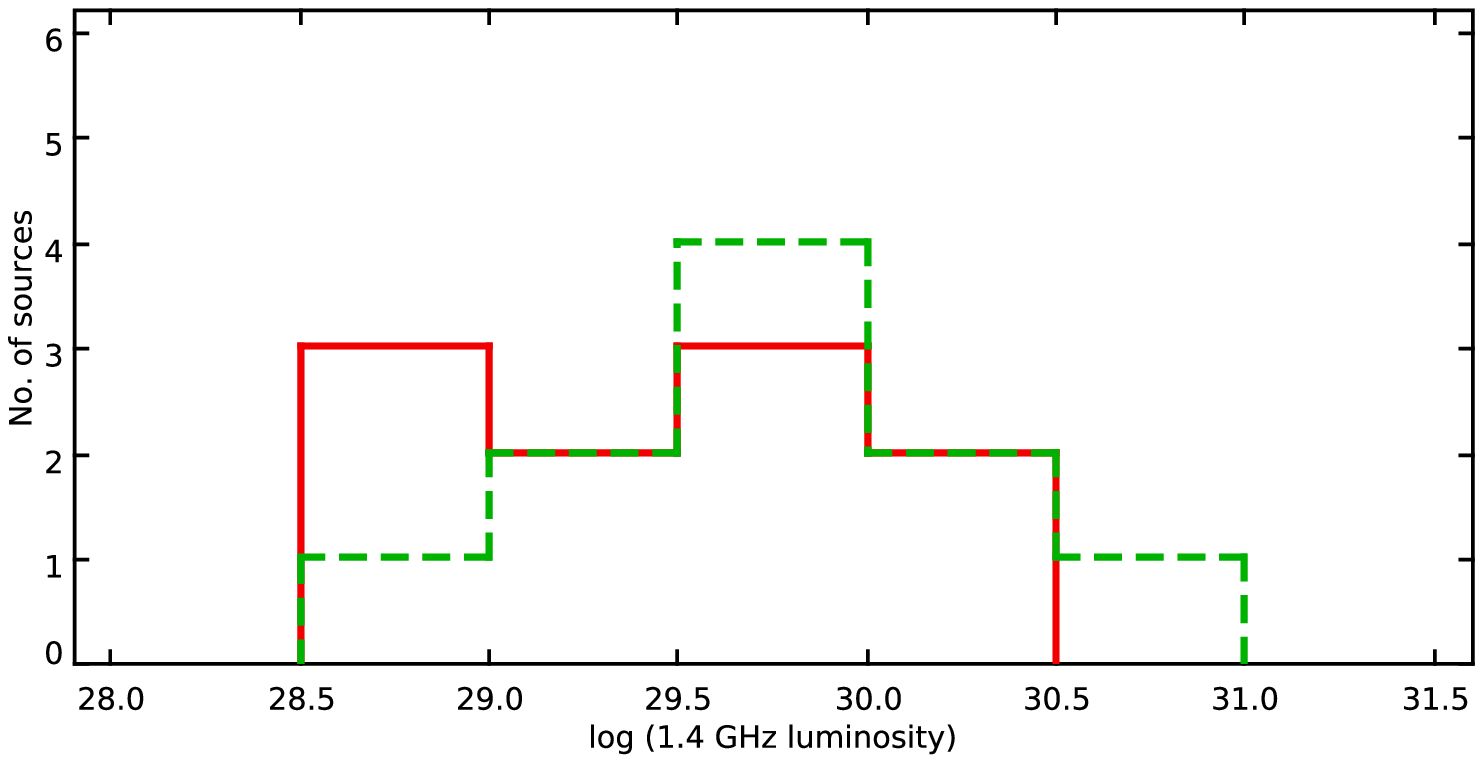}{\includegraphics[angle=0,width=7.5cm,height=0.26\textheight]{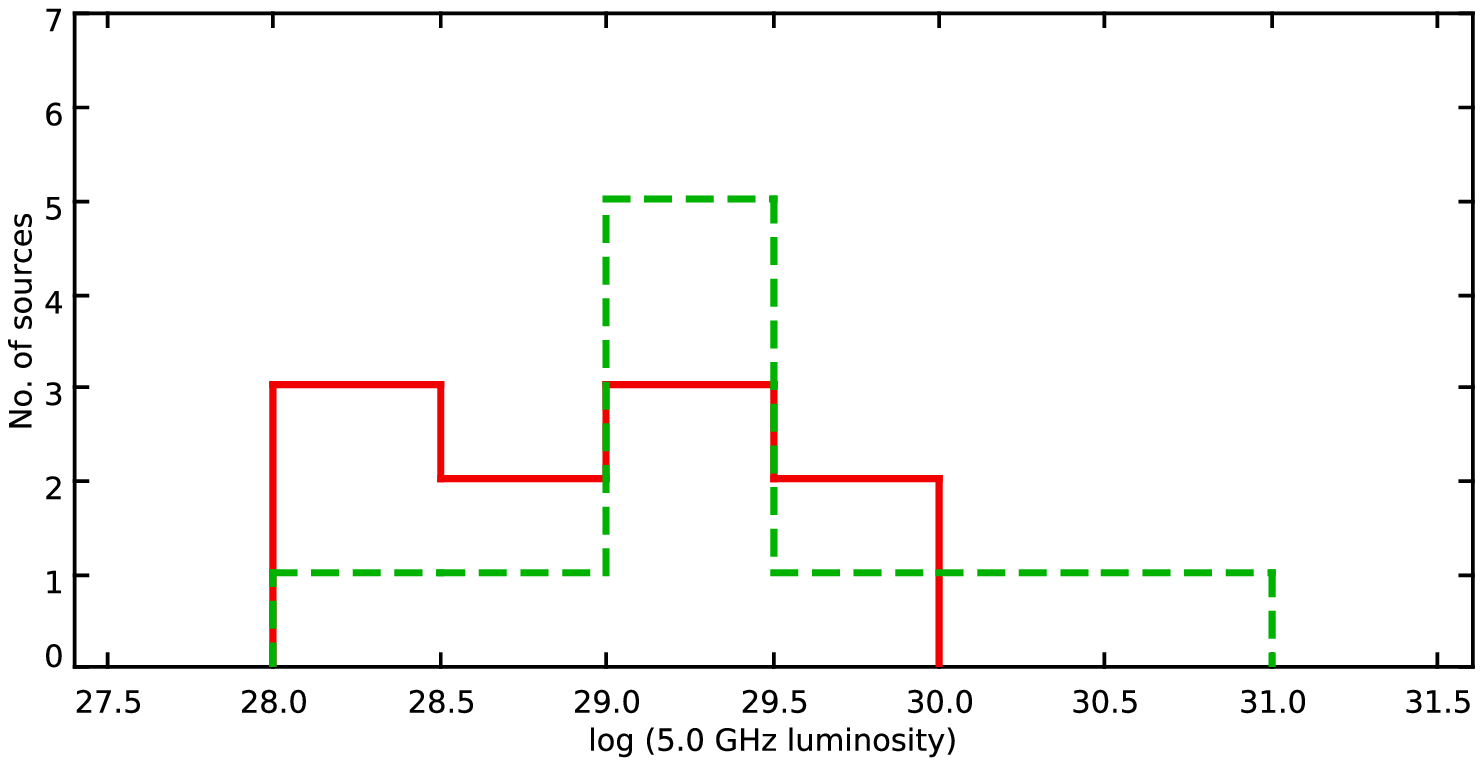}}
\caption{Histograms of radio luminosities at 240 MHz, 610 MHz, 1.4 GHz and 5.0 GHz for Seyfert type 1s and type 2s. 
The histograms for type 1s and type 2s are plotted with `Red colored solid lines' and `Green colored dashed lines', respectively.}
\end{figure*}
\subsection{Radio spectra of Seyfert type 1s and type 2s}
Most of the previous studies on the spectral properties of Seyfert galaxies have been limited mainly to high frequency regime ($\geq$ 1.4 GHz) 
\citep{Barvainis96,Rush96,Morganti99}. Therefore, we aim to explore the nature of the radio spectra of Seyfert galaxies at relatively lower frequencies 
extending down to 240 MHz. And, we compare the radio spectra of Seyfert type 1s and type 2s in the framework of the unification scheme. 
We discussed in the previous section that the radio luminosity/flux in Seyfert galaxies is independent to the orientation of the obscuring torus. 
This result implies that radio spectra should also be similar for the Seyfert type 1s and type 2s.  
This prediction of the unification scheme on the radio spectra have been examined in some of the previous studies reporting varying results. 
For instance, \cite{Edelson87} measured three point (1.4 GHz, 5.0 GHz and 20 GHz) radio spectra of Seyfert galaxies and 
reported that type 1s and 2s show steep ($\alpha$ $\sim$ $-$0.7, S$_{\nu}$ $\propto$ ${\nu}^{\alpha}$) radio spectra 
with type 1s occasionally tend to show flatter or inverted spectra. However, \cite{Rush96} reported that there is no significant 
difference between the average 1.4 GHz - 5.0 GHz spectral indices of Seyfert type 1s and type 2s.
\cite{Barvainis96} studied four point (1.5 GHz, 4.9 GHz, 8.4 GHz and 14.9 GHz) radio spectra and reported that the radio spectral shapes of 
Seyfert galaxies and radio-quiet quasars are quite heterogeneous.
\\
We obtain two point spectral indices and integrated radio spectra over 240 MHz to 5.0 GHz, of our sample Seyfert galaxies using 
240 MHz, 610 MHz flux densities from our GMRT observations, 1.4 GHz flux densities from NVSS \citep{Condon98} and 5.0 GHz flux densities from 
the literature ({\eg}\cite{Gallimore06,Edelson87,Griffith95}). 
Since GMRT observations at 240 MHz, 610 MHz and NVSS observations at 1.4 GHz are of relatively low resolution, 
we considered 5 GHz flux density measured with low resolution observations, 
{\ie}VLA in `D' configuration \citep{Gallimore06,Edelson87} or from single dish Green Bank and Parkes radio telescopes \citep{Griffith95,Gregory91}.
\\
Table 5 lists the two point radio spectral indices measured between 240 MHz to 610 MHz (${\alpha}^{\rm 610 MHz}_{\rm 240 MHz}$), 
610 MHz to 1.4 MHz (${\alpha}^{\rm 1.4 GHz}_{\rm 610 MHz}$) and 1.4 GHz to 5.0 GHz (${\alpha}^{\rm 5.0 GHz}_{\rm 1.4 GHz}$) 
as well as the integrated radio spectral indices (${\alpha}_{\rm int}$) estimated using four flux density points.
Figure 4.0 shows the four point (240 MHz, 610 MHz, 1.4 GHz and 5.0 GHz) radio spectra of all our sample Seyfert galaxies.
The integrated radio spectral index (${\alpha}_{\rm int}$) is obtained by fitting all four spectral points with a linear chi-square fit. 
The slope of the line fit in logarithmic scale gives the index of the powerlaw spectrum (S$_{\nu}$ $\propto$ ${\nu}^{\alpha}$).
Figure 5.0 shows the distributions of two points ${\alpha}^{\rm 610 MHz}_{\rm 240 MHz}$, ${\alpha}^{\rm 1.4 GHz}_{\rm 610 MHz}$,
${\alpha}^{\rm 5.0 GHz}_{\rm 1.4 GHz}$ and integrated ${\alpha}_{\rm int}$ spectral indices for Seyfert type 1s and type 2s of our sample.
We note that spectral index distributions for the two subtypes span over similar range with median spectral indices 
${\alpha}^{\rm 610 MHz}_{\rm 240 MHz}$ $\simeq$ -0.83, ${\alpha}^{\rm 1.4 GHz}_{\rm 610 MHz}$ $\simeq$ -0.59, 
${\alpha}^{\rm 5.0 GHz}_{\rm 1.4 GHz}$ $\simeq$ -0.77, ${\alpha}_{\rm int}$ $\simeq$ -0.74 for type 1s and 
${\alpha}^{\rm 610 MHz}_{\rm 240 MHz}$ $\simeq$ -0.52, ${\alpha}^{\rm 1.4 GHz}_{\rm 610 MHz}$ $\simeq$ -0.59, 
${\alpha}^{\rm 5.0 GHz}_{\rm 1.4 GHz}$ $\simeq$ -0.67, ${\alpha}_{\rm int}$ $\simeq$ -0.72 for type 2s (\cf Table 6.0).
The statistical comparison using two sample KS test shows that the distributions of spectral indices for the two Seyfert subtypes are not 
very different (\cf Table 6).
\\
Since the synthesized beam size is different at different frequencies ({\eg}$\sim$ 20$^{\prime\prime}$ - 40$^{\prime\prime}$ at 240 MHz, 
$\sim$ 8$^{\prime\prime}$ - 10$^{\prime\prime}$ at 610 MHz, $\sim$ 45$^{\prime\prime}$ at 1.4 GHz 
and $\sim$ 20$^{\prime\prime}$ or larger at 5.0 GHz), it may result an error in the estimated spectral index values. 
For example, the larger NVSS synthesized beam ($\sim$45$^{\prime\prime}$) at 1.4 GHz compared to the GMRT synthesized beam 
($\sim$ 8$^{\prime\prime}$ - 10$^{\prime\prime}$) at 610 MHz may result less steep spectrum than actual. 
We assume a conservative fiducial error values of 15$\%$ at 240 MHz flux density, 10$\%$ at 610 MHz flux density, 7$\%$ at 1.4 GHz flux density 
and 7$\%$ at 5.0 GHz flux density. 
The assumed errors in flux densities can result 
$\sim$ 8$\%$, $\sim$ 5$\%$, $\sim$ 4$\%$ and $\sim$ 7$\%$ errors in ${\alpha}^{\rm 610 MHz}_{\rm 240 MHz}$, 
${\alpha}^{\rm 1.4 GHz}_{\rm 610 MHz}$, ${\alpha}^{\rm 5.0 GHz}_{\rm 1.4 GHz}$ and ${\alpha}_{\rm int}$ respectively. 
Also, non-simultaneous observations may contribute to error in ${\alpha}^{\rm 1.4 GHz}_{\rm 610 MHz}$, ${\alpha}^{\rm 5.0 GHz}_{\rm 1.4 GHz}$ and 
${\alpha}_{\rm int}$ estimates, if the source flux density varies in between two observations. Moreover, the estimates of 
${\alpha}^{\rm 610 MHz}_{\rm 240 MHz}$ are free from the error due to non-simultaneity since 240 MHz and 610 MHz flux densities are 
from simultaneous GMRT observations. 
\\
In MRK 348, the 5.0 GHz flux density ($\sim$ 807.1 mJy from VLA `D' array observations \citep{Gallimore06}) is much higher than expected 
from powerlaw spectral shape determined by 240 MHz, 610 MHz and 1.4 GHz flux densities. Therefore, we consider 5.0 GHz flux density as an outlier 
in fitting the radio spectrum of MRK 348. The unusual high flux density at 5.0 GHz can be attributed to strong variability 
as the core of MRK 348 is variable at 5.0 GHz on a scale of months \citep{Neff83,Ulvestad99}.
However, it is worth noting that except in few cases \citep{Neff83,Wrobel2000,Falcke2000}, 
most of the Seyfert galaxies show little radio variability over the period of few years \citep{Edelson87,Mundell09} 
and therefore our statistical results are not expected to be affected much by variability. 
\\
We note that most of the Seyfert galaxies in our sample have steep integrated radio spectra (${\alpha}_{\rm int}$ $\sim$ -0.65 to -0.85), 
except NGC 5252 and NGC 5728 which show flat and inverted spectrum, respectively. 
Since NGC 5728 shows inverted spectrum over 240 MHz to 5.0 GHz, we do not obtain integrated spectral 
index measured by linear chi-square fit. 
There are a few sources {\eg}NGC 3227, MRK 1066, NGC 5548 which show hint of spectral flattening at lower frequency. 
While a few cases ({\eg}NGC 3516, MRK 530) show hint for spectral steepening at lower frequencies.
The steep radio spectrum can be interpreted as emission produced via synchrotron radiation by a population of relativistic electrons and 
are typically associated to relatively extended emission. 
% While flat or inverted spectral indices can be ascribed to a synchrotron self absorption
% mechanism or to a thermal process like Bremsstrahlung.
% 
The integrated radio spectral shape of Seyfert galaxy can be attributed to the combined contributions of all the radio emitting components.
In low resolution radio observations, the radio emission from Seyfert galaxies can have contributions from four components, 
{\ie}a partially opaque synchrotron emission from a compact parsec-scale nuclear core, optically thin synchrotron emission from an extended component 
powered by AGN, optically thin synchrotron emission from star forming regions present in the host galaxy disk and 
a circumnuclear starburst emission \citep{Wilson91}.
The relative fraction of these emitting components may vary from one source to another and in turn, may change the spectral shape. 
Previous radio studies have shown that the total radio emission in Seyfert galaxies may have contributions from host galaxy disk and 
from starburst regions but the radio emission in most of the Seyfert galaxies is dominated by nuclear radio emission 
characterized with high brightness temperature, steep spectrum and non-thermal emission \citep{Kukula98}.
The extended emission powered by AGN as well as star formation give rise the steep spectrum and therefore it is difficult to 
conclude whether emission is either powered by AGN or starburst by using only spectral shape information. 
In case of sources showing flat or inverted spectra, the total radio emission 
is likely to be dominated by the compact nuclear core which is partially opaque to synchrotron emission \citep{Kukula98}. 
The compact nuclear radio emission characterized with high brightness temperature ($\sim$ 10$^{8}$ K) and 
inverted spectrum seen in some of Seyfert galaxies 
is indicative of the synchrotron self-absorption close to the jet-emanating region \citep{Mundell2000}, however, 
the free-free absorption by thermal, ionized gas in the vicinity of the nucleus might also be sufficient to flatten the 
intrinsically steeper synchrotron spectra \citep{Ho01}.
\begin{figure*}[!htbp]
\centering
\includegraphics[angle=0,width=7.2cm,height=0.23\textheight]{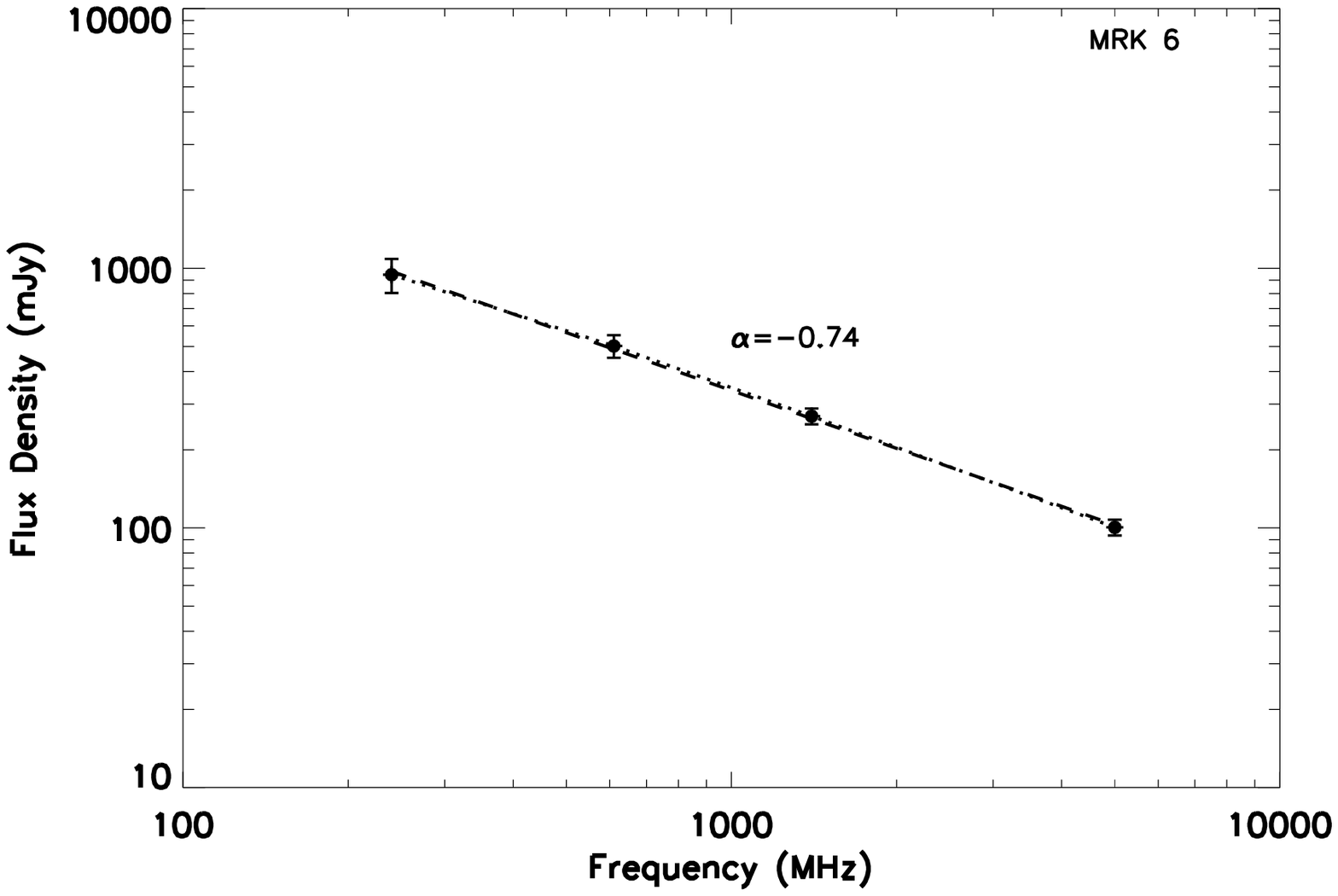}{\includegraphics[angle=0,width=7.2cm,height=0.23\textheight]{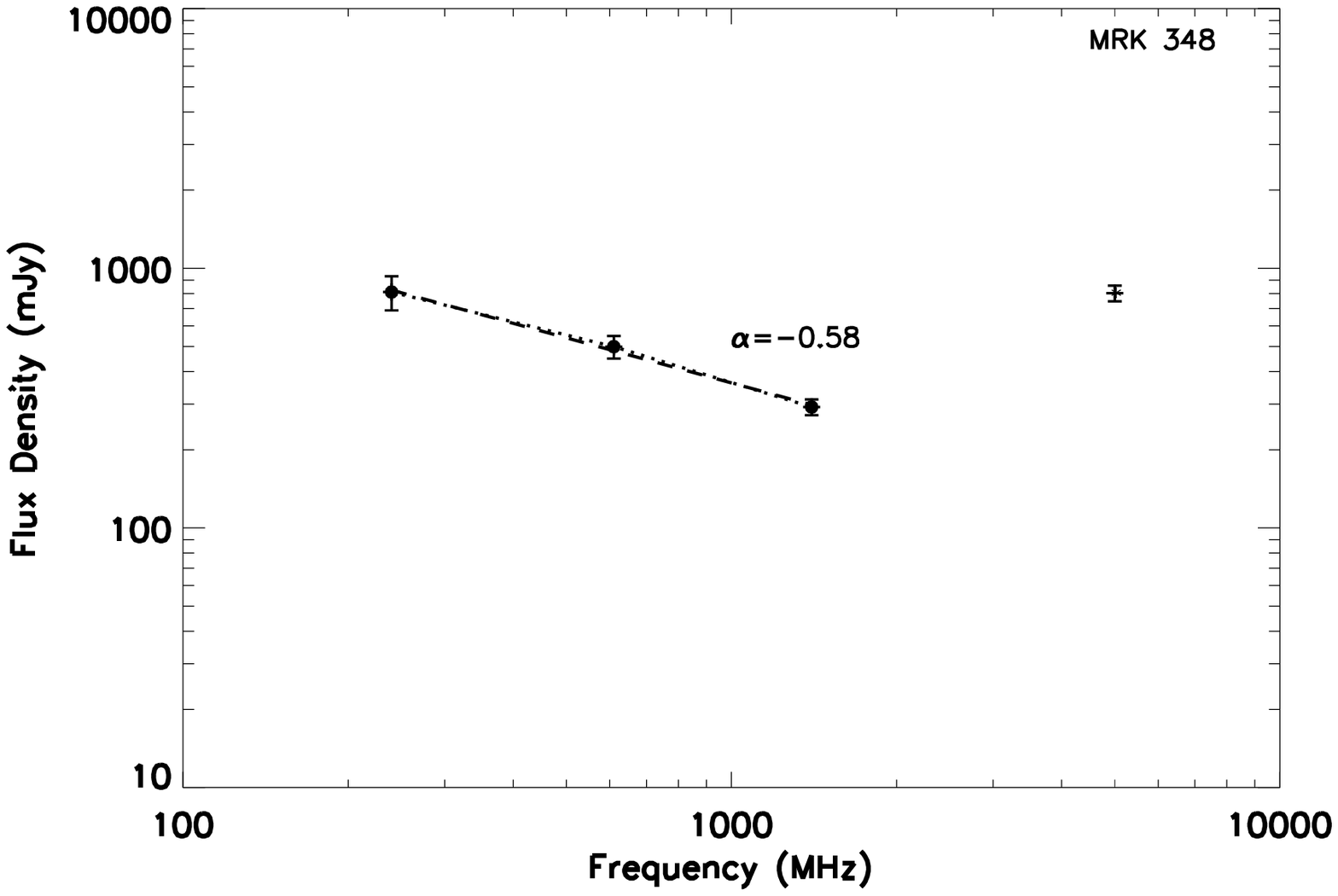}}
\includegraphics[angle=0,width=7.2cm,height=0.23\textheight]{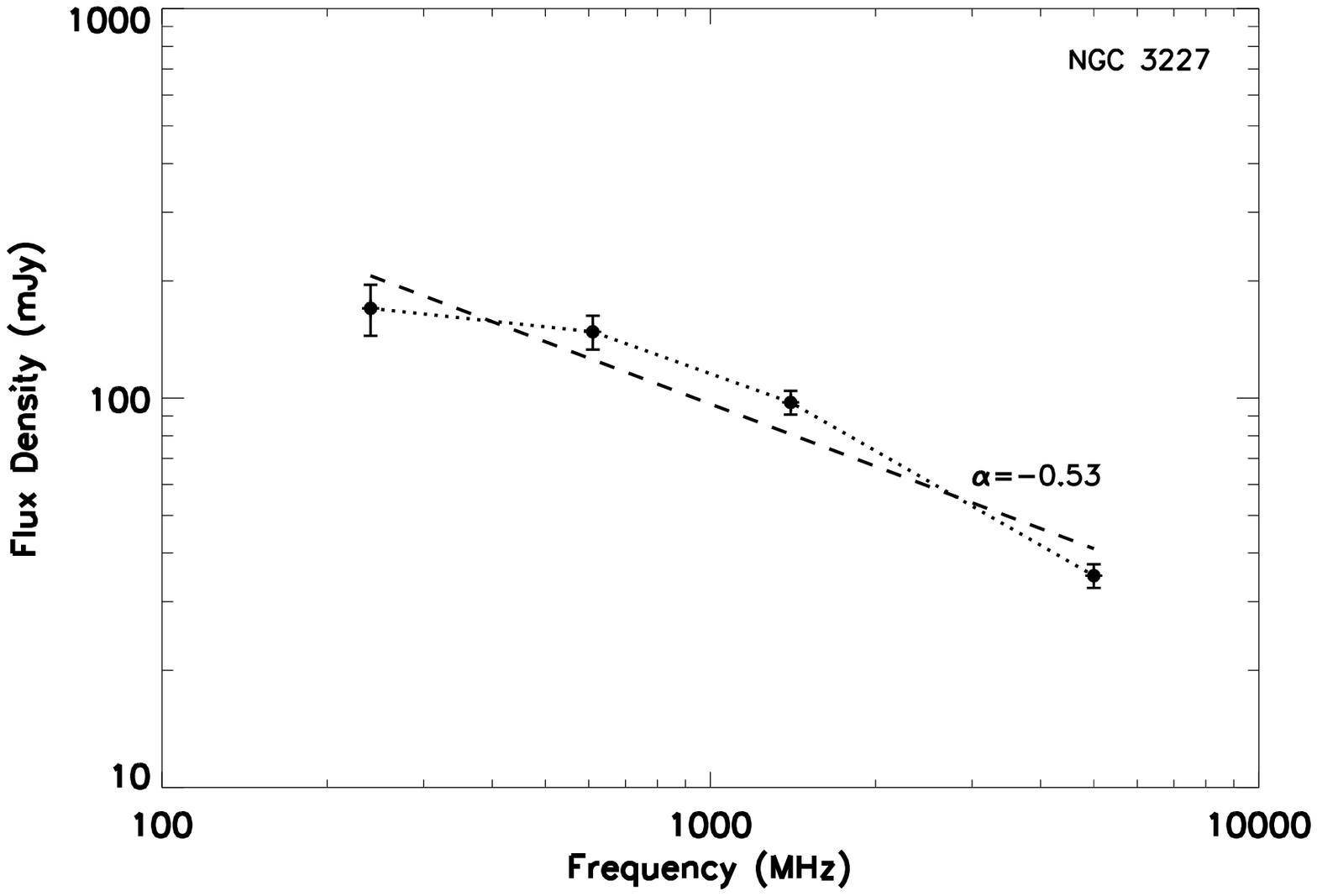}{\includegraphics[angle=0,width=7.2cm,height=0.23\textheight]{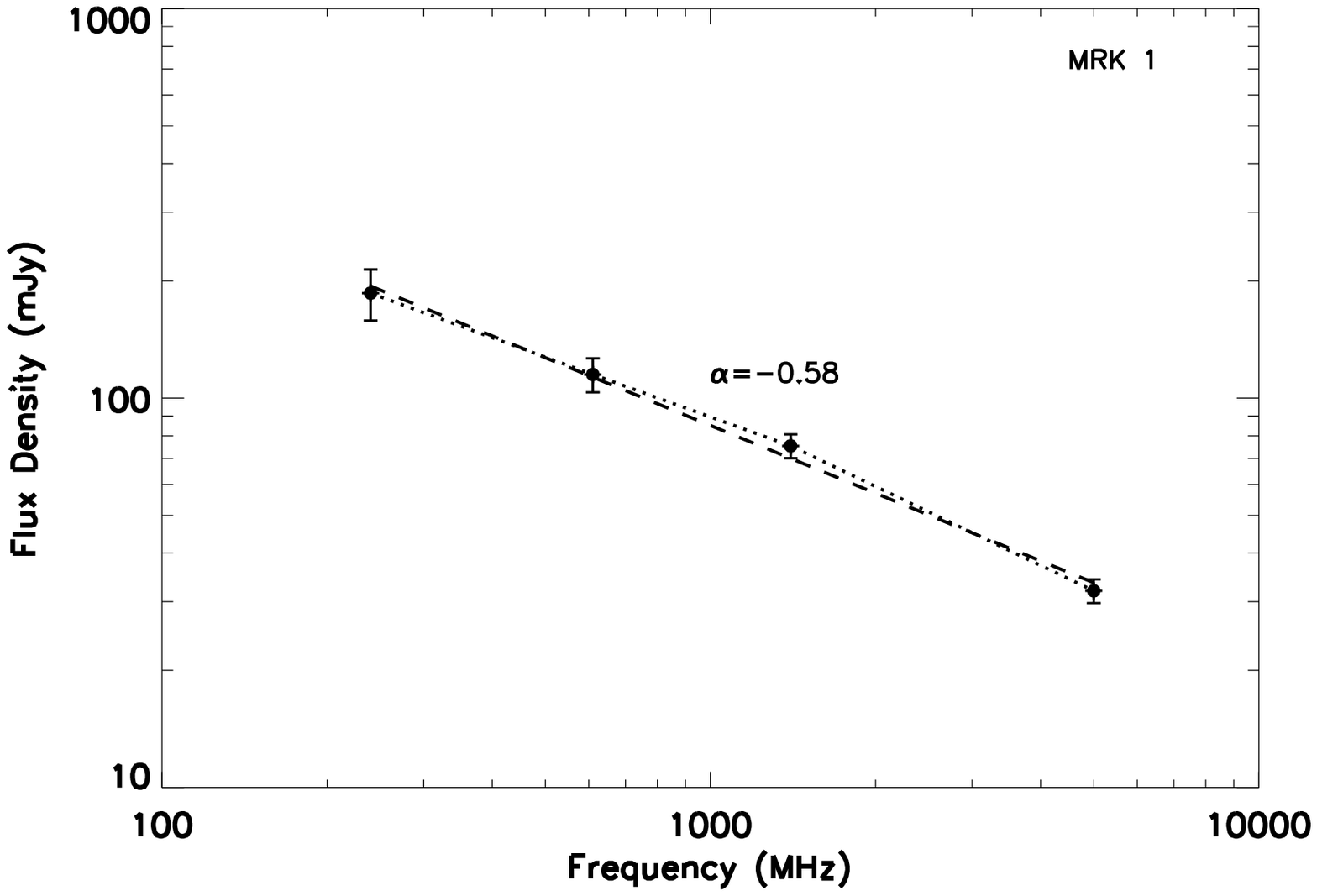}}
\includegraphics[angle=0,width=7.2cm,height=0.23\textheight]{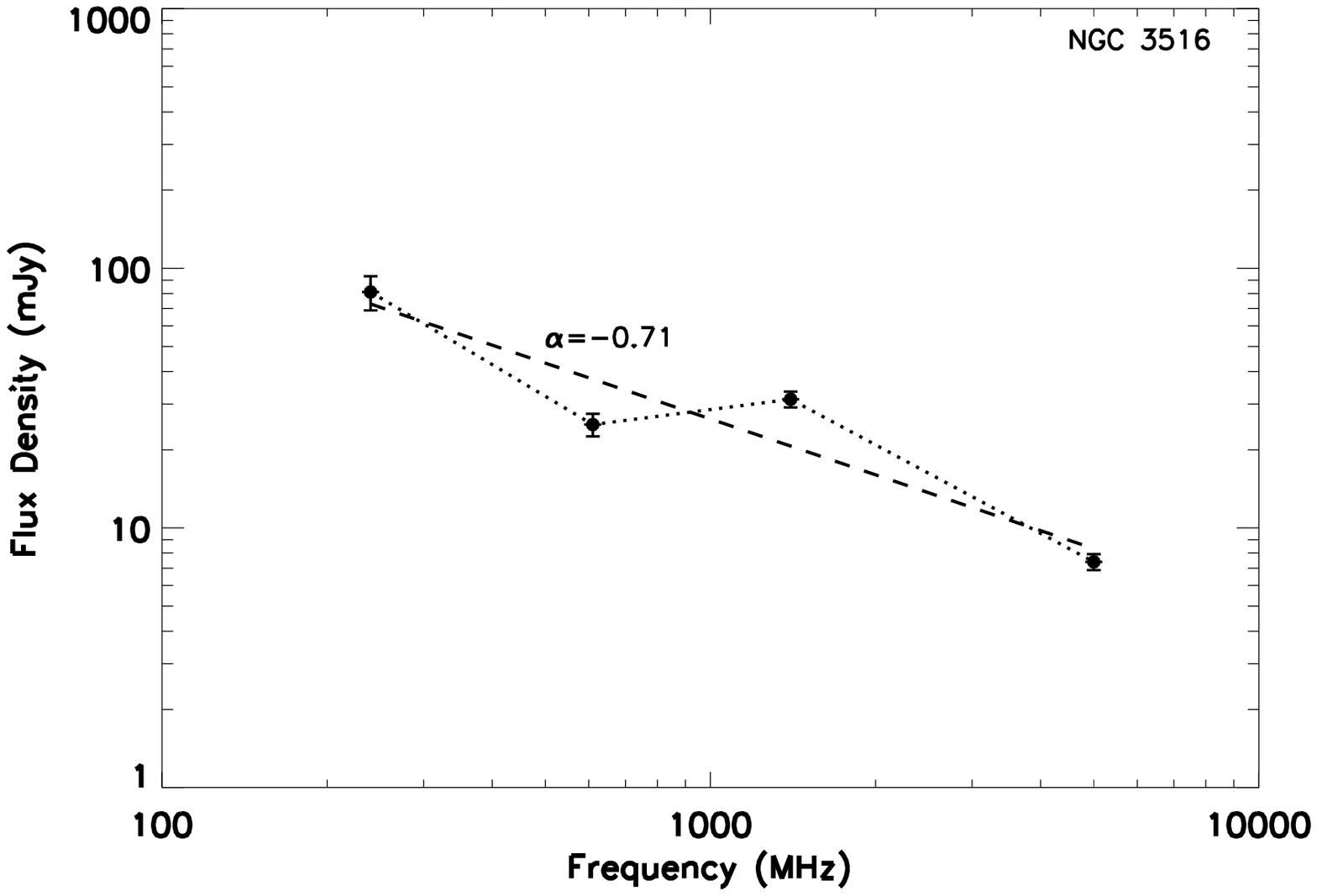}{\includegraphics[angle=0,width=7.2cm,height=0.23\textheight]{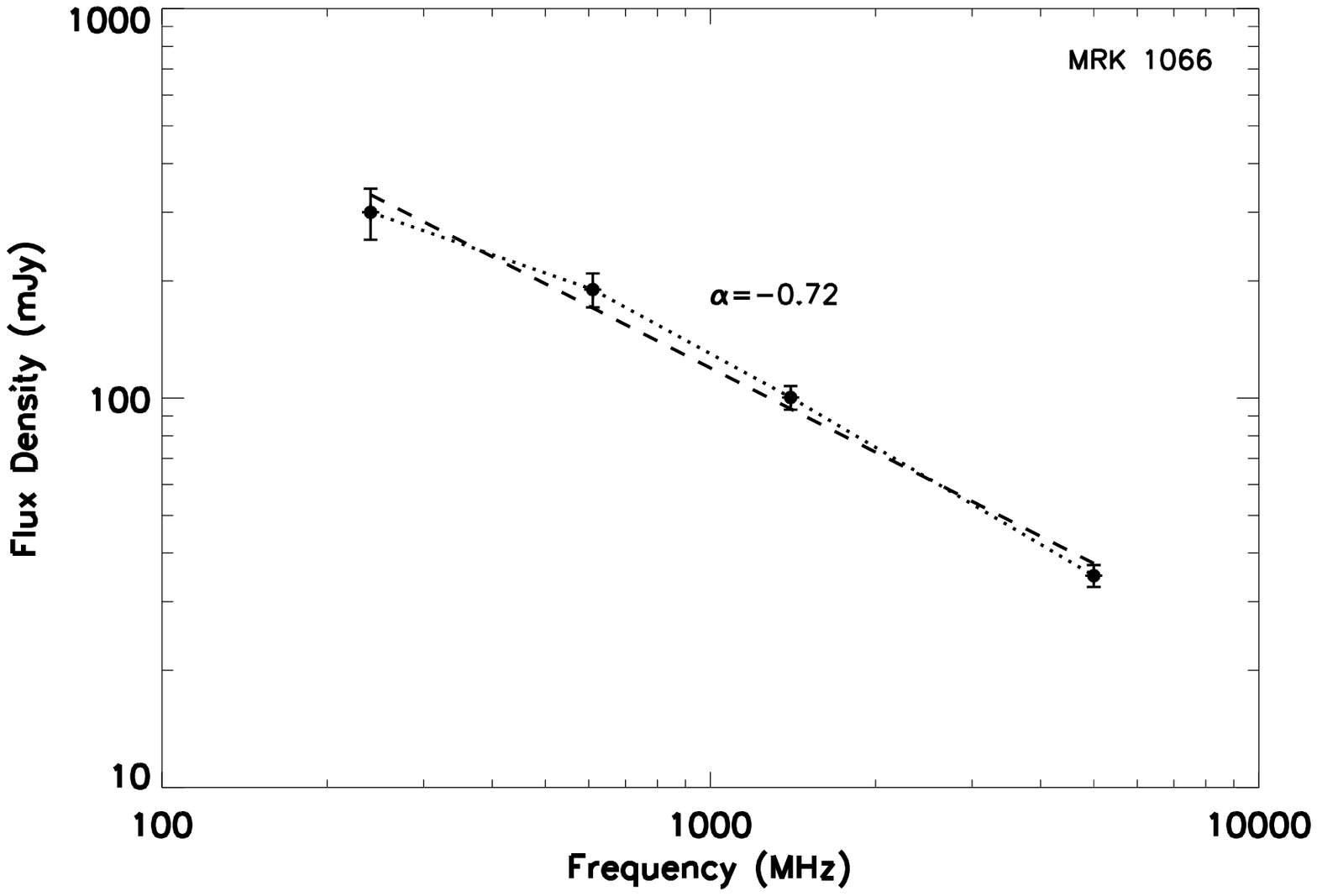}}
\includegraphics[angle=0,width=7.2cm,height=0.23\textheight]{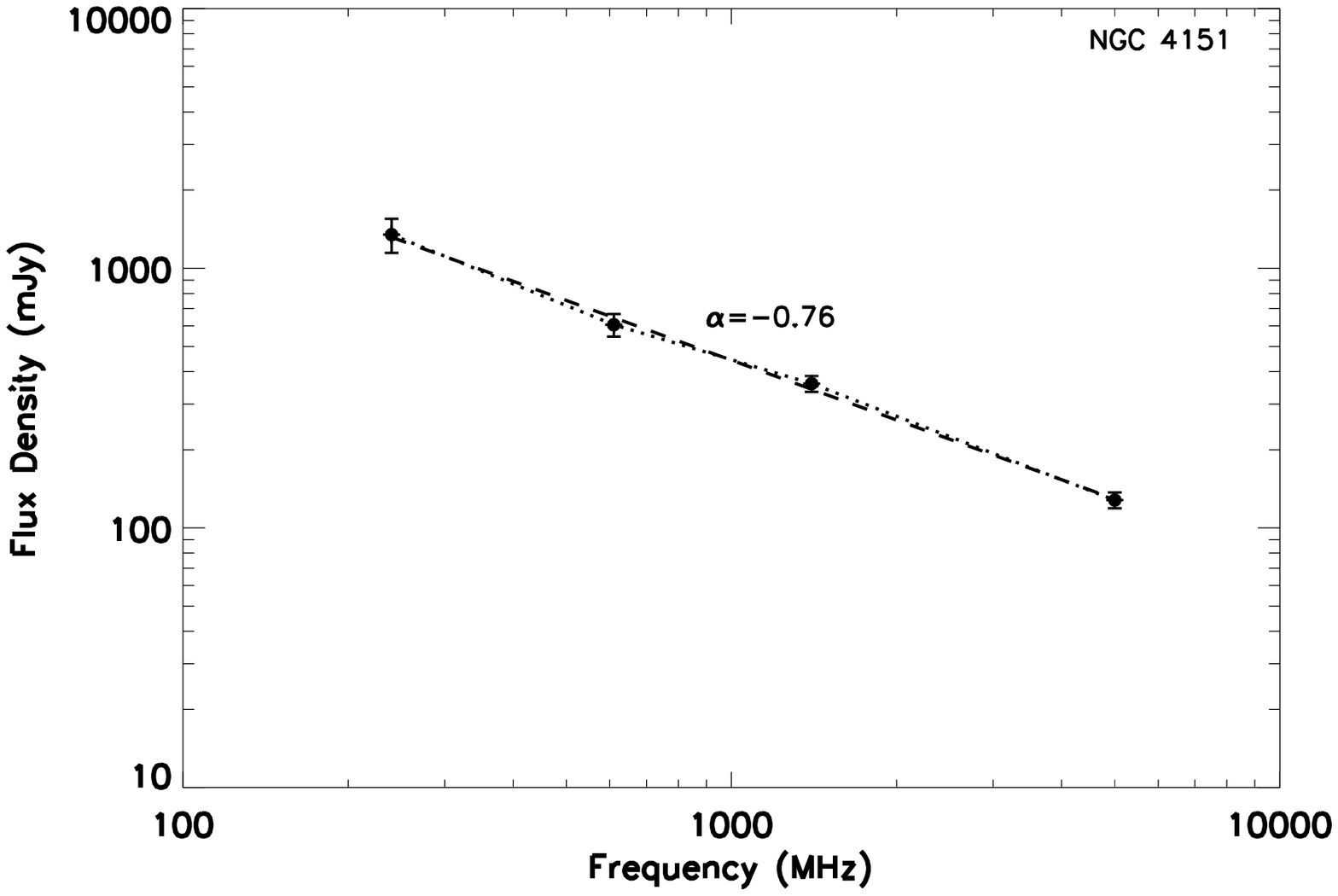}{\includegraphics[angle=0,width=7.2cm,height=0.23\textheight]{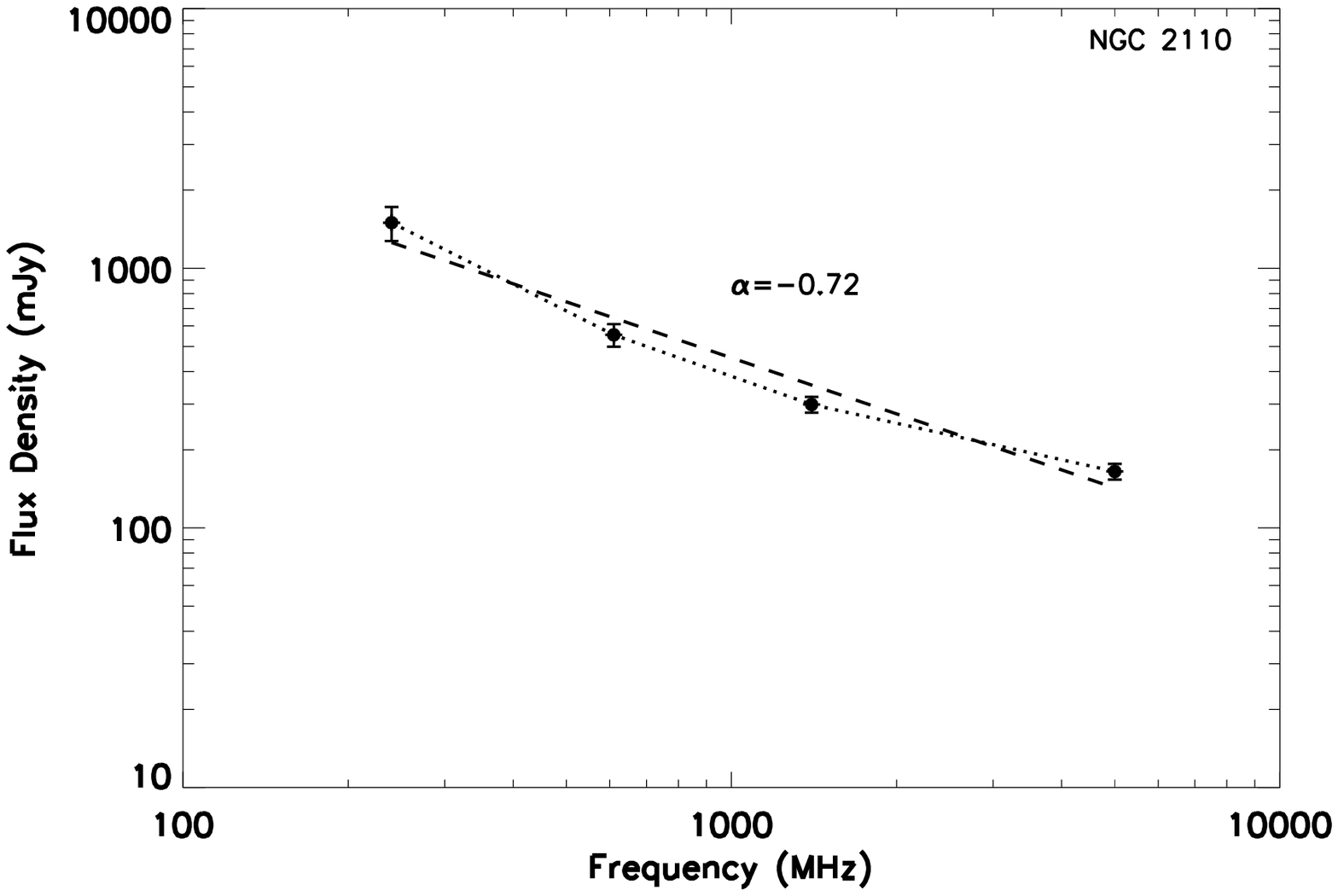}}
\caption{Four point (240 MHz, 610 MHz, 1.4 GHz and 5.0 GHz) radio spectra of Seyfert galaxies of our sample. 
The dotted line connects the flux density points and dashed line shows the least chi-square linear fit to the spectral points. 
Radio spectra of Seyfert type 1s and type 2s are shown in left and right panel, respectively.}
\end{figure*}
\addtocounter{figure}{-1}
\begin{figure*}[!htbp]
\centering
\includegraphics[angle=0,width=7.2cm,height=0.23\textheight]{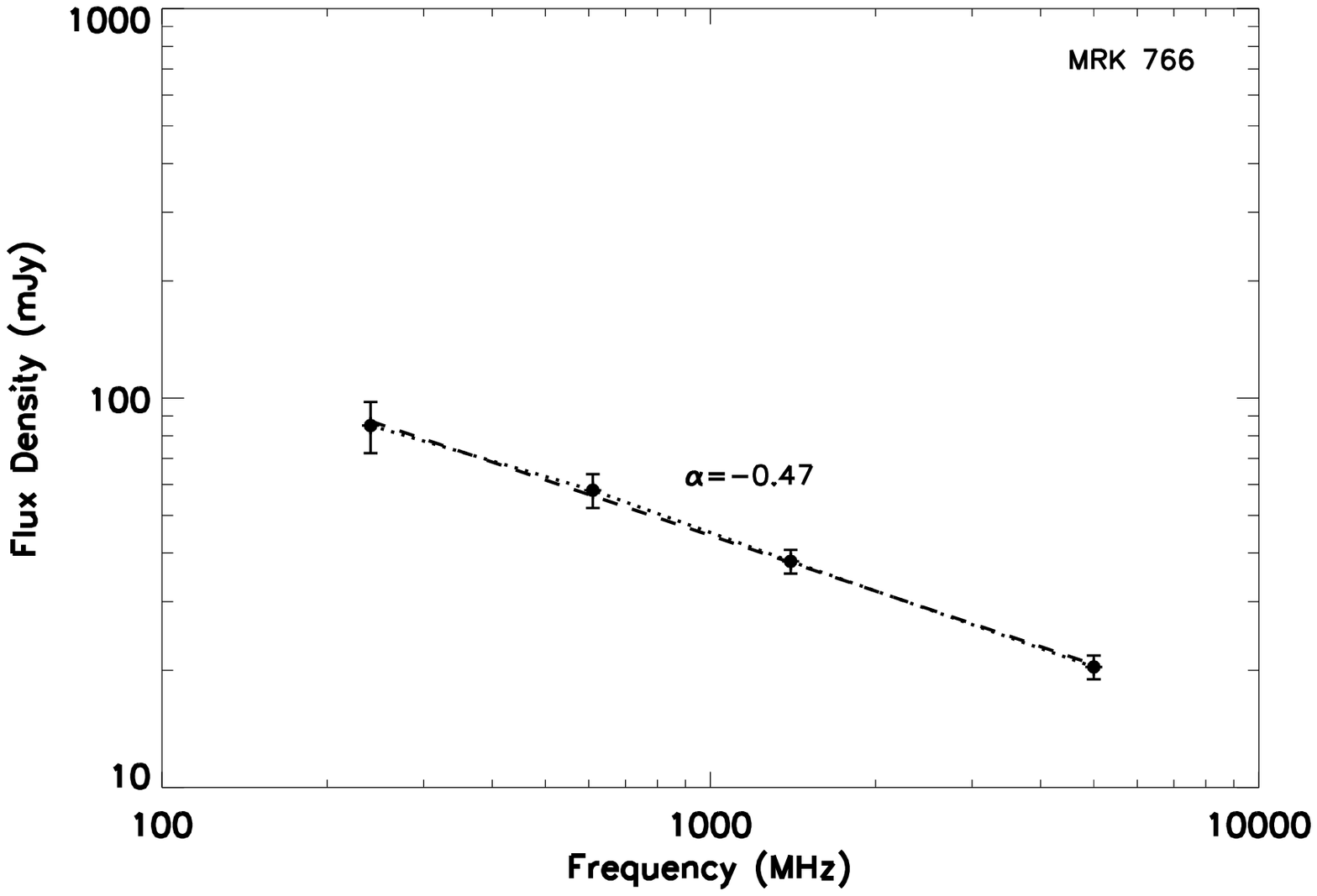}{\includegraphics[angle=0,width=7.2cm,height=0.23\textheight]{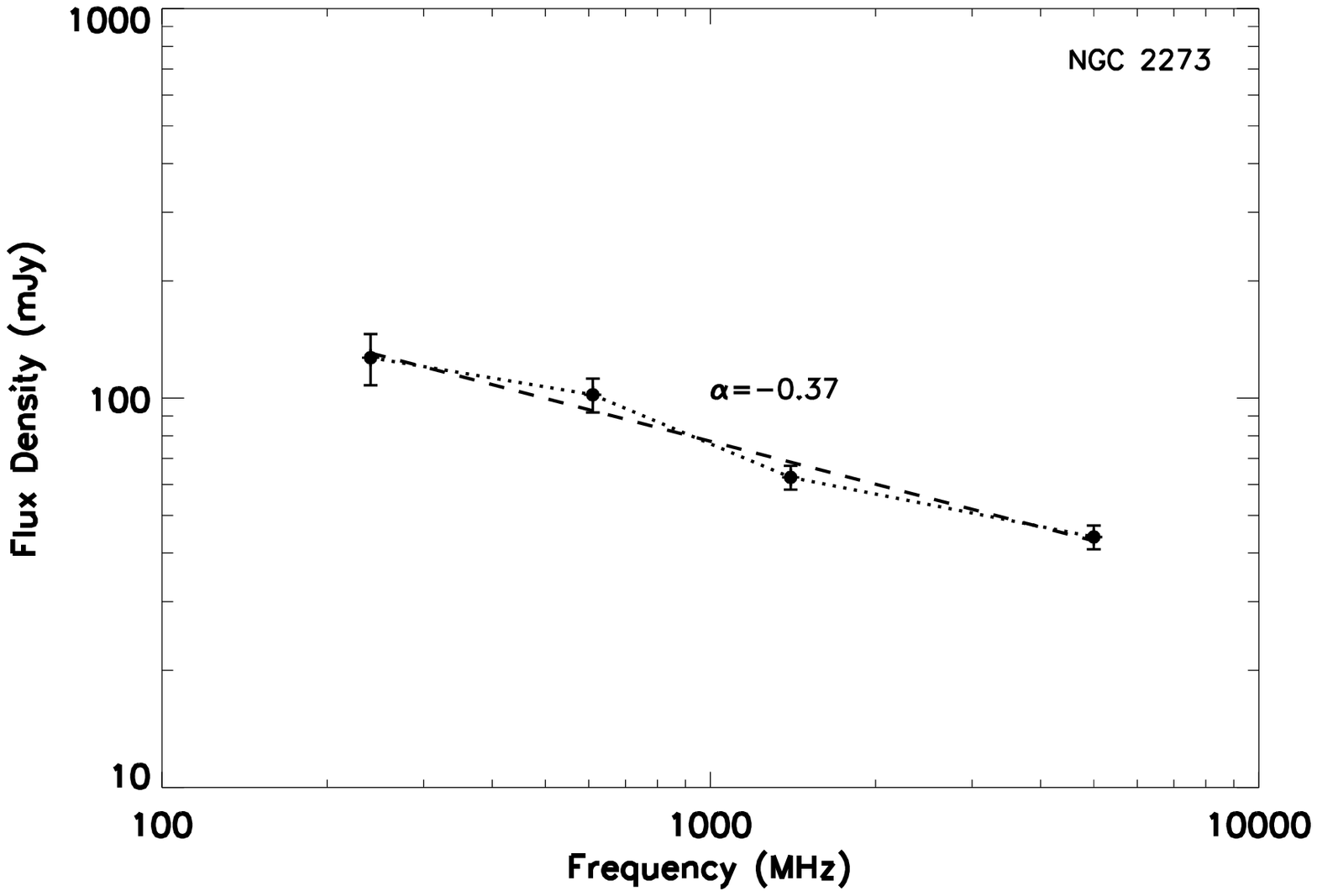}}
\includegraphics[angle=0,width=7.2cm,height=0.23\textheight]{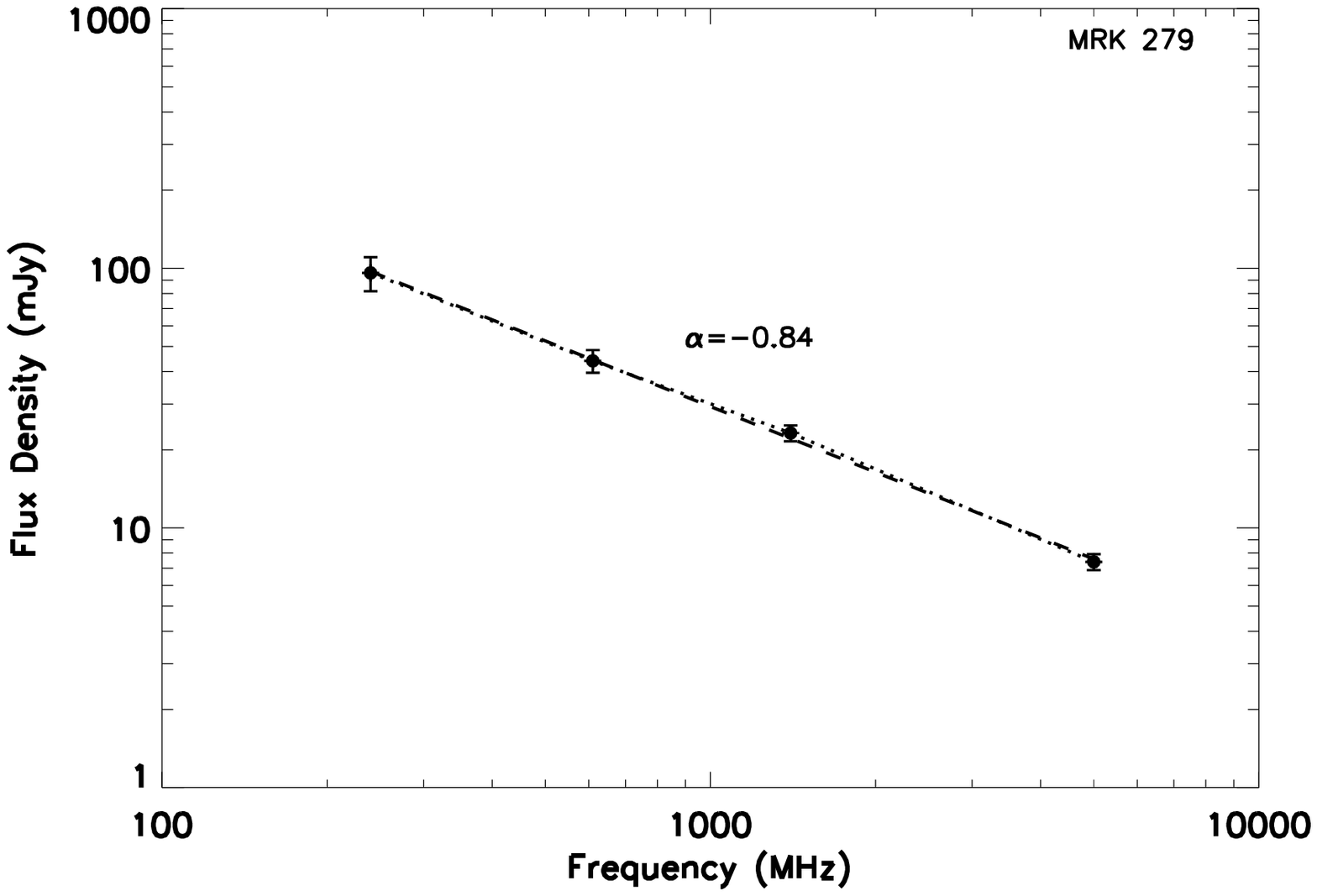}{\includegraphics[angle=0,width=7.2cm,height=0.23\textheight]{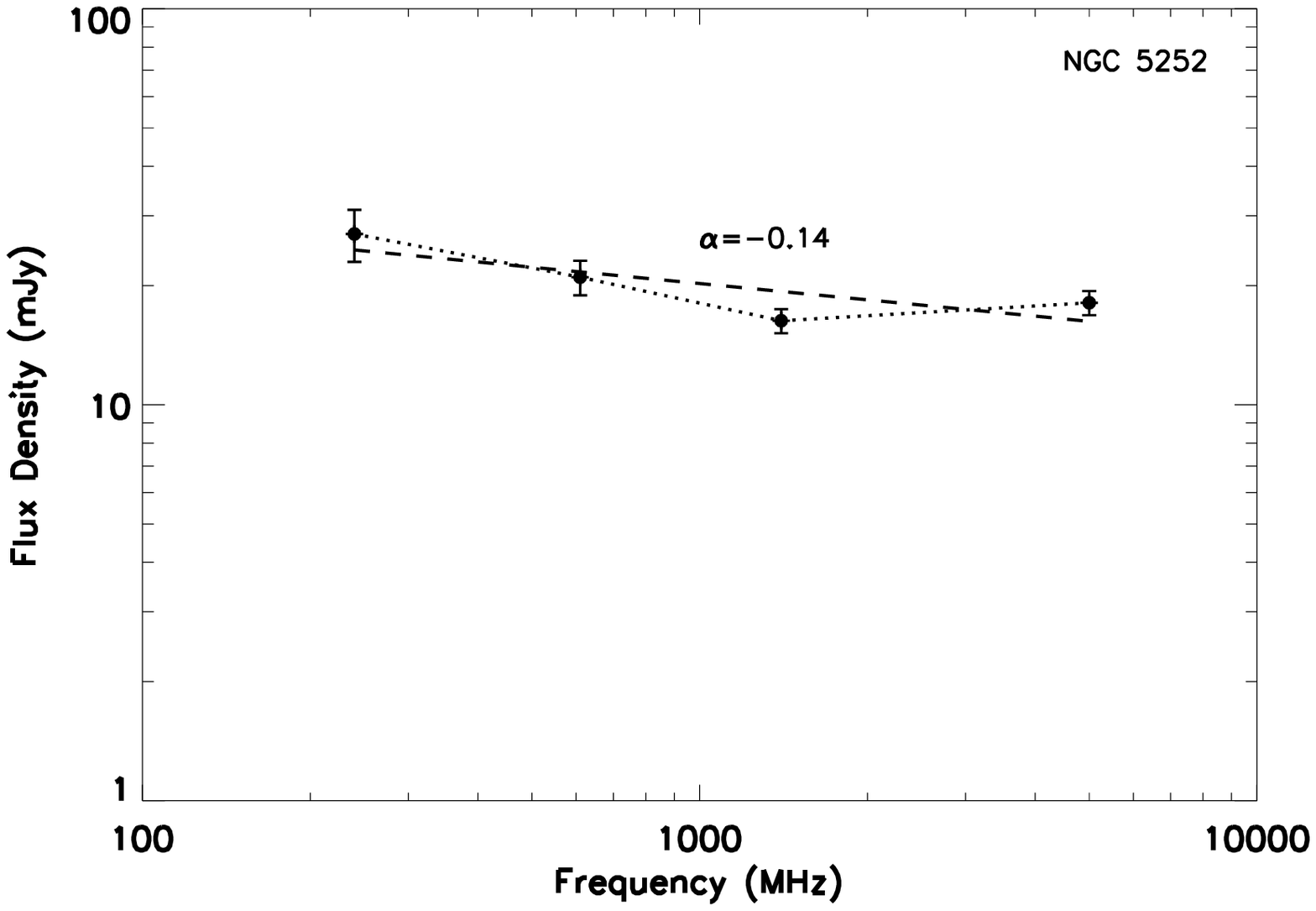}}
\includegraphics[angle=0,width=7.2cm,height=0.23\textheight]{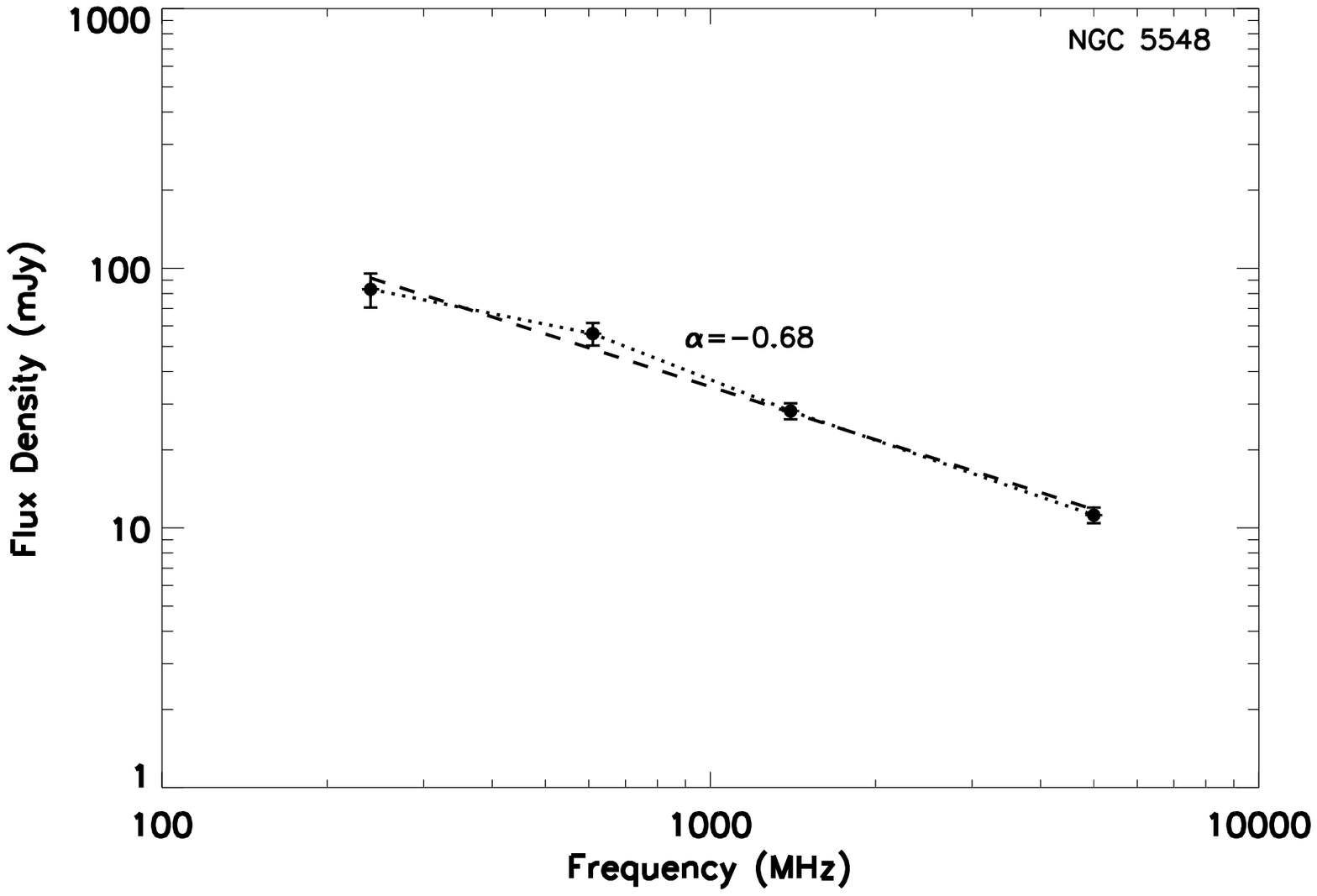}{\includegraphics[angle=0,width=7.2cm,height=0.23\textheight]{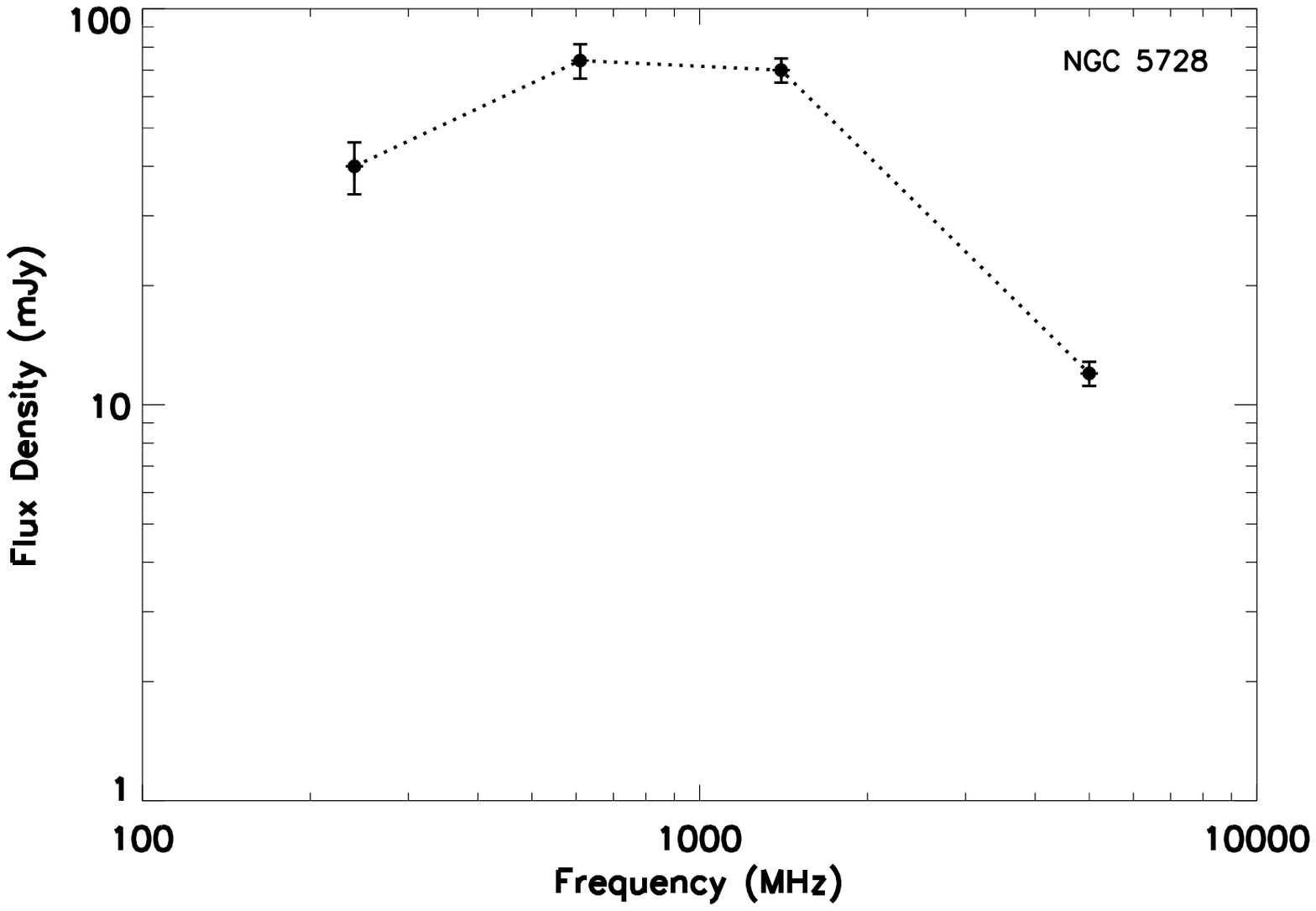}}
\includegraphics[angle=0,width=7.2cm,height=0.23\textheight]{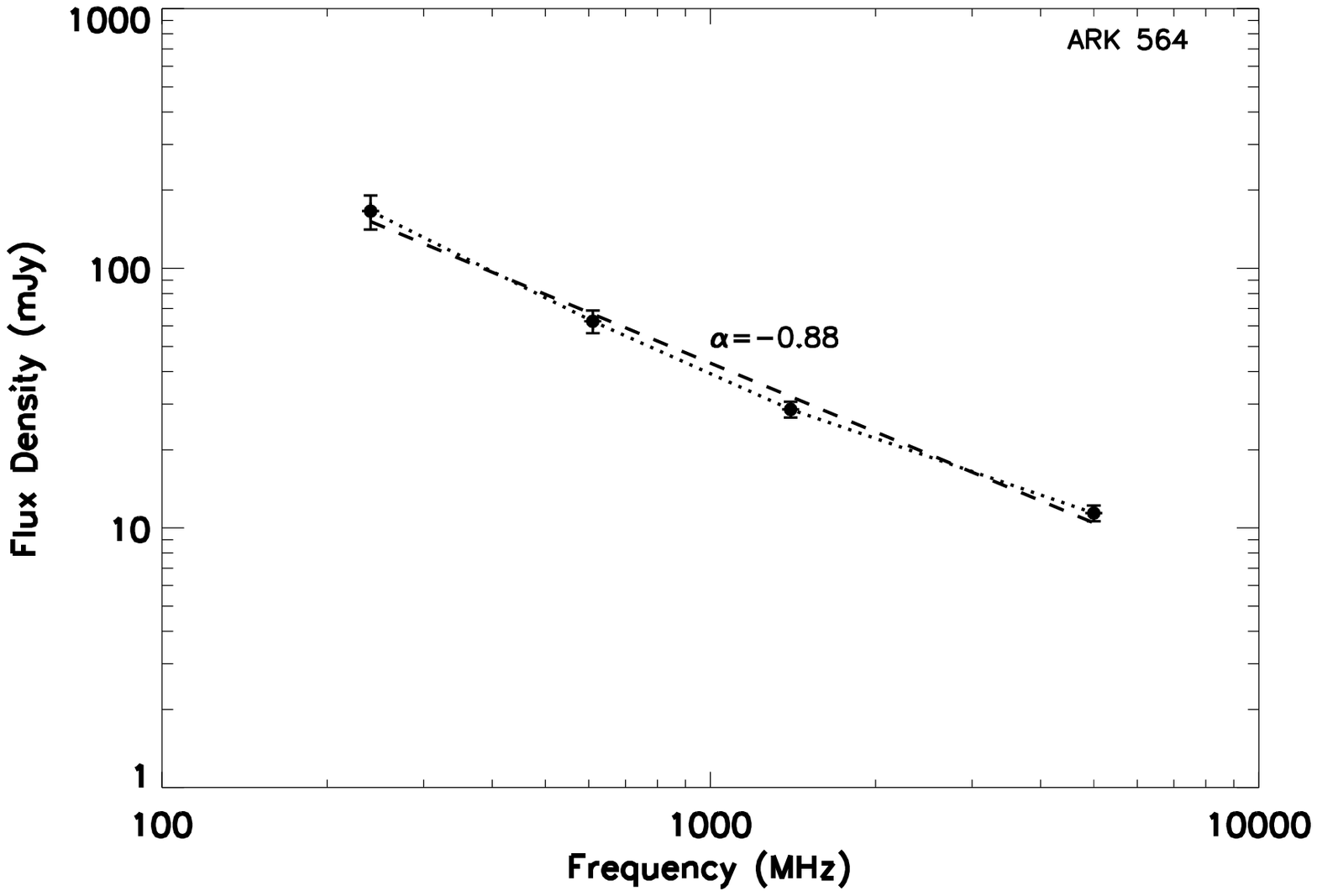}{\includegraphics[angle=0,width=7.2cm,height=0.23\textheight]{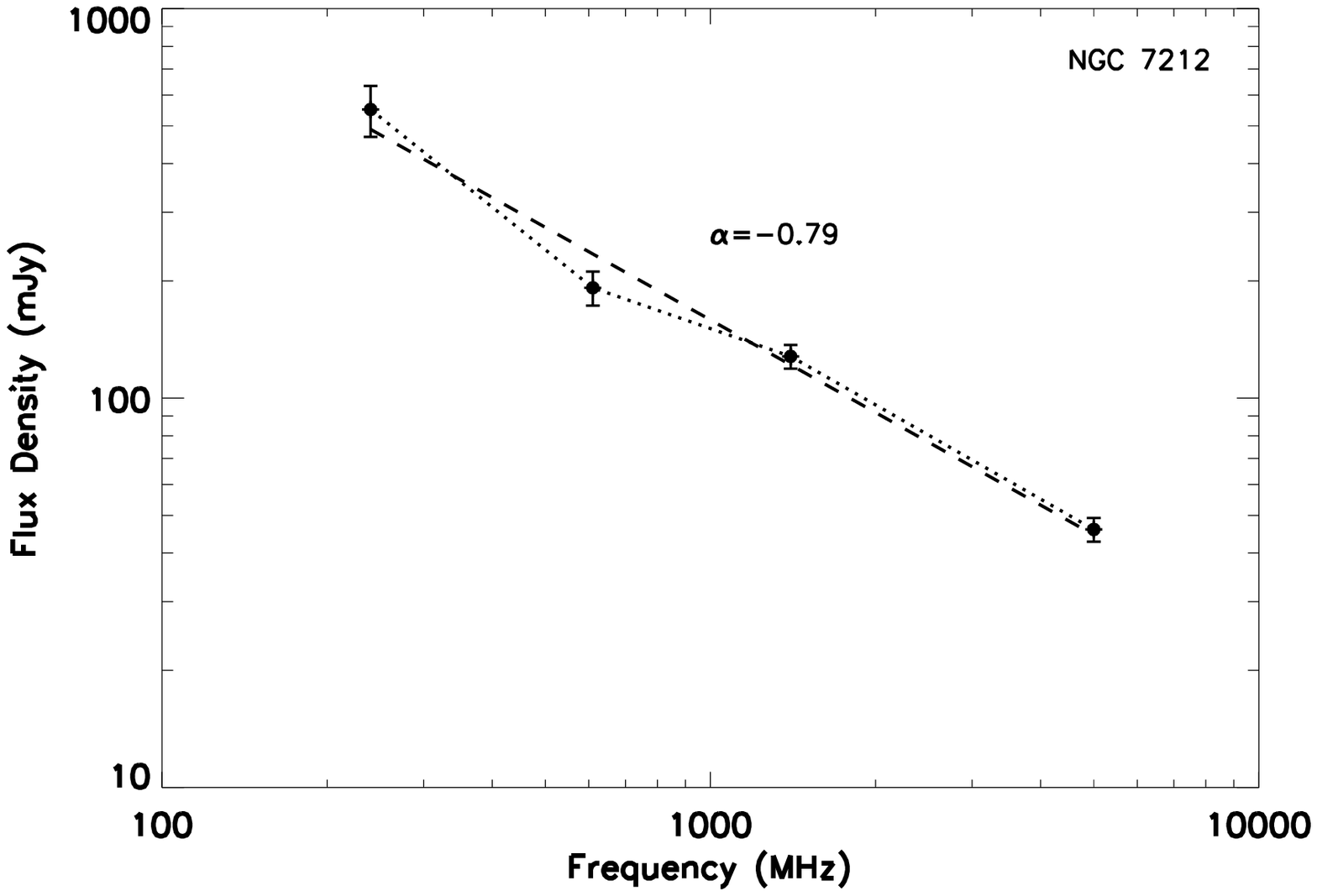}}
\caption{-{\it continued} : Radio spectra of Seyfert type 1s and type 2s are shown in left and right panel, respectively.}
\end{figure*}
\addtocounter{figure}{-1}
\begin{figure*}[!htbp]
\centering
\includegraphics[angle=0,width=7.2cm,height=0.23\textheight]{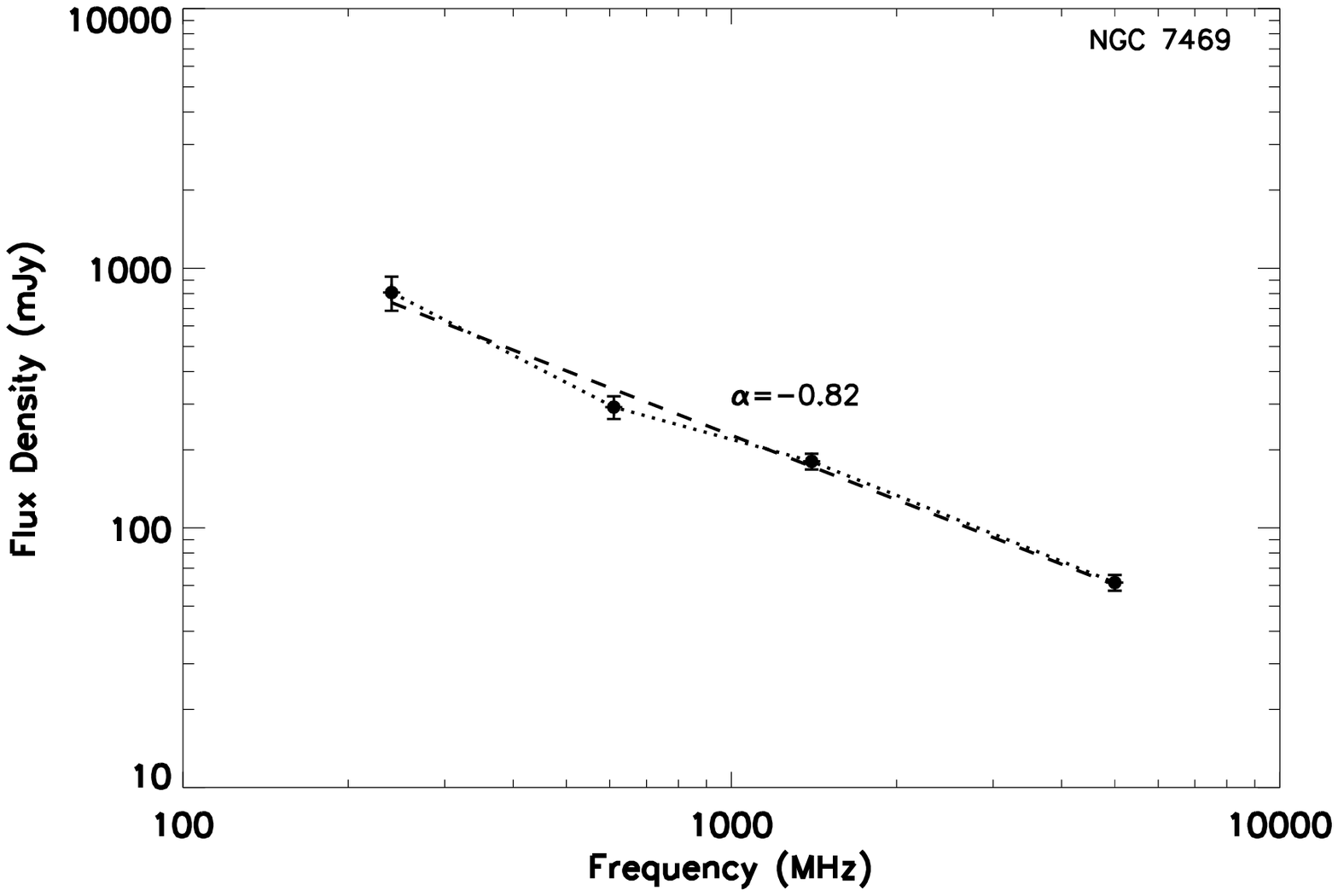}{\includegraphics[angle=0,width=7.2cm,height=0.23\textheight]{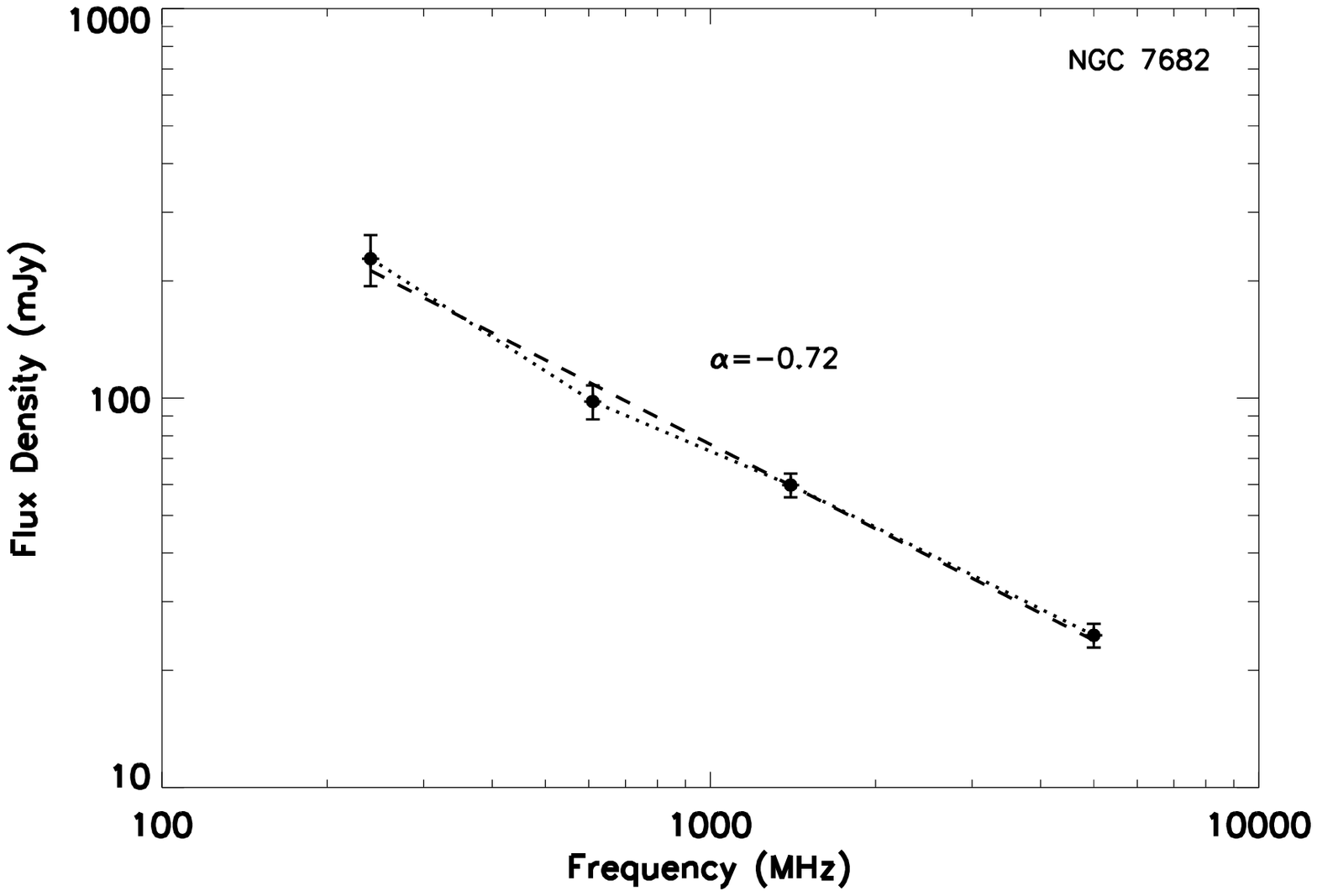}}
\includegraphics[angle=0,width=7.2cm,height=0.23\textheight]{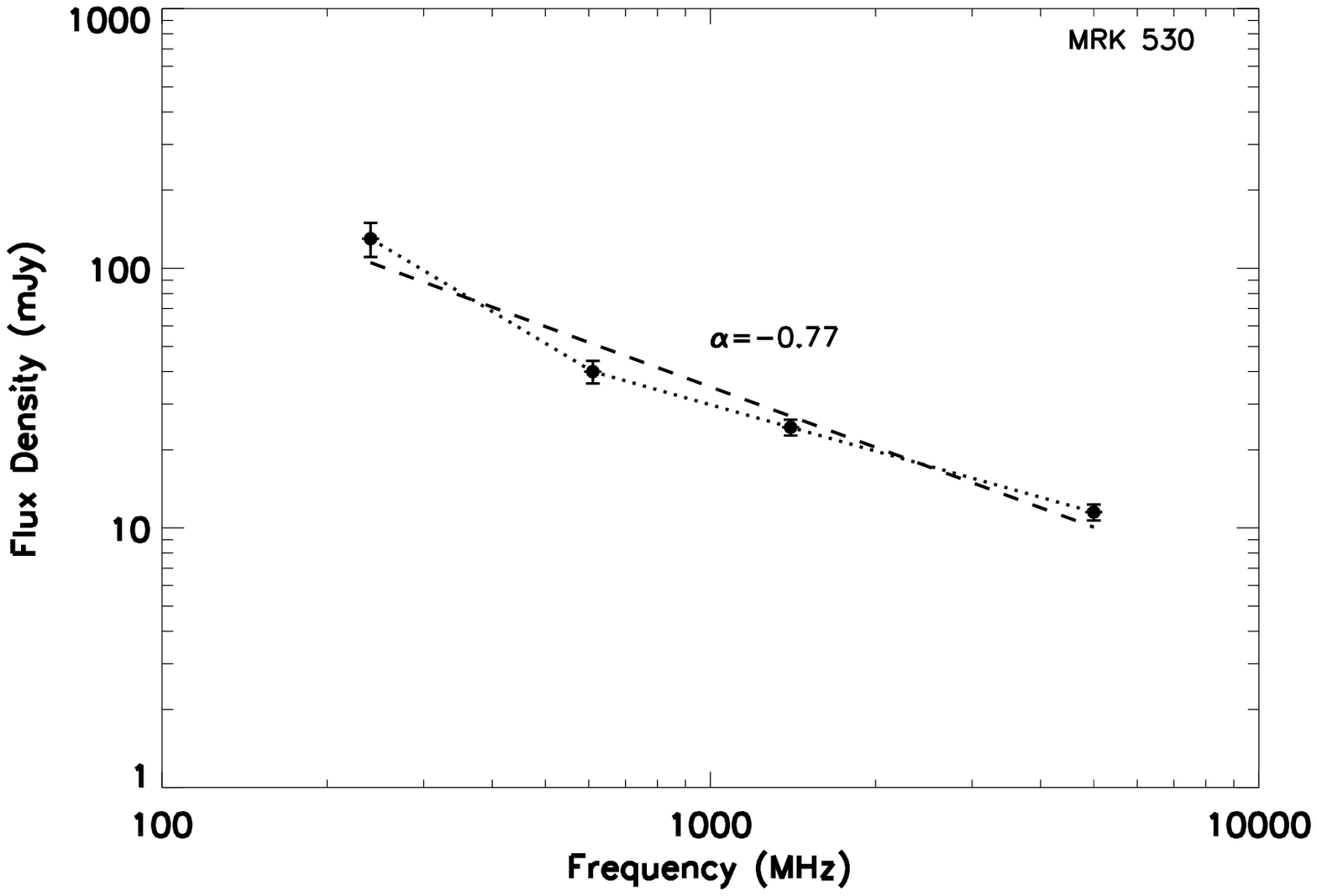}{\includegraphics[angle=0,width=7.2cm,height=0.23\textheight]{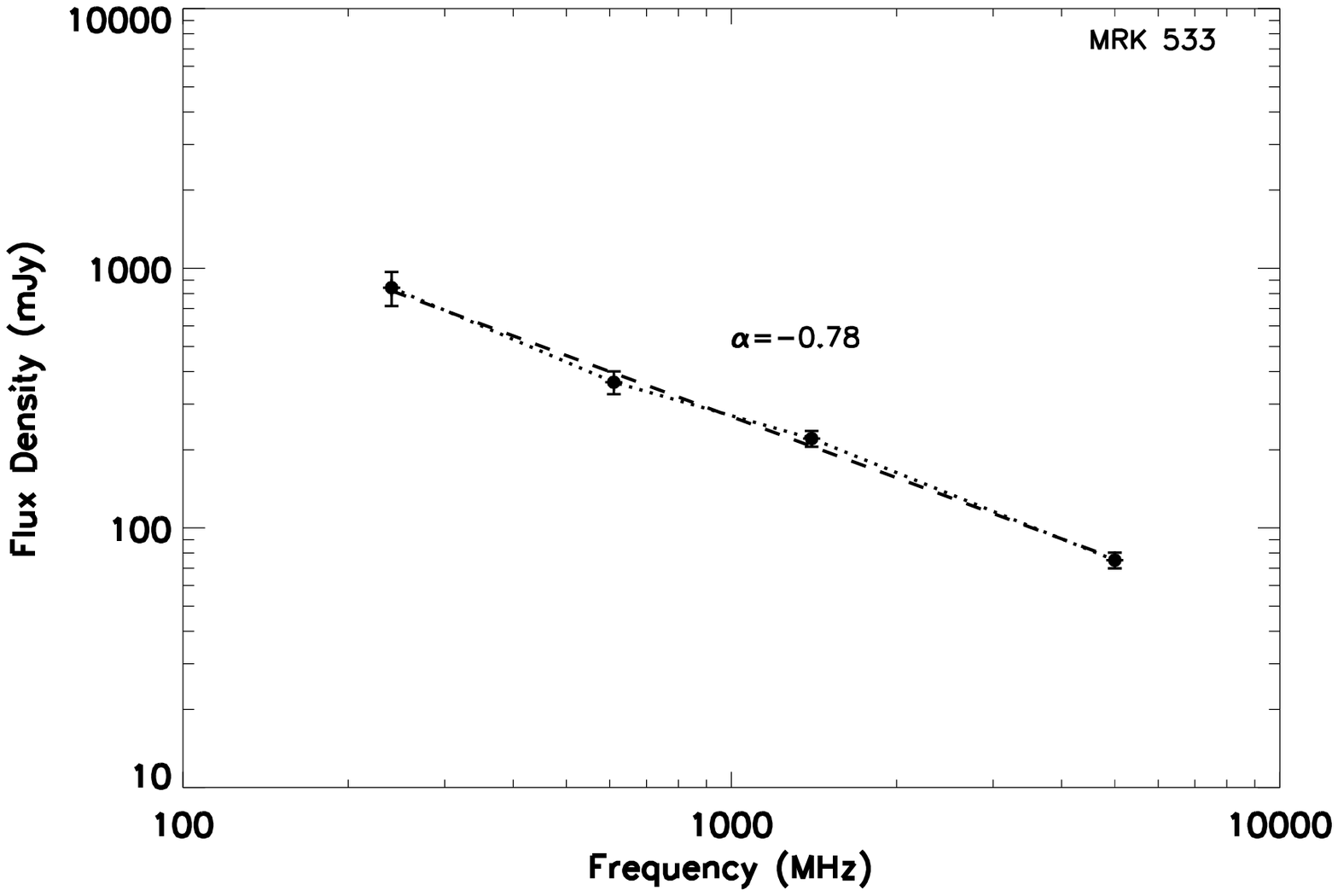}}
\caption{-{\it continued} : Radio spectra of Seyfert type 1s and type 2s are shown in left and right panel, respectively.}
\end{figure*}
\begin{figure*}[!htbp]
\centering
\includegraphics[angle=0,width=7.5cm,height=0.26\textheight]{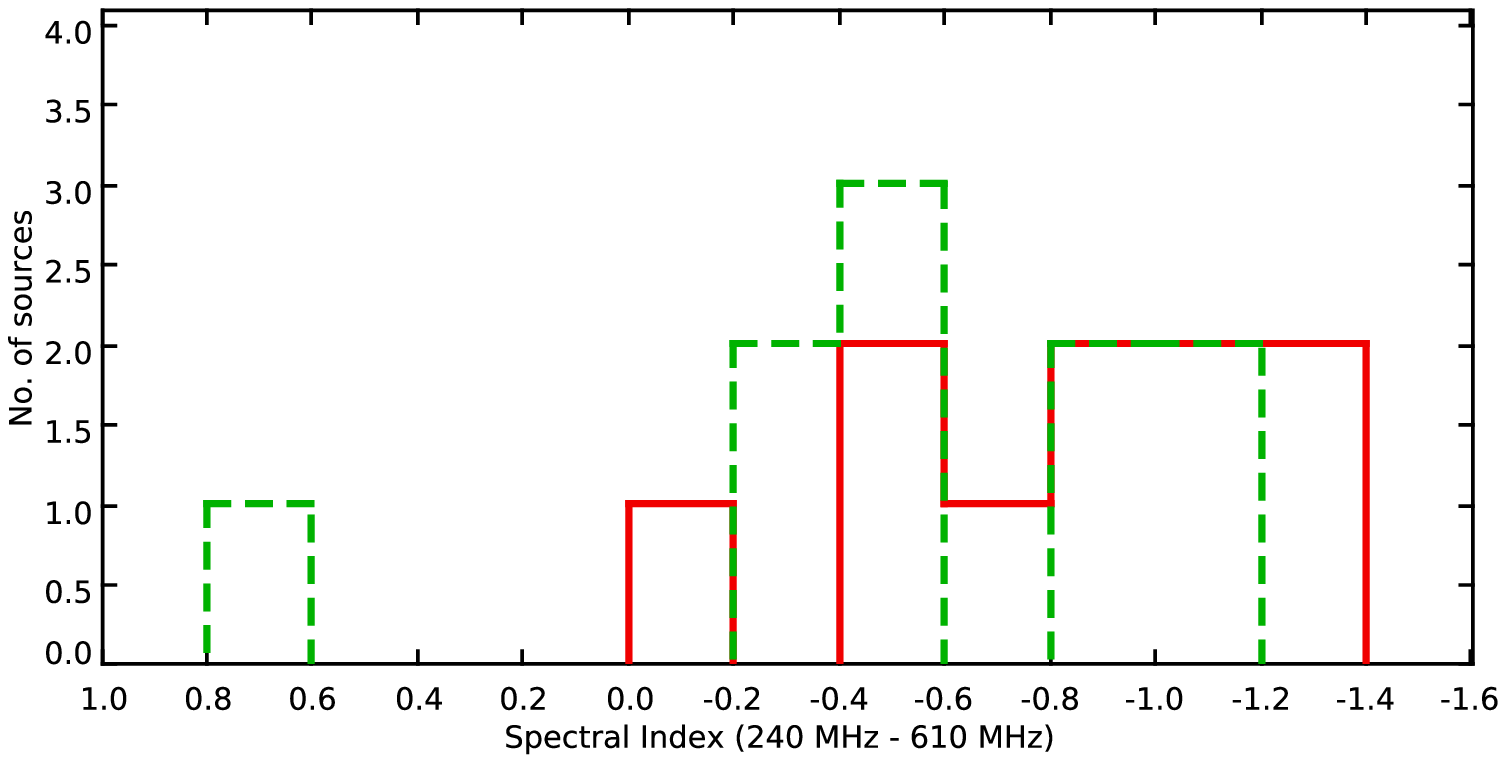}{\includegraphics[angle=0,width=7.5cm,height=0.26\textheight]{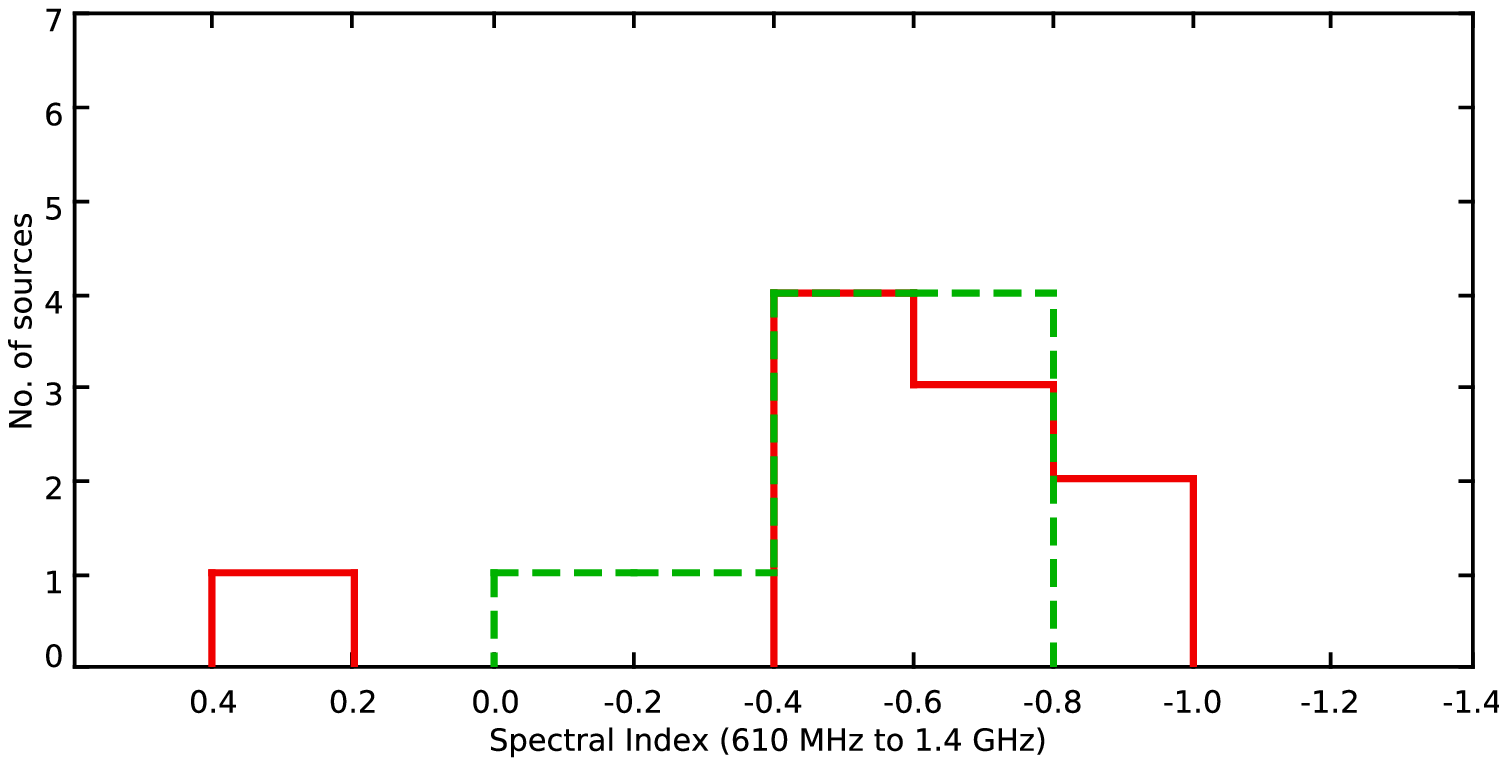}}
\includegraphics[angle=0,width=7.5cm,height=0.26\textheight]{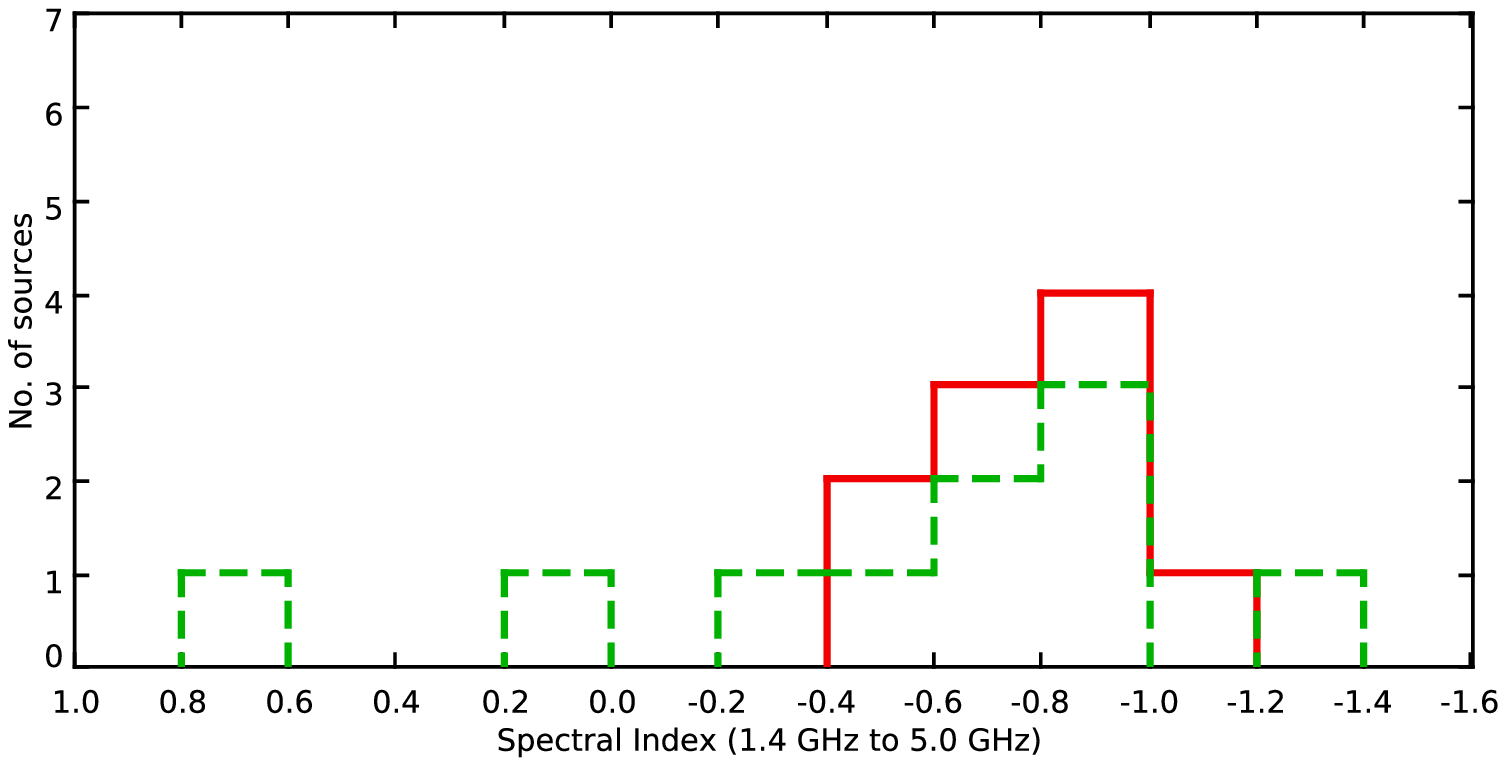}{\includegraphics[angle=0,width=7.5cm,height=0.26\textheight]{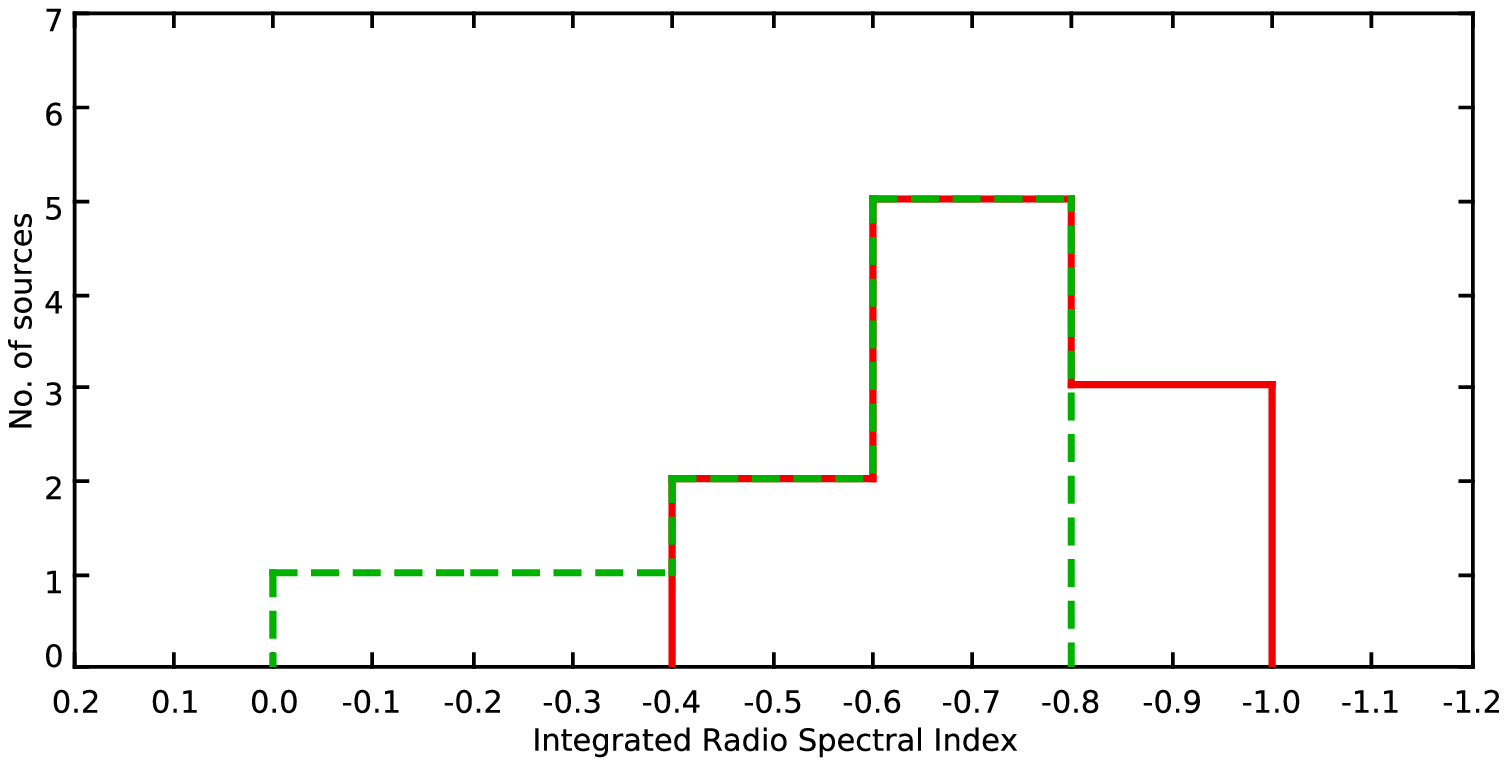}}
\caption{Histograms of two-point spectral indices ${\alpha}^{\rm 610 MHz}_{\rm 240 MHz}$, ${\alpha}^{\rm 1.4 GHz}_{\rm 610 MHz}$, ${\alpha}^{\rm 1.4 GHz}_{\rm 5.0 GHz}$
 and integrated spectral index (${\alpha}_{\rm int}$) for the two subtypes of our sample Seyfert galaxies. 
The histograms for type 1s and type 2s are plotted with `Red colored solid lines' and `Green colored dashed lines', respectively.}
\end{figure*}
\subsection{Radio morphologies of Seyfert type 1s and type 2s}
High resolution radio observations of Seyfert galaxies show parsec-scale nuclear emission 
often accompanied with jet-like elongated feature, which is believed to represent outflowing 
radio emitting plasma from AGN, in the form of jets, bubbles or plasmoids 
\citep{Wilson82,Thean01,Lal04}. These are thought to be small-scale, low-power versions of the large-scale jets seen in radio loud AGNs.
Radio observations of relatively lower resolution reveal that several Seyfert galaxies exhibit extended radio emission up to 
few kpc \citep{Baum93,Colbert96,Gallimore06}.
The comparison of the radio sizes of two Seyfert subtypes can be used to test the unification scheme, provided that the radio emission is primarily due to 
linear jet-like outflows emanating from AGN and radio structures are well resolved. 
According to the orientation based unification scheme, Seyfert type 1s are expected to show smaller projected radio source sizes than type 2s, since 
radio jets in type 1s are expected to lie along the line-of-sight to the observer, and hence to be foreshortened. 
Early studies attempted to test this prediction reported that Seyfert type 2s are larger in radio than type 1s ({\eg}\cite{Ulvestad84a}). 
Although, after controlling for the strengths of the radio sources \cite{Ulvestad89} found that the differences 
in radio sizes of two subtypes are not statistically significant.
Using a sample of Seyfert galaxies based on 60 $\mu$m, \cite{Schmitt01b} reported the radio sizes of type 2s are systematically larger than type 1s.
While, \cite{Ulvestad01} reported that type 1s appears to be larger than type 2s, which is inconsistent to the unification scheme. 
\par
We carried out our GMRT observations with the primary aim to study the radio spectra of Seyfert galaxies at lower frequencies. 
Moreover, we attempt to study the low-frequency radio morphologies of our sample Seyfert galaxies.
Figure 2.0 shows the 240 MHz and 610 MHz radio contour images overlaid on their Digital Sky Survey (DSS) optical images for all of 
our sample Seyfert galaxies.
Table 3.0 and Table 4.0 list the map parameters ({\ie}synthesized beam size, its position angle and noise rms) and 
source parameters ({\ie}total flux density (S$_{\rm int}$), peak flux density (S$_{\rm peak}$) and source fitted sizes) for all the 20 Seyfert galaxies 
at 610 MHz and 240 MHz, respectively.
In our 610 MHz radio maps the typical noise rms and resolution are $\sim$ 1.0 mJy/beam and $\sim$ 8$^{\prime\prime}$ - 10$^{\prime\prime}$, respectively, 
while at 240 MHz, the typical noise rms is $\sim$ 7.5 mJy/beam and resolution is $\sim$ 18$^{\prime\prime}$ - 40$^{\prime\prime}$.
We use AIPS task `JMFIT' to measure the angular size of the source. 
Most of our sample sources are fitted with only one Gaussian component. 
The radio emission at 610 MHz is fitted with a single Gaussian in all of our sample sources except for NGC 4151 and NGC 5728. 
In NGC 4151 and NGC 5728, the 610 MHz radio emission has two distinct components which are fitted with two elliptical Gaussian components. 
At 610 MHz, several of our sample sources, {\ie}NGC 3516, NGC 4151, NGC 5548, NGC 7469, MRK 1066 and NGC 5728 have 
fitted Gaussian sizes larger than their synthesized beams which indicate the possibility of the existence of kpc-scale extended radio emission 
in these sources. Indeed, some of our sample sources {\eg}NGC 4151, MRK 766, MRK 348, NGC 5548, NGC 7469 are reported to possess kpc-scale 
extended emission at 5.0 GHz \citep{Gallimore06}.
At 240 MHz, all our sources are fitted with a single Gaussian component with sizes similar to their synthesized beams. 
Thus all our sample Seyferts can be interpreted as unresolved point sources at 240 MHz. 
We do not compare the radio sizes of the two Seyfert subtypes as most of our sample sources are seen as unresolved point sources at 240 MHz and 610 MHz.
Moreover, both Seyfert type 1s and type 2s in our sample show similar likelihood of being represented as unresolved point sources at the given 
sensitivity and resolution of our 240 MHz and 610 MHz GMRT observations.   
The low-frequency radio emission in Seyfert galaxies may have contributions from star-forming regions or starburst other than the AGN.
But our marginally resolved or unresolved radio images do not allow us to put any constrain on the relative 
contributions from different emitting components.
% Our less sensitive radio images are likely to miss very faint extended emission. 
% Therefore, more sensitive low-frequency radio imaging of Seyfert galaxies is required to detect weaker radio emitting features of 
% low-surface-brightness emission and to understand their origin. 
% 
\begin{table*} 
% \centering
\caption{Comparison of radio luminosities and spectral indices of Seyfert type 1s and type 2s}
\small
% \label{tab_stats}
\begin{minipage}{160mm}
% \centering
%\vspace {0.2cm}
%\hspace {3.8cm} 
% \resizebox{14.0cm}{!}{\begin{tabular} {@{}cccccc@{}}
\begin{tabular} {@{}cccccccccccc@{}}
\hline
Distribution             & \multicolumn{8}{c}{statistical parameters} & \multicolumn{2}{c}{KS test}  & F.D. Ref. \\  
                         & \multicolumn{4}{c}{Seyfert type 1s}  &  \multicolumn{4}{c}{Seyfert type 2s} &   &   &  \\ \cline{2-5} \cline{6-9} 
                     &   min  &  max  &  SD &   median &  min  &  max  &  SD &   median &  D  &  p-value  &   \\ \hline
logL$_{\rm 240~MHz}$    & 28.78 & 30.90 & 0.66 & 30.31 & 28.87 & 31.22 & 0.73 & 30.18 & 0.2 & 0.99 & 1 \\
logL$_{\rm 610~MHz}$    & 28.66 & 30.62 & 0.61 & 29.90 & 28.92 & 30.86 & 0.58 & 29.83 & 0.2 & 0.99 & 1 \\
logL$_{\rm 1.4~GHz}$    & 28.54 & 30.35 & 0.55 & 29.62 & 28.70 & 30.64 & 0.55 & 29.64 & 0.3 & 0.79 & 2 \\
logL$_{\rm 5.0~GHz}$    & 28.10 & 29.92 & 0.57 & 29.19 & 28.51 & 30.62 & 0.63 & 29.35 & 0.3 & 0.79 & 3 \\
${\alpha}^{\rm 610 MHz}_{\rm 240 MHz}$  & -1.26 & -0.15 & 0.36 & -0.83  & -1.13 & +0.66 & 0.50 & -0.52  & 0.3  &  0.79 & 1  \\
${\alpha}^{\rm 1.4 GHz}_{\rm 610 MHz}$  & -0.94 & +0.27 & 0.31 & -0.59  & -0.77 & -0.07 & 0.20 & -0.59  & 0.3  &  0.79 & 1,2  \\
${\alpha}^{\rm 5.0 GHz}_{\rm 1.4 GHz}$  & -1.13 & -0.49 & 0.16 & -0.77  & -1.38 & +0.79 & 0.57 & -0.67 & 0.4  &  0.42  & 2,3 \\ 
${\alpha}_{\rm Int.}$                   &  0.88 & -0.47 & 0.12 & -0.74  & -0.79 & -0.14 & 0.20 & -0.72 & 0.38 &  0.51 & 1,2,3  \\ \hline
\end{tabular}
%\label{table:nolin} (is used to refer this table in the text "table:nolin")
% \justify
\vspace {0.05cm} \\
\footnotesize{{\bf Notes.} Flux density (F.D.) references: (1) our GMRT observations; (2) NRAO VLA Sky Survey (NVSS); (3) Literature 
\citep{Gallimore06,Edelson87,Gregory91,Griffith95}. The adopted conservative error values of 15$\%$, 10$\%$, 7$\%$ and 7$\%$ for S$_{\rm 240 MHz}$, 
S$_{\rm 610 MHz}$, S$_{\rm 1.4 GHz}$ and S$_{\rm 5.0 GHz}$, respectively, render 
$\sim$ 8$\%$, $\sim$ 5$\%$, $\sim$ 4$\%$ and $\sim$ 7$\%$ errors in ${\alpha}^{\rm 610 MHz}_{\rm 240 MHz}$, 
${\alpha}^{\rm 1.4 GHz}_{\rm 610 MHz}$, ${\alpha}^{\rm 5.0 GHz}_{\rm 1.4 GHz}$ and ${\alpha}_{\rm int}$, respectively. 
Radio luminosities are in units of ergs s$^{-1}$ Hz$^{-1}$. 
Kolmogorov - Smirnov (KS) two sample test examines the hypothesis that two samples comes from same distribution.
D = Sup x $|$S1(x) - S2(x)$|$ is the maximum difference between the cumulative distributions of two samples S1(x) and S2(x), respectively.}
\end{minipage}
\end{table*}
\section{Conclusions}
We present low-frequency radio images and spectra of our sample of 20 Seyfert galaxies wherein the sample is based on orientation independent 
isotropic properties. In our sample the two subtypes have matched distributions in orientation-independent parameters which 
allow us to assume that the two subtypes are intrinsically similar within the framework of the unification scheme. 
% {\bf Our restrictive sample selection criteria and observational constraints resulted a small sample and in future, we plan to extend our analysis to a 
% larger sample}.  
Here we outline following conclusions from our study.
\begin{enumerate}
\item This work is the first attempt, to our knowledge, for a systematic study of low-frequency radio imaging and spectral properties 
of a well defined sample of Seyfert galaxies.     

\item The 240 MHz, 610 MHz, 1.4 GHz and 5.0 GHz radio luminosities of our sample Seyfert galaxies are in the range of 
$\sim$ 10$^{28}$ - 10$^{31}$ erg s$^{-1}$. The 240 MHz, 610 MHz, 1.4 GHz and 5.0 GHz radio luminosity distributions 
of Seyfert type 1s and type 2s span over similar range with similar median values at the respective frequencies. 
The two sample KS test shows that 240 MHz, 610 MHz, 1.4 GHz and 5.0 GHz radio luminosity distributions of the two Seyfert subtypes are similar 
with statistically significant probability.
\item We obtain integrated radio spectra over 240 MHz to 5.0 GHz, of all our 20 Seyfert galaxies.
We find that the distributions of two point spectral indices (${\alpha}^{\rm 610 MHz}_{\rm 240 MHz}$, ${\alpha}^{\rm 1.4 GHz}_{\rm 610 MHz}$, 
${\alpha}^{\rm 5.0 GHz}_{\rm 1.4 GHz}$) as well as integrated spectral index for the two subtypes span over similar range with median values 
${\alpha}^{\rm 610 MHz}_{\rm 240 MHz}$ $\simeq$ -0.83, ${\alpha}^{\rm 1.4 GHz}_{\rm 610 MHz}$ $\simeq$ -0.59, 
${\alpha}^{\rm 5.0 GHz}_{\rm 1.4 GHz}$ $\simeq$ -0.77, ${\alpha}_{\rm int}$ $\simeq$ -0.74 for type 1s, and 
${\alpha}^{\rm 610 MHz}_{\rm 240 MHz}$ $\simeq$ -0.52, ${\alpha}^{\rm 1.4 GHz}_{\rm 610 MHz}$ $\simeq$ -0.59, 
${\alpha}^{\rm 5.0 GHz}_{\rm 1.4 GHz}$ $\simeq$ -0.67, ${\alpha}_{\rm int}$ $\simeq$ -0.72 for type 2s.
The two sample KS test shows that the distributions of spectral indices for the two Seyfert subtypes are not statistically different.
\item We note that most of the Seyfert galaxies in our sample have steep integrated radio spectra (${\alpha}_{\rm int}$ $\sim$ -0.7) 
except NGC 5252 and NGC 5728 which show flat and inverted spectrum, respectively.    
The average steep radio spectral index is consistent with the previous studies \citep{Morganti99} and can be explained as optically thin 
synchrotron emission. 
\item 610 MHz radio images of our sample sources can generally be represented as unresolved point sources 
wherein the radio emission is fitted with a single Gaussian component in all of our sample sources except for NGC 4151 and NGC 5728. 
In NGC 4151 and NGC 5728, the 610 MHz radio emission shows two distinct components.
\item At 240 MHz, all our sample sources show radio emission as unresolved point source as the radio emission is fitted with a single Gaussian 
component with sizes similar to their synthesized beams. 
\item Radio images in our snap-shot GMRT 240/610 MHz observations remain mostly unresolved. This does not allow us 
to compare the radio sizes of the two Seyfert subtypes in the framework of the unification scheme. Moreover, both Seyfert type 1s and type 2s
in our sample show similar likelihood of being represented as unresolved point sources at the given sensitivity and resolution of our 240 MHz 
and 610 MHz GMRT observations.
\item In our study we have shown that the multifrequency radio properties {\ie}luminosity and spectral distributions over 240 MHz to 5 GHz of our 
sample Seyfert galaxies are not inconsistent with the orientation and obscuration based unification scheme.
In the appendix, we discuss 240/610 MHz GMRT radio properties and its comparison to the radio observations reported in the literature for 
all the individual sources in our sample. Our results on the low-frequency radio properties are complementary and consistent with the high-frequency 
radio observations reported in the literature.  

\end{enumerate}
\begin{acknowledgements}
We thank the staff of GMRT who have made these observations possible. 
GMRT is run by the National Centre for Radio Astrophysics of the Tata Institute of Fundamental Research.
VS would like to thank Dr. Chiranjib Konar for helpful discussions on 240 MHz GMRT data reduction. 
Also, this research has made use of the NASA/IPAC Extragalactic Database (NED) which is operated by the Jet Propulsion Laboratory, 
California Institute of Technology, under contract with the National Aeronautics and Space Administration.
\end{acknowledgements}
% 
% \clearpage
% 
\begin{appendix}
\section{Notes on individual sources}
In this section we describe the radio properties of individual sources of our sample. 
\subsection{MRK 6}
\cite{Kharb06} presented a detailed study on MRK 6 radio morphology and reported that the radio emission is seen on three different spatial scales 
{\ie}$\sim$~7.5 kpc bubbles, $\sim$~1.5 kpc bubbles lying nearly orthogonal to each other, and a $\sim$~1 kpc radio jet 
lying orthogonal to the 1.5 kpc bubble. They suggested that the existence of kpc scale radio bubbles/lobes can be attributed to the radio relics related to 
past episodic nuclear activity. 
Our radio images at 610 MHz and 240 MHz show only single component that is fitted with elliptical Gaussian of 10$\arcsec$.42$\times$5$\arcsec$.06 and 
24$\arcsec$.14$\times$11$\arcsec$.76, respectively. The radio emission at 610 MHz and 240 MHz are likely to include all the emitting 
components detected by \cite{Kharb06} as the combined 5.0 GHz flux density of all the components in \cite{Kharb06} follow the spectral shape 
determined by our 240 MHz, 610 MHz GMRT and 1.4 GHz NVSS flux densities.    
\subsection{NGC 3227} 
8.4 GHz VLA observations show north-south elongated nuclear radio emission and at sub-arcsec 
resolution, the core is resolved into 0$^{\prime\prime}$.4 double sources \citep{Mundell95}.
5.0 GHz VLA `D' array observations show the kpc-scale extended radio emission which is interpreted as the emission from star-forming host galaxy 
disk \citep{Gallimore06}. 
Our GMRT images at 610 MHz and 240 MHz show radio emission as nearly unresolved point sources which are fitted with an 
elliptical Gaussian of 8$\arcsec$.85$\times$6$\arcsec$.70 and 32$\arcsec$.61$\times$18$\arcsec$.02, respectively. 
The fitted powerlaw radio spectrum over 240 MHz to 5.0 GHz renders spectral index (${\alpha}_{\rm int}$) $\sim$ -0.53, 
while there is an indication of spectral flattening toward lower frequencies. 
\subsection{NGC 3516}
High resolution VLA observations of NGC 3516 show a compact flat-spectrum core with a $\sim$ 0$^{\prime\prime}$.7 ($\sim$ 120 pc) 
one-sided northern extension along P.A. $\simeq$ 8$^\circ$ \citep{Nagar99}. 
At arcsec resolution, additional radio components are seen that extend out to $\sim$ 4$^{\prime\prime}$ ($\sim$ 740 pc) and align with the co-spatial 
S-shaped optical emission-line structure \citep{Miyaji92,Ferruit98}. 
At coarser resolution, a linear radio structure spanning $\sim$ 45$^{\prime\prime}$ ($\sim$ 8.3 kpc), oriented along 
P.A. $\sim$ 44$^\circ$, across the nucleus is also seen \citep{Baum93}. 
In our GMRT observations, the 610 MHz radio emission is detected as marginally resolved point source fitted with an elliptical Gaussian of 
10$^{\prime\prime}$.44$\times$7$^{\prime\prime}$.38. The integrated flux density at 610 MHz is substantially higher than the peak flux density 
and also, the fitted Gaussian size is larger than the synthesized beam size, and this may be indicative of the presence of extended emission at the kpc-scale.
We fit the 240 MHz radio emission with a large Gaussian of 37$^{\prime\prime}$.09$\times$14$^{\prime\prime}$.31. 
There is hint of second component along the north-east direction from the central nuclear component which 
seems to be consistent with the radio morphology reported by \cite{Baum93}.   
NGC 3516 shows inverted spectrum between 1.4 GHz and 610 MHz while a steep spectrum between 610 MHz and 240 MHz. 
The changing spectral behavior can be interpreted as the presence of at least two emitting component. 
Possibly 5.0 GHz to 610 MHz is dominated by compact AGN core having self absorbed synchrotron emission and low frequency emission at 610 MHz to 240 
MHz is dominated by optically thin synchrotron emission from extend component. 
\subsection{NGC 4151}
The nuclear radio source in NGC 4151 resolves into compact knots aligned in a slightly curved jet structure spanning 
$\sim$ 5$^{\prime\prime}$ ($\sim$ 340 parsec) across the nucleus and oriented along P.A. $\sim$ 80$^\circ$ \citep{Ulvestad81,Booler82,Wilson82,Pedlar93,Mundell03}. 
The WSRT observations of NGC 4151 revealed a $\sim$ 15$^{\prime\prime}$ ($\sim$ 1 kpc) linear structure, 
roughly aligned with the inner jet, and a bracketing pair of radio continuum arcs located 
at $\sim$ 45$^{\prime\prime}$ ($\sim$ 3 kpc) from the radio nucleus \citep{Baum93}. 
In our GMRT observations, the 610 MHz radio emission is double peaked with less stronger peak residing 
at a distance of $\sim$ 8$^{\prime\prime}$.54 along the P.A. $\sim$ 65$^{\circ}$.2 from the central component.
The 610 MHz radio emission can be fitted with two elliptical Gaussian components with central component having 
peak flux density $\sim$ 312.9 mJy/beam and convolved size $\sim$ 8$^{\prime\prime}$.12$\times$5$^{\prime\prime}$.28, 
while the second Gaussian component has peak flux density $\sim$ 218.1 mJy/beam and convolved size $\sim$ 8$^{\prime\prime}$.30$\times$
5$^{\prime\prime}$.74.
Owing to its larger beam-size, the 240 MHz radio image shows a nearly unresolved point source fitted with a Gaussian of $\sim$ 17$^{\prime\prime}$.10$\times$11$^{\prime\prime}$.64.
The four point radio spectrum remains fairly steep (${\alpha}_{\rm int}$ $\sim$ -0.76) over 240 MHz to 5.0 GHz. 
\subsection{MRK 766}
8.4 GHz VLA `A' and `C' array observations of MRK 766 show an unresolved point source emission and 
no evidence for emission at intermediate scale \citep{Kukula95}.
In our GMRT observations, the 610 MHz and 240 MHz radio emission is seen as unresolved point sources that are fitted with Gaussian of 
$\sim$ 7$^{\prime\prime}$.92$\times$4$^{\prime\prime}$.81 and $\sim$ 16$^{\prime\prime}$.20$\times$9$^{\prime\prime}$.90, respectively.
The four point integrated radio spectral index (${\alpha}_{\rm int}$ $\sim$ -0.47) is less steep than average spectral index for our sample 
Seyfert galaxies.
\subsection{MRK 279}
VLA observations of MRK 279 at 8.4 GHz show an unresolved point source emission in both `A' and `C' array configurations 
and no evidence for emission at intermediate scale \citep{Kukula95}. Our GMRT observations at 610 MHz and 240 MHz show radio emission as 
unresolved point source fitted with the Gaussian of sizes 
$\sim$ 7$^{\prime\prime}$.86$\times$5$^{\prime\prime}$.54 and $\sim$ 17$^{\prime\prime}$.69$\times$11$^{\prime\prime}$.52, respectively. 
The four point radio spectrum of MRK 279 is fairly steep with integrated spectral index (${\alpha}_{\rm int}$) $\sim$ -0.88.
\subsection{NGC 5548}
VLA observations of NGC 5548 revealed a compact core between two diffuse lobes separated by 
$\sim$ 15$^{\prime\prime}$ \citep{Wilson82a}. 
Using VLA `D' array observations \cite{Gallimore06} detected marginally resolved lobes along with the nuclear point source emission. 
In our GMRT observations, we fit the 610 MHz radio emission with a Gaussian of $\sim$ 14$^{\prime\prime}$.33$\times$7$^{\prime\prime}$.11. 
We note that the 610 MHz total flux density is significantly higher than the peak flux density and also the fitted Gaussian size is larger 
than the synthesized beam size. This may infer the presence of kpc-scale extended radio emission.
At 240 MHz, the radio emission is nearly an unresolved point source fitted with a Gaussian of $\sim$ 23$^{\prime\prime}$.54$\times$14$^{\prime\prime}$.83. 
The four point radio spectrum of NGC 5548 is fitted with spectral index of (${\alpha}_{\rm int}$) $\sim$ -0.68.  
\subsection{ARK 564}
8.4 GHz VLA observations of ARK 564 show a triple radio source along the north-south 
direction (P.A. $\sim$ 6$^\circ$), extended to $\sim$ 320 pc \citep{Moran2000a}.
In our GMRT observations, the radio emission at 610 MHz and 240 MHz is nearly point source emission fitted with Gaussian of 
$\sim$ 7$^{\prime\prime}$.62 $\times$ 4$^{\prime\prime}$.91 and $\sim$ 24$^{\prime\prime}$.29$\times$12$^{\prime\prime}$.92, respectively.
The four point radio spectrum over 240 MHz to 5.0 GHz is fairly steep with integrated spectral index (${\alpha}_{\rm int}$) $\sim$ -0.88.  
\subsection{NGC 7469}
1.6 GHz VLBI observations resolved the core-jet structure into five different components, 
lying in an east-west line which extends up to $\sim$ 55 pc \citep{Lonsdale03}.
A variety of observations at high spatial resolution indicate that the AGN in NGC 7469 is surrounded by a ring of starburst 
\citep{Mauder94,Genzel95} with radio observations showing an unresolved central component surrounded by the ring-like structure \citep{Wilson91}. 
In our GMRT observations, the radio emission at 610 MHz and 240 is fitted with single Gaussian components with sizes of 
$\sim$ 10$^{\prime\prime}$.43$\times$6$^{\prime\prime}$.70 and $\sim$ 37$^{\prime\prime}$.86$\times$20$^{\prime\prime}$.68, respectively. 
The integrated flux density at 610 MHz is substantially higher than the peak flux density and also the fitted Gaussian size is larger than the 
synthesized beam size. And, this may be considered as an indication for the presence of kpc-scale extended emission in NGC 7469. 
The four point radio spectrum of NGC 7469 is fairly steep with integrated spectral index (${\alpha}_{\rm int}$) $\sim$ -0.82.
\subsection{MRK 530}
VLA observations of MRK 530 at 8.4 GHz show a compact point source emission with a slight extension \citep{Kukula95}. 
The milli-arcsec resolution VLBI observations showed a compact core emission with extension along east-west 
direction at fainter levels \citep{Lal04}. 
In our GMRT observations, the radio emission at 610 MHz and 240 MHz is seen as nearly point source that is fitted with single Gaussian components of 
sizes $\sim$ 7$^{\prime\prime}$.28$\times$7$^{\prime\prime}$.18 and $\sim$ 46$^{\prime\prime}$.92$\times$21$^{\prime\prime}$.46, respectively. 
The four point radio spectrum of MRK 530 is fairly steep (${\alpha}_{\rm int}$) $\sim$ -0.77. 
\subsection{MRK 348}
In MRK 348, the radio continuum emission is dominated by a variable, subparsec-scale ($\sim$ 0.5 pc) jet that 
feeds into a larger ($\sim$ 60 pc) linear radio structure oriented roughly north-south \citep{Neff83,Ulvestad99,Anton02}. 
\cite{Baum93} observed large-scale radio lobes ($\sim$ 6 kpc extent) that roughly align with the small-scale jet structure. 
Using 5.0 GHz VLA `D' array observations, \cite{Gallimore06} reported the presence of extended emission such that the large-scale lobes 
are marginally resolved after the nuclear point source subtraction. 
In our GMRT observations, the radio emission 610 MHz and 240 MHz is seen as nearly point source that we fit with single Gaussian components of sizes 
$\sim$ 6$^{\prime\prime}$.34$\times$5$^{\prime\prime}$.81 and $\sim$6$^{\prime\prime}$.34$\times$ 5$^{\prime\prime}$.81, respectively.
The radio spectrum of MRK 348 fitted over 240 MHz to 1.4 GHz has index (${\alpha}_{\rm int}$) $\sim$ -0.58. 
The 5.0 GHz flux density appears to be dominated by variable AGN core component and does not fit with the spectral index measured over 240 MHz to 1.4 GHz.  
\subsection{MRK 1}
VLA observations of MRK 1 show an unresolved point source radio emission from AGN \citep{Kinney2000}.
In our GMRT observations, 610 MHz and 240 MHz radio images are like point source emission and are fitted with single Gaussian components of sizes 
$\sim$ 6$^{\prime\prime}$.31 $\times$ 5$^{\prime\prime}$.32 and $\sim$ 17$^{\prime\prime}$.57$\times$12$^{\prime\prime}$.36, respectively. 
The four point radio spectrum of MRK~1 is fitted with an index (${\alpha}_{\rm int}$) $\sim$ -0.58.  
\subsection{MRK 1066}
4.9 GHz VLA observations of MRK 1066 show a linear, probably triple, source extending $\sim$ 2$^{\prime\prime}$.8 along 
P.A. $\sim$ 134$^\circ$ \citep{Ulvestad89}. 
Using 1.4 GHz and 8.4 GHz VLA observations \cite{Nagar99} reported that the 1.4 GHz radio emission can be fitted with a single Gaussian component 
and 8.4 GHz radio emission displays the central source and northwest extension along the position angle $\sim$ 305$^\circ$ with respect to the core. 
A more diffuse southeast extension along the position angle $\sim$ 140$^\circ$ with respect to the core also seems to be present \citep{Nagar99}.
In our GMRT observations, 610 MHz and 240 MHz radio images show nearly point source emission and are fitted with single Gaussian components of 
sizes $\sim$ 6$^{\prime\prime}$.67$\times$5$^{\prime\prime}$.63 and $\sim$ 17$^{\prime\prime}$.25$\times$13$^{\prime\prime}$.53, respectively. 
The integrated radio spectral index fitted over 240 MHz to 5.0 GHz is $\sim$ -0.72. 
\subsection{NGC 2110}
VLA observations show a symmetrical, jet-like radio emission, extending $\sim$ 4$^{\prime\prime}$ 
in the north-south direction and straddling a central compact core \citep{Ulvestad84b}.
Using 1.4 GHz and 8.4 GHz VLA (A and AnB hybrid array configuration) observations \cite{Nagar99} reported 1.4 GHz 
emission as a point source emission, while at 8.4 GHz a linear jet-like structure was noticed along the position angle $\sim$ 9 $^\circ$. 
In our GMRT observations, the radio emission at 610 MHz and 240 MHz are are seen as nearly point source emission. We fit 610 MHz and 240 MHz emission 
with single Gaussian components of sizes $\sim$ 7$^{\prime\prime}$.68$\times$5$^{\prime\prime}$.11 and 
$\sim$ 38$^{\prime\prime}$.85$\times$32$^{\prime\prime}$.47, respectively. 
The four point radio spectrum of NGC 2110 is steep with integrated spectral index (${\alpha}_{\rm int}$) $\sim$ -0.72.
\subsection{NGC 2273}
5.0 GHz sub-arcsec resolution VLA observations of NGC 2273 show an unequal double with a separation of $\sim$ 1$^{\prime\prime}$.2 
($\sim$ 145 pc) at P.A. 90$^\circ$. While observations at 1.4 GHz show the double structure embedded in a more 
amorphous structure with an extent of 2$^{\prime\prime}$.5 ($\sim$ 300 pc) along P.A. 20$^\circ$ \citep{Ulvestad84b}. 
5.0 GHz WSRT observations show an additional amorphous structure on lager scale and suggest an extended emission with a 
total extent of $\sim$ 9$^{\prime\prime}$ ($\sim$ 1 kpc) along P.A. 160$^\circ$ \citep{Baum93}.
In our GMRT observations, the radio images at 610 MHz and 240 MHz show nearly point source emission which is fitted with single Gaussian components 
of $\sim$ 8$^{\prime\prime}$.32$\times$5$^{\prime\prime}$.34 and $\sim$ 20$^{\prime\prime}$.84$\times$11$^{\prime\prime}$.53, respectively. 
The 610 MHz radio image appears to be marginally resolved.
The four point radio spectrum of NGC 2273 is relatively flat with integrated spectral index (${\alpha}_{\rm int}$) $\sim$ -0.37.
\subsection{NGC 5252}
1.4 GHz and 5 GHz VLA observations show a radio structure consisting of a central, 
compact core, with a relatively flat spectrum and a weaker emission extending $\sim$ 2$^{\prime\prime}$ north and south of the core \citep{Wilson94}. % 
\cite{Nagar99} confirm the overall flatter spectrum of the core (${\alpha}^{\rm 1.4 GHz}_{\rm 5.0 GHz}$ $\sim$ -0.32) and 
the radio continuum features seen in the earlier observations.
In our GMRT observations, we fit the radio emission at 610 MHz and 240 MHz with single Gaussian components of sizes 
$\sim$ 10$^{\prime\prime}$.56$\times$5$^{\prime\prime}$.66 and $\sim$ 46$^{\prime\prime}$.20$\times$19$^{\prime\prime}$.31, respectively.
The four point radio spectrum of NGC 5252 is flat with integrated spectral index (${\alpha}_{\rm int}$) $\sim$ -0.14, 
suggesting that the radio emission over 240 MHz to 5.0 GHz is dominated by compact AGN core. 
\subsection{NGC 5728}
5.0 GHz and 14.9 GHz VLA observations of NGC 5728 show a compact core and a faint radial feature extending along north-east \citep{Schommer88}.
% The later radio continuum observations of NGC 5728 at 5 GHz and 1.4 GHz revealed that the nuclear emission is fairly co-spatial with 
% the bi-conical ionization cone \citep{Wilson93}.
In our 610 MHz GMRT image we noticed a double peaked emission wherein the second off-nuclear component may correspond 
to the nuclear star-forming region reported in previous studies \citep{Schommer88,Mazzuca08}. 
The peaks of the two emitting components are separated by $\sim$ 10$^{\prime\prime}$.7 ($\sim$ 2.0 kpc) 
with the off-nuclear component residing along P.A. $\sim$ 67$^{\circ}$ from the central nuclear component.
The two components have peak flux densities $\sim$ 25.6 mJy/beam and 17.5 mJy/beam and are fitted with Gaussian of sizes 
$\sim$ 12$^{\prime\prime}$.60$\times$7$^{\prime\prime}$.46 and $\sim$ 16$^{\prime\prime}$.10$\times$9$^{\prime\prime}$.39, respectively.
The 240 MHz radio image show one component of emission (possibly due larger synthesized beam-size) 
fitted with a Gaussian of $\sim$ 37$^{\prime\prime}$.62$\times$19$^{\prime\prime}$.20.
The four point radio spectrum have inverted shape with turnover between 610 MHz and 1.4 GHz. 
\subsection{NGC 7212}
VLA observations of NGC 7212 show a compact double source separated by $\sim$ 0$^{\prime\prime}$.7 in the 
north-south direction and the northern blob appears to be slightly elongated \citep{Falcke98}.
In our GMRT observations, the 610 MHz image shows two emission component. 
The radio contours overlaid on DSS optical image clearly show that the second radio component is associated with the neighboring galaxy NGC 7213.
The radio emission component associated with NGC 7212 is fitted with a single Gaussian of $\sim$ 8$^{\prime\prime}$.40$\times$4$^{\prime\prime}$.85. 
Due to large synthesized beam size ({\ie}$\sim$ 40$^{\prime\prime}$.16$\times$19$^{\prime\prime}$.61) the 240 MHz radio image 
shows only one emission component that includes emission from NGC 7213 as well.
The 240 MHz radio emission is fitted with a Gaussian of $\sim$ 32$^{\prime\prime}$.85$\times$25$^{\prime\prime}$.79.
The large beam size at 240 MHz does not allow us to estimate the flux densities of NGC 7212 and NGC 7213 separately.
Therefore, 240 MHz flux density measurement of NGC 7212 is contaminated and over-estimated.
The four point integrated radio spectral index of NGC 7212 is $\sim$ -0.79.
We note that spectral steepening seen over 610 MHz to 240 MHz is likely to be caused by over-estimated 240 MHz flux density.
It can also give rise the steeper value of the integrated spectral index than the actual value.
\subsection{NGC 7682}
8.4 GHz VLA observations of NGC 7682 show an unresolved point source nuclear emission \citep{Kukula95}, however, 
the milli-arcsec 5.0 GHz VLBI observations show a point source with extension along the south (P.A. $\sim$ 180$^\circ$) and the 
south-east (P.A. $\sim$ 120$^\circ$) direction \citep{Lal04}.
Our GMRT images at 610 MHz and 240 MHz are fitted with single Gaussian components of sizes $\sim$ 10$^{\prime\prime}$.21$\times$6$^{\prime\prime}$.82 
and $\sim$ 38$^{\prime\prime}$.68$\times$29$^{\prime\prime}$.23, respectively.
The four point spectrum of NGC 7682 is steep with spectral index (${\alpha}_{\rm int}$) $\sim$ -0.72.
\subsection{MRK 533}
VLA `C' array observations of MRK 533 at 8.4 GHz show a slightly extended nuclear emission 
which is seen resolved into 0.5 arcsec double structure in VLA `A' array observations \citep{Kukula95}. 
VLA and EVN observations reveal a linear triple radio source of $\sim$ 0$^{\prime\prime}$.7 angular extent with 
the components at $\sim$ 0$^{\prime\prime}$.5 west and  $\sim$ 0$^{\prime\prime}$.15 east of the main peak \citep{Unger88}. 
\cite{Momjian03} made more sensitive observations of MRK 533 using VLBA, phased VLA and Arecibo at 1.4 GHz and report 
the triple source and additional low-surface-brightness emission forming an S-shaped structure.
Our GMRT images at 610 MHz and 240 MHz show nearly point source radio emission which are fitted with single Gaussian components of sizes 
$\sim$ 6$^{\prime\prime}$.98$\times$6$^{\prime\prime}$.26 and $\sim$ 20$^{\prime\prime}$.01$\times$19$^{\prime\prime}$.04, respectively.
The radio spectrum of MRK 533 over 240 MHz to 5.0 MHz is fairly steep with spectral index (${\alpha}_{\rm int}$) $\sim$ -0.78.
\end{appendix} 
\bibliographystyle{aa}
\bibliography{myNov09}
\end{document}